\documentclass[11pt,letterpaper]{article}
\pdfoutput=1
\usepackage{jheppub}
\usepackage{color}
\usepackage{graphicx}
\usepackage{tabularx}
\usepackage{xspace}
\usepackage{empheq}
\usepackage{verbatim}
\usepackage{amsmath}
\usepackage{amssymb}
\usepackage[caption=false]{subfig}
\usepackage{url}
\usepackage{bbold}
\usepackage{slashed}
\usepackage{array}

\usepackage{multirow}
\usepackage{threeparttable}
\usepackage{paralist}
\usepackage{xspace}
\usepackage{upgreek}
\usepackage{lipsum}
\usepackage[T1]{fontenc}

\usepackage{tcolorbox}
\usepackage[framemethod=TikZ]{mdframed}
\usepackage[framemethod=TikZ]{mdframed}

\usepackage{setspace}

\usepackage{mathtools}

\allowdisplaybreaks

\newcommand\colvec[3][]{\begin{pmatrix}\ifx\relax#1\relax\else#1\\\fi#2\\#3\end{pmatrix}}

\newcommand{\beq}{\begin{equation}}
\newcommand{\beqn}{\begin{eqnarray}}
\newcommand{\eeq}{\end{equation}}
\newcommand{\eeqn}{\end{eqnarray}}

\DeclareRobustCommand{\Eq}[1]{Eq.~(\ref{#1})}
\DeclareRobustCommand{\Eqs}[2]{Eqs.~(\ref{#1}) and (\ref{#2})}

\DeclareRobustCommand{\Sec}[1]{Sec.~\ref{#1}}

\DeclareRobustCommand{\Fig}[1]{Fig.~\ref{#1}}

\newcommand{\be}{\begin{equation}}
\newcommand{\ee}{\end{equation}}

\def\nn{{\nonumber}}

\newcommand{\fd}[2]{\parbox{#1}{\includegraphics[width=#1]{#2}}}

\def\cB{\mathcal{B}}

\def\cM{\mathcal{M}}

\def\cO{\mathcal{O}}

\def\cP{\mathcal{P}}

\newcommand{\zcut}{z_\mathrm{cut}}

\newcommand{\SCETi}{\mbox{${\rm SCET}_{\rm I}$}\xspace}
\newcommand{\SCETii}{\mbox{${\rm SCET}_{\rm II}$}\xspace}

\newcommand{\cusp}{\mathrm{cusp}}

\newcommand{\df}{\mathrm{d}}

\newcommand\bn{{\bar n}}

\newcommand{\la}{\lambda}

\newcommand{\bea}{\begin{eqnarray}}
\newcommand{\eea}{\end{eqnarray}}

\newcommand{\aW}{\alpha_{\scriptscriptstyle W}}
\newcommand{\TaW}{\tilde{\alpha}_{\scriptscriptstyle W}}
\newcommand{\mW}{m_{\scriptscriptstyle W}}

\newcommand{\thetaW}{\theta_{\scriptscriptstyle W}}
\newcommand{\sW}{s_{\scriptscriptstyle W}}

\newcommand{\id}{\mathbf{1}}

\newcommand{\eq}[1]{Eq.~\eqref{eq:#1}}
\newcommand{\eqs}[2]{Eqs.~\eqref{eq:#1} and \eqref{eq:#2}}

\begin{document}

\preprint{\vbox{\hbox{LA-UR-18-25972}\hbox{MIT-CTP 5025}}}

\title{
Precision Photon Spectra for Wino Annihilation
}

\author[1]{\small Matthew Baumgart,}
\author[2]{\small Timothy Cohen,}
\author[3]{\small Emmanuel Moulin,}
\author[4,5]{\small Ian Moult,}
\author[3]{\small Lucia Rinchiuso,}
\author[4,5,6]{\small \\[3pt] Nicholas L. Rodd,}
\author[6]{\small Tracy R. Slatyer,}
\author[6]{\small Iain W. Stewart,}
\author[7]{\small and Varun Vaidya}

\affiliation[1]{\footnotesize Department of Physics, Arizona State University, Tempe, AZ 85287, USA}
\affiliation[2]{\footnotesize Institute of Theoretical Science, University of Oregon, Eugene, OR 97403, USA}
\affiliation[3]{\footnotesize IRFU,\! CEA,\! D\'epartement de Physique des Particules, Universit{\'{e}} Paris-Saclay,\! F-91191\! Gif-sur-Yvette,\! France}
\affiliation[4]{\footnotesize Berkeley Center for Theoretical Physics, University of California, Berkeley, CA 94720, USA}
\affiliation[5]{\footnotesize Theoretical Physics Group, Lawrence Berkeley National Laboratory, Berkeley, CA 94720, USA}
\affiliation[6]{\footnotesize Center for Theoretical Physics, Massachusetts Institute of Technology, Cambridge, MA 02139, USA}
\affiliation[7]{\footnotesize Theoretical Division, MS B283, Los Alamos National Laboratory, Los Alamos, NM 87545, USA}

\date{\today}

\abstract{We provide precise predictions for the hard photon spectrum resulting from neutral SU$(2)_W$ triplet (wino) dark matter annihilation.  Our calculation is performed utilizing an effective field theory expansion around the endpoint region where the photon energy is near the wino mass.  This has direct relevance to line searches at indirect detection experiments. We compute the spectrum at next-to-leading logarithmic (NLL) accuracy within the framework established by a factorization formula derived previously by our collaboration.  This allows simultaneous resummation of large Sudakov logarithms (arising from a restricted final state) and Sommerfeld effects. Resummation at NLL accuracy shows good convergence of the perturbative series due to the smallness of the electroweak coupling constant -- scale variation yields uncertainties on our NLL prediction at the level of $5\%$. We highlight a number of interesting field theory effects that appear at NLL associated with the presence of electroweak symmetry breaking, which should have more general applicability. We also study the importance of using the full spectrum as compared with a single endpoint bin approximation when computing experimental limits.  Our calculation provides a state of the art prediction for the hard photon spectrum that can be easily generalized to other DM candidates, allowing for the robust interpretation of data collected by current and future indirect detection experiments.}

\maketitle
\setcounter{page}{2}

\section{Introduction}

Indirect detection is critical to the hunt for multi-TeV WIMP dark matter (DM). New data are continually being collected by current experiments, \emph{e.g.}~H.E.S.S.~\cite{Hinton:2004eu,Abramowski:2013ax,Abdallah:2018qtu}, HAWC~\cite{Sinnis:2004je,Harding:2015bua,Pretz:2015zja}, VERITAS~\cite{Weekes:2001pd,Holder:2006gi,Geringer-Sameth:2013cxy}, and MAGIC~\cite{FlixMolina:2005hv,Ahnen:2016qkx}, and a number of dedicated line searches for photons have been performed~\cite{Abramowski:2011hc,Abdallah:2018qtu}. Future experiments such as CTA~\cite{Consortium:2010bc,Acharya:2017ttl} will provide even greater sensitivity. Deriving the experimental ramifications these data will have on the parameter space of DM models requires precise predictions for the hard photon spectrum.  Due to finite resolution effects inherent to the relevant experiments, a reliable prediction for not only the rate but also the shape of the spectrum is required to derive robust comparisons between theory and experiment~\cite{Baumgart:2017nsr}.

The annihilation of TeV-scale DM is a multi-scale problem which is amenable to the application of
effective field theory (EFT) techniques. In particular, non-relativistic EFTs can be used to treat the annihilating DM, and Soft-Collinear Effective Theory (SCET)~\cite{Bauer:2000ew,Bauer:2000yr, Bauer:2001ct, Bauer:2001yt} can be used for the final state radiation. The combination of these two EFTs~\cite{Baumgart:2014vma,Bauer:2014ula,Ovanesyan:2014fwa,Baumgart:2014saa,Baumgart:2015bpa,Ovanesyan:2016vkk} allows for the simultaneous resummation of Sudakov logarithms $\aW^n\log^{m}\big(M_\chi/\mW\big)$, with $m\leq 2\,n-1$ in the differential spectrum~\cite{Hryczuk:2011vi}, and Sommerfeld enhancement effects $\big( \aW M_\chi/\mW \big )^k$~\cite{Hisano:2003ec,Hisano:2004ds,Cirelli:2007xd,ArkaniHamed:2008qn,Blum:2016nrz}. In~\cite{Baumgart:2017nsr} we extended these EFT approaches to allow for the calculation of the hard photon spectrum in the endpoint region, where the photon energy $E_\gamma$ is near the DM mass $M_\chi$, as is relevant for line searches. Our framework additionally allows for the resummation of resolution effects $\aW^n\log^m\big(1-z\big)$ with $m\leq 2\,n-1$, where $z= E_{\gamma}/M_{\chi}$. Such logarithms are directly related to the experimental energy resolution, since $z$ quantifies the distance from the exclusive case, given by a line at $z=1$. A finite experimental resolution smears photons with a small $1-z$ into the expected exclusive event rate, and our calculation is able to realistically incorporate such effects.

\begin{figure}
\begin{center}
\includegraphics[width=0.65\columnwidth]{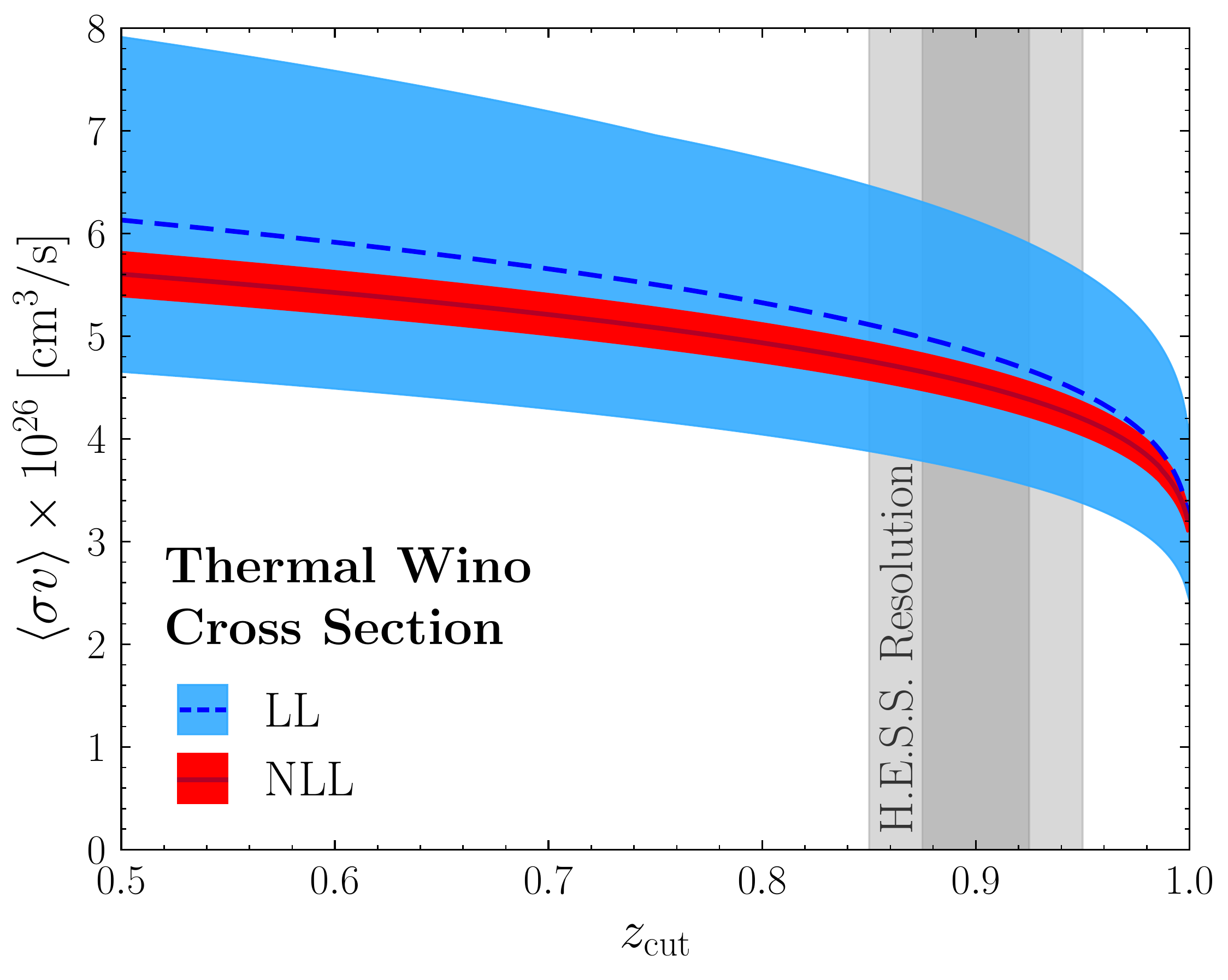}
\end{center}
\vspace{-15pt}
\caption{The cross section for a thermal wino, \emph{i.e.}, with mass $M_\chi=2.9$ TeV~\cite{Beneke:2016ync}, as a function of resolution parameter $z_{\rm cut}$, showing our results at both LL and NLL. The NLL calculation significantly reduces uncertainties as compared to LL. A region appropriate for the H.E.S.S. experimental resolution, which is $\sim10\%$ at these energies~\cite{deNaurois:2009ud}, is shown in the grey band. This band is representative of the range of values that will contribute when our spectrum is convolved with the H.E.S.S. energy resolution function.
}
\label{fig:intro_fig}
\end{figure}

In this paper we extend this result to next-to-leading logarithmic (NLL) accuracy. To understand the importance of including the NLL corrections, we note that for the situation of interest here the logarithms $L$ become large enough so that $L^2\sim 1/\aW$. A leading logarithmic (LL) calculation then captures all terms scaling as $1$, and so should provide a good description of the shape of the distribution. However, a LL calculation does not probe higher order radiative corrections, and therefore typically has large uncertainties. On the other hand, an NLL calculation captures the first radiative corrections scaling like $\aW L$, and therefore typically provides a large reduction of the theoretical uncertainties. This reduction of uncertainties is clearly illustrated in \Fig{fig:intro_fig}, which shows a comparison of our earlier LL calculation with the NLL result achieved here. With NLL accuracy, the theory uncertainties become a subdominant contribution to the total uncertainty relevant experimentally.

While our calculational framework is generally applicable to any heavy WIMP candidate, we will often specify to the case where the DM candidate is a wino, the neutral component of a triplet of SU$(2)_W$ with zero hypercharge. In addition to allowing us to illustrate our formalism, the wino is well motivated phenomenologically (see \emph{e.g.}~\cite{Giudice:1998xp,Randall:1998uk,ArkaniHamed:2004fb, ArkaniHamed:2004yi, Giudice:2004tc, Wells:2004di, Pierce:2004mk, Arvanitaki:2012ps, ArkaniHamed:2012gw, Hall:2012zp}), making our results of interest to current experiments. In a companion paper~\cite{Rinchiuso:2018ajn} we have used these results in a realistic H.E.S.S. forecast analysis to study the impact of having a complete description of the  shape of the photon spectrum for wino searches. Here we provide the details of our NLL calculation, and perform a numerical study demonstrating that the theoretical uncertainty is significantly reduced when compared with the LL result, achieving an uncertainty from higher order corrections at the level of $5\%$. 
The cumulative cross section for $z\geq \zcut$ is shown in \Fig{fig:intro_fig} for a wino with a mass of $2.9$ TeV, which corresponds to the case where the thermal relic density matches the measured one~\cite{Beneke:2016ync}.  Here $z_{\rm cut}$ restricts the cross section by allowing only photons with $z\ge \zcut$, see \Eq{eq:cumulative_def_intro}. We additionally show a band depicting the approximate values of $\zcut$ that correspond to the H.E.S.S. energy resolution.  As can be seen, our calculation significantly reduces errors associated with the particle physics component of the annihilation cross section, which allows for the robust interpretation of experimental results in terms of DM model and astrophysical parameters.    We also study the importance of using the full spectrum as compared with a single endpoint bin approximation when computing experimental limits by performing a mock H.E.S.S. analysis, and using the results of our forecasted H.E.S.S.~limits~\cite{Rinchiuso:2018ajn}.  We find that the use of the full spectrum near the endpoint is crucial to preserve the desired accuracy, emphasizing the importance of our EFT formalism.

Although the primary goal of this work is to provide a precision prediction for heavy WIMP annihilation in the endpoint region, a number of interesting features of SCET with broken gauge symmetry that have not previously appeared in the literature arise in our NLL calculation, and we devote several sections to their discussion.

An outline of this paper is as follows. In \Sec{sec:review} we review the structure of the factorization formula for the endpoint region derived in \cite{Baumgart:2017nsr}, and prove that it remains valid at NLL accuracy. In \Sec{sec:NLL} we discuss the necessary formalism for achieving resummation at NLL accuracy, give explicit results for all one-loop anomalous dimensions, and solve the relevant renormalization group (RG) equations. In \Sec{sec:answer} we present analytic results for the cumulative and differential spectra at NLL accuracy, and comment on some interesting aspects of their structure. Numerical results and a study of the related theoretical uncertainties are given in \Sec{sec:results}. In \Sec{sec:compare} we compare the use of the full spectrum with a single endpoint bin in a mock H.E.S.S. analysis and with our forecasted limits, emphasizing the importance of having a complete description of the shape of the photon spectrum in the endpoint region. We conclude in \Sec{sec:conc}. In App.~\ref{app:explicitxsec} we collect all ingredients required for the cumulative and differential spectra, which are otherwise scattered throughout the paper. Finally in App.~\ref{app:intbrem} we demonstrate that our result matches existing fixed order calculations in the appropriate limit, and briefly comment on how the internal bremsstrahlung contribution, which is widely discussed in the literature, is reproduced in our result.

\section{Factorization Formula for the Endpoint Region}\label{sec:review}

\subsection{Factorization at LL order} 

In this section we provide a brief review of the factorization formula used to describe the photon spectrum in the endpoint region at LL order. A complete description including all notational conventions followed here, as well as a derivation of the formula through a multi-stage matching procedure, can be found in~\cite{Baumgart:2017nsr}. 

\begin{figure}
\begin{center}
\subfloat[]{\label{fig:csoft_intro_a}
\includegraphics[width=5.5cm]{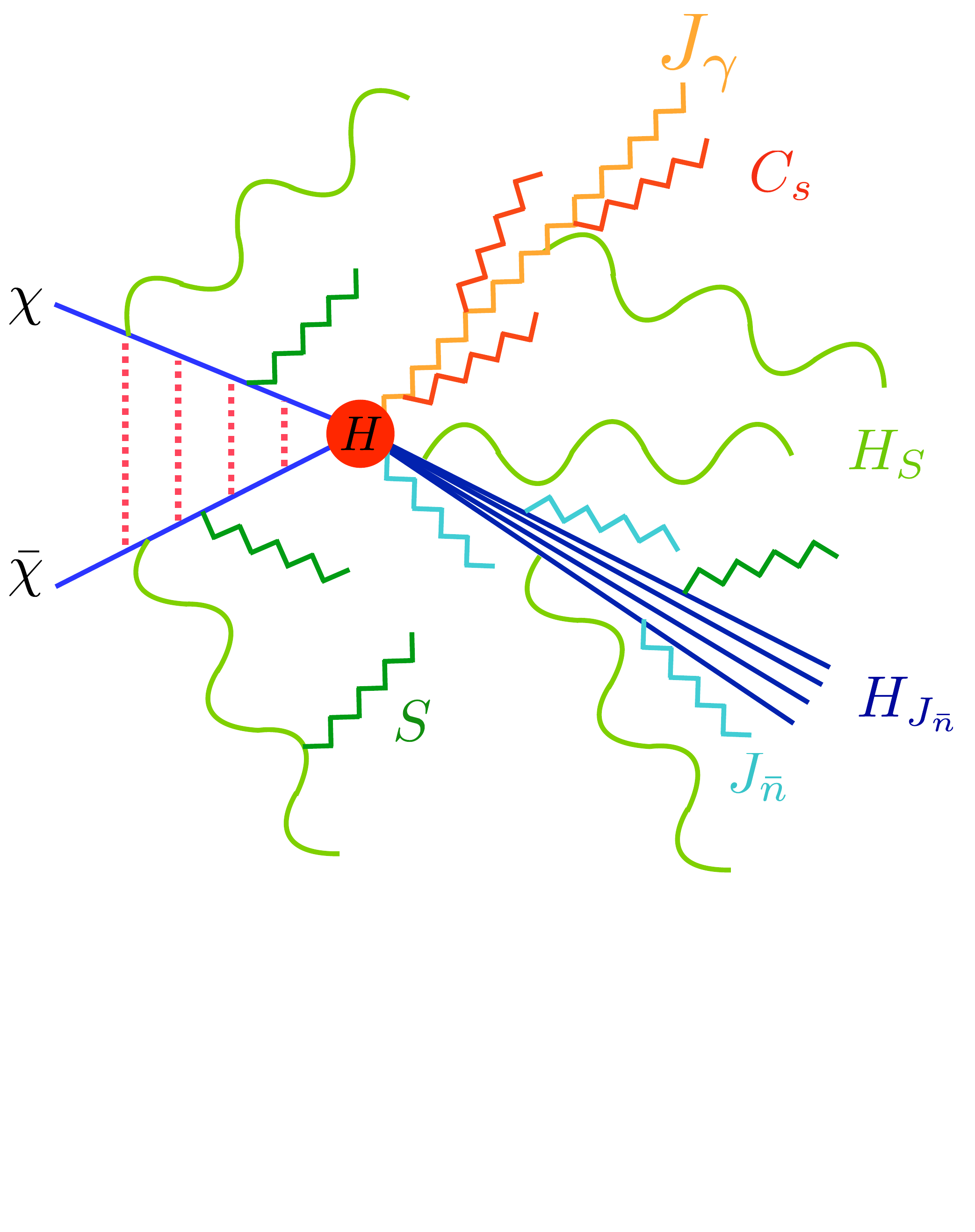}    
}\hspace{40pt}
\subfloat[]{\label{fig:csoft_intro_b}
\includegraphics[width=7.5cm]{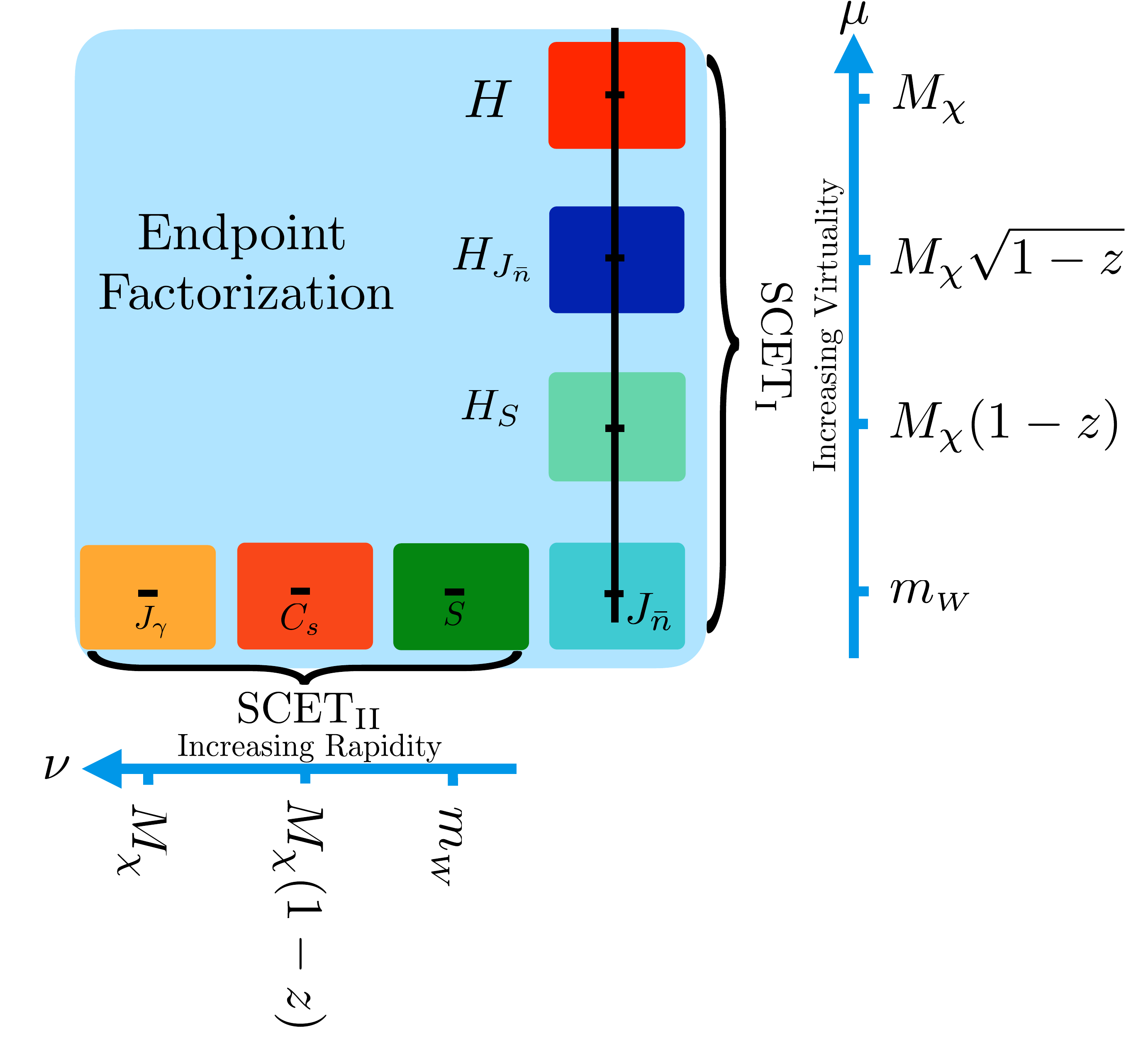}
}
\end{center}
\vspace{-0.2cm}
\caption{A review of the factorization formula derived in \cite{Baumgart:2017nsr}. (a) A schematic depiction of the relevant modes. (b) The virtuality and rapidity of the different modes.
}
\label{fig:csoft_intro}
\end{figure}

The essential process of interest is $\chi\,\chi \rightarrow \gamma \,X$, where $\chi$ is the DM which annihilates to a hard photon $\gamma$, which is detected by the experiment, and additional radiation $X$.  We will characterize the photon energy spectrum with the dimensionless variable 
\begin{align}
z = \frac{E_\gamma}{M_\chi} \in [0,1]\,.
\end{align}
We will be interested both in the differential spectra $\text{d} \sigma/\text{d}z$ as a function of $z$, as well as the cumulative spectra as a function of $\zcut$
\begin{align}\label{eq:cumulative_def_intro}
\sigma(\zcut) = \int\limits_{\zcut}^{1}\text{d}z\, \frac{\text{d} \sigma}{\text{d}z}\,,
\end{align}
which is shown in \Fig{fig:intro_fig}. In the fully exclusive case $(z=1)$, only two bosons are produced in the final state.  This is the relevant configuration for a line search with 
perfect
resolution. However, given the finite energy resolution of real experiments,  the region relevant for line searches is the so-called endpoint region, characterized by an energy resolution $M_\chi (1-\zcut) \ll M_\chi$~\cite{Baumgart:2017nsr}. In this case, additional radiation beyond two bosons is present in the final state. To be near $z=1$, this radiation must be either low energy, or collimated along the direction of the boson recoiling against the detected photon. This configuration is illustrated in \Fig{fig:csoft_intro_a}. The collimated spray of radiation is referred to as a jet. Due to the phase space restrictions, large logarithms, $\log(M_\chi/\mW)$ and $\log(1-z)$, appear in the perturbative calculation of the spectrum. In this paper we focus on improving the precision of the calculation relevant to the endpoint region.

Our approach combines a number of different EFTs including Non-Relativistic DM and SCET as in~\cite{Baumgart:2014vma,Bauer:2014ula,Ovanesyan:2014fwa}, but with additional formalism to describe the photon endpoint energy spectrum. The formalism to describe the endpoint region developed in~\cite{Baumgart:2017nsr} makes use of the limit $\mW \ll M_\chi(1-z)\ll M_\chi$. This enables us to factorize the cross section into a number of 
different functions, each describing the dynamics at a particular scale. Log enhanced contributions to the cross section can then be resummed using
RG
techniques. The resulting resummed functions are then recombined to produce a precise prediction for the cross section. In particular, since the DM velocity $v/c\sim 10^{-3}\ll 1$ in the DM halo, it is appropriate to use a non-relativistic DM EFT (analogous to NRQCD \cite{Caswell:1985ui,Bodwin:1994jh,Luke:1999kz}) to describe the annihilating initial state.\footnote{The NRDM formalism has also been applied to the scattering of DM with nucleon targets~\cite{Fan:2010gt,Hill:2011be,Fitzpatrick:2012ix,Hill:2013hoa}.} For the radiation in the final state, we use SCET~\cite{Bauer:2000ew,Bauer:2000yr, Bauer:2001ct, Bauer:2001yt}, including its generalizations to massive gauge bosons~\cite{Chiu:2007yn,Chiu:2008vv,Chiu:2007dg}, and its multi-scale extensions \cite{Bauer:2011uc,Larkoski:2014tva,Procura:2014cba,Larkoski:2015zka,Pietrulewicz:2016nwo} (although we will not emphasize it in detail, we use a combination of $\SCETi$ and $\SCETii$ \cite{Bauer:2002aj}).

As is well known, the Sommerfeld effect is relevant to heavy WIMP annihilation~\cite{Hisano:2003ec,Hisano:2004ds,Cirelli:2007xd,ArkaniHamed:2008qn,Blum:2016nrz}.  In our framework, the Sommerfeld effect can be factorized out of the cross section using
\begin{align}\label{eq:cross1}
\frac{\text{d}\sigma}{\text{d}z} = \sum_{a'b'ab} F^{a'b'ab}\,\frac{\text{d}\hat{\sigma}^{a'b'ab}}{\text{d}z}\,,
\end{align}
where it is captured by the matrix elements
 \begin{align}\label{eq:Lfunction}
F^{a'b'ab} = \Big\langle \big(\chi^0 \chi^0\big)_S \Big| \big( \chi_v^{a' T}\, i\hspace{0.6pt} \sigma_2\, \chi^{b'}_v\big)^\dagger \Big|0 \Big\rangle \Big\langle 0 \Big| \big(\chi_v^{aT}\, i\hspace{0.6pt} \sigma_2\, \chi_v^b \big) \Big|\big(\chi^0 \chi^0\big)_S \Big\rangle \,.
\end{align}
In particular, we will require the following two matrix elements
\begin{align} \label{eq:wavefunction}
 &  \Big\langle 0 \Big|\, \chi_v^{3\,T}\, i\hspace{0.6pt}\sigma_2 \,\chi_v^{3\,\,} \,\Big| \big(\chi^0 \chi^0\big)_S \Big\rangle = 4\, \sqrt{2} \, M_\chi\, s_{00} \,, \\[5pt]
 & \Big\langle 0 \Big|\, \chi_v^{+T}\, i\hspace{0.6pt}\sigma_2 \,\chi_v^-\, \Big| \big(\chi^0 \chi^0\big)_S \Big\rangle= 4\, M_\chi \,s_{0\pm} \,,\nn
\end{align}
where $\chi^0 = \chi^3$ and $\chi^\pm = \big(\chi^1 \mp i \hspace{0.6pt}\chi^2\big)/\sqrt{2}$\,.
These matrix elements are obtained by numerically solving the associated Schr\"odinger problem using the approach detailed in App. A of~\cite{Cohen:2013ama}. 
We include a mass splitting, which for winos we take to be $\delta = M_{\chi^\pm} - M_{\chi^0} \simeq 164.4 \text{ MeV}$~\cite{Ibe:2012sx}.

The focus of this paper is on extending the description of the radiation in the final state to a higher perturbative order.   The starting point is the LL factorization theorem derived in~\cite{Baumgart:2017nsr}: 
\begin{align}
\label{eq:factorization}
\left(\frac{\text{d} \hat{\sigma}}{\text{d}z}\right)^{\text{LL}}=~&H\big(M_\chi, \mu\big)\, J_\gamma\big(\mW,\mu, \nu\big)\, J_\bn\big(\mW,\mu, \nu\big)\, S \big(\mW,\mu, \nu\big) \nn \\[3pt]
&\times H_{J_\bn}\big(M_\chi, 1-z,\mu\big) \otimes H_{S}\big(M_\chi, 1-z,\mu\big) \otimes C_S\big(M_\chi, 1-z,\mW,\mu, \nu\big)  \,,\\[-10pt] \nn
\end{align}
where the dependence on the gauge indices $a'b'ab$ is suppressed. 
Here we have explicitly added a superscript LL to emphasize that in \cite{Baumgart:2017nsr} it was only shown that this factorization holds in this form at LL order. However, in the next section we will show that this formula holds also at NLL order, so that this higher order calculation can be achieved by improving the precision of the individual functions in \eq{factorization}.  Here $\mu$ and $\nu$ are the virtuality and rapidity scales 
respectively, and we have defined the following shorthand notation for the convolution in the $z$ variable
\begin{align}
F(1-z)\otimes G(1-z)=\int \!\text{d}z_1 \,\text{d}z_2 \, \delta (1+z-z_1-z_2)\, F(1-z_1)\, G(1-z_2)\,.
\end{align}
Operator definitions of the relevant functions  will be given in \Sec{sec:NLL} along with their anomalous dimensions. Here we restrict ourselves to providing a physical description of the different functions that appear in the factorization formula in \Eq{eq:factorization}. This factorization, along with the physical radiation described by each of the functions appearing in the factorization formula, is illustrated in  \Fig{fig:csoft_intro_b}. The functions appearing in the factorization formula naturally divide up into two groups, those describing the dynamics above the electroweak symmetry breaking scale, which can be evaluated in the unbroken theory: 
\begin{itemize}
\setlength{\itemindent}{-1.4em}
\item[$\bullet$] $H\big(M_\chi, \mu\big)$: this hard function captures virtual corrections for $\chi\, \chi \to \gamma\, \gamma, \gamma\,Z$.  
\item[$\bullet$] $H_{J_\bn}\big(M_\chi, 1-z,\mu\big)$: this jet function captures the collinear radiation at the scale $M_\chi \sqrt{1-z}$.
\item[$\bullet$] $H_{S}\big(M_\chi, 1-z,\mu\big)$: this soft function captures soft radiation at the scale $M_\chi (1-z)$. \end{itemize}
and those that depend on the gauge boson masses and must be evaluated in the broken theory:
\begin{itemize}
\setlength{\itemindent}{-1.4em}
\item[$\bullet$] $J_\gamma\big(\mW,\mu, \nu\big)$: this photon jet function captures the virtual corrections to the outgoing $\gamma$. 
\item[$\bullet$] $S\big(\mW,\mu, \nu\big)$: this soft function captures wide angle soft radiation at the electroweak scale. 
\item[$\bullet$] $J_\bn\big(\mW,\mu, \nu\big)$:  this jet function captures collinear radiation in $X$ at the electroweak scale.  
\item[$\bullet$] $C_S\big(M_\chi, 1-z,\mW,\mu, \nu\big)$:  this collinear-soft function captures radiation along the  $\gamma$ direction.
\end{itemize}
Overlap between the different functions is removed using the zero bin procedure \cite{Manohar:2006nz}.

 In \cite{Baumgart:2017nsr} each of these functions was computed to LL accuracy, and the consistency of the factorization was shown at that order.  After showing that this factorization also holds at NLL, we will calculate each of these functions to NLL accuracy.

\subsection{Validity of the Factorization at NLL}\label{sec:no_NGL}

Various aspects of the LL factorization formula derived in~\cite{Baumgart:2017nsr} and displayed in \eq{factorization} obviously 
remain valid at NLL (and higher) orders. This includes the factorization of hard-collinear and soft-collinear modes in the intermediate EFT, as well as the refactorization of the jet sector to separate $\mW\ll M_\chi \sqrt{1-z}$. This suffices to define the functions $H$, $J_\gamma$, $J_\bn$, and $H_{J_\bn}$ appearing in \eq{factorization}, but leaves a soft function $S'$ that is not fully factorized.  Thus the key non-trivial aspect of the factorization formula in \eq{factorization} is the refactorization of the soft function $S'$, which describes low energy (soft) radiation. 

\begin{figure}[t!]
\begin{center}
\subfloat[]{\label{fig:soft_refactorization_b}
\includegraphics[trim=0cm -5cm 0cm 0cm,clip=true,width=6.0cm]{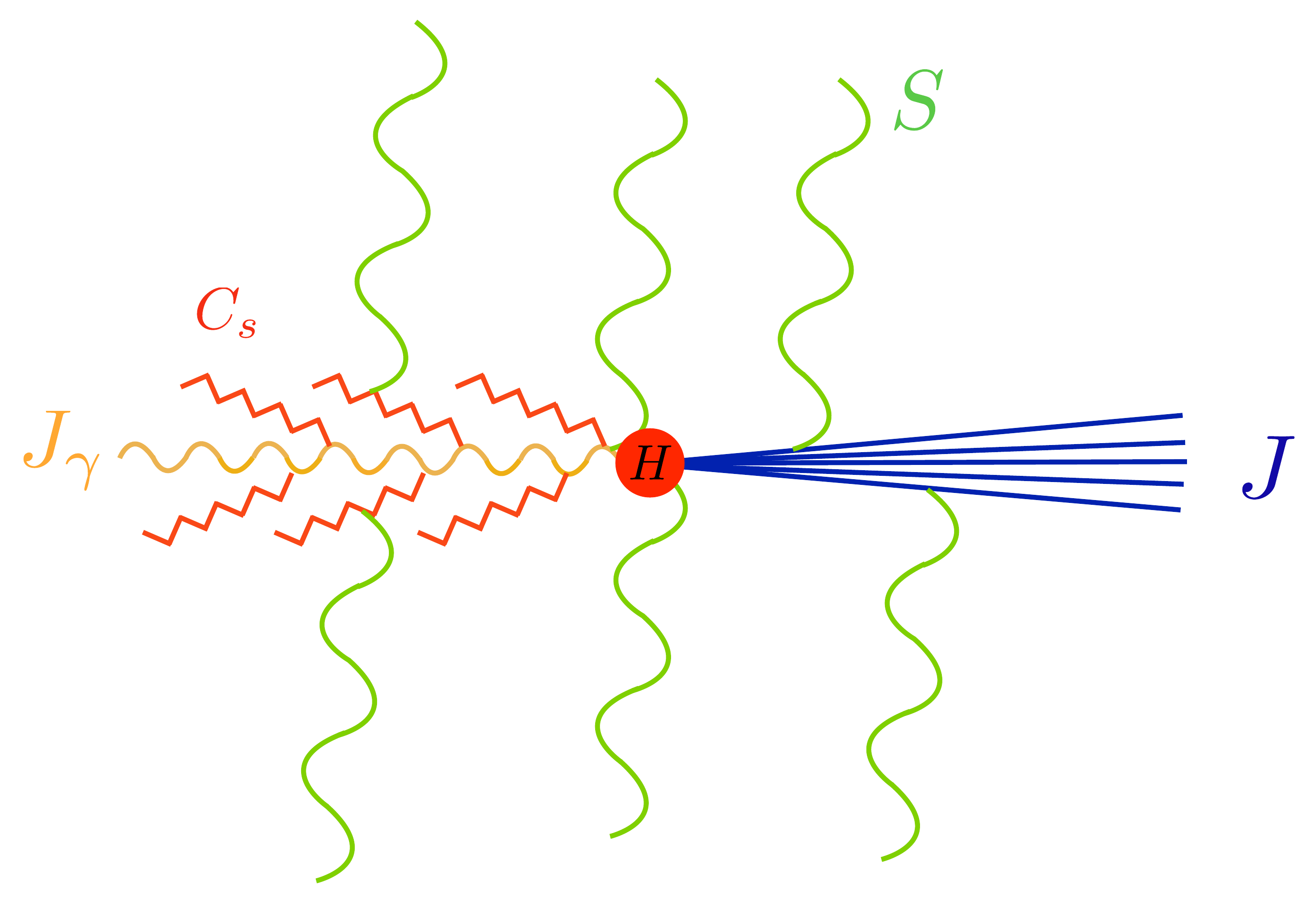}
}\hspace{5pt}
\subfloat[]{\label{fig:soft_refactorization_a}
\includegraphics[width=8.2cm]{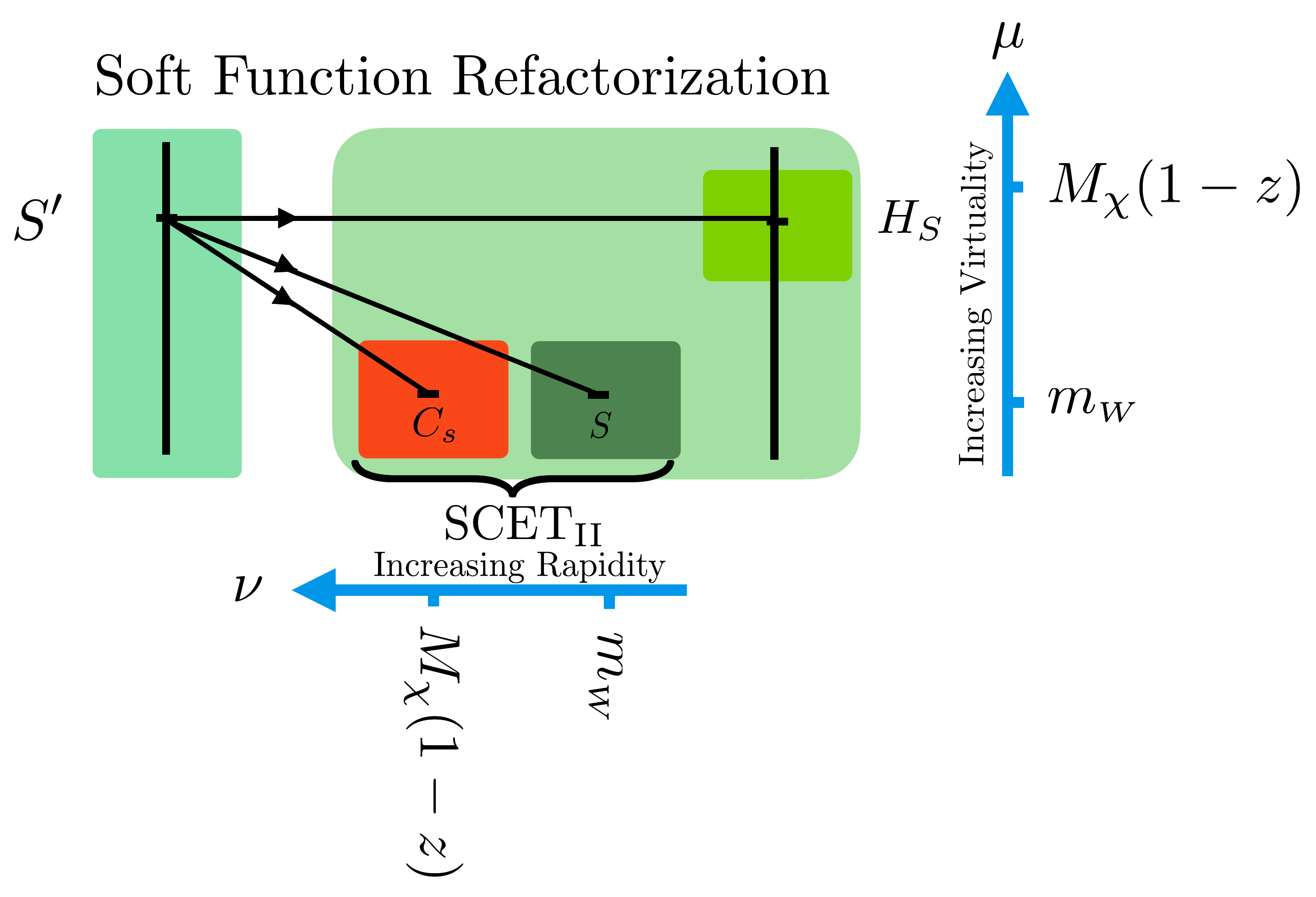}    
}
\end{center}
\caption{
A review of the refactorization of the soft function derived at LL accuracy in \cite{Baumgart:2017nsr}. In this paper we prove that this refactorization is also valid at NLL. (a) A schematic depiction of the refactorization of the soft function into different modes, illustrating their physical interpretation. 
(b) The virtuality and rapidity of the different modes appearing in the factorization formula.
}
\label{fig:soft_refactorization}
\end{figure}

In  \cite{Baumgart:2017nsr} it was shown that $S'$ could be factorized at LL into a hard matching coefficient $H_S$, a collinear soft function $C_S$ (describing soft radiation collimated along the direction of the photon), and a soft function $S$. Explicitly, 
\begin{align}\label{eq:genrefactorize}
S^{\prime \ aba'b'}_{i} \big(M_\chi, 1-z, \mW,\mu,\nu \big) = H_{S, ij}\big(M_\chi, 1-z, \mu\big) \,& \Big[ C_{S}\big(M_\chi, 1-z, \mW , \mu, \nu\big)  S\big( \mW, \mu\big) \Big]^{ab a'b'}_j \nn \\[5pt]
&\times \left[ 1+ \cO\left( \frac{\mW}{M_\chi (1-z)} \right) \right ]\,.
\end{align}
This refactorization is shown schematically in \Fig{fig:soft_refactorization}. It involves splitting the soft contributions into modes for $S$ at the scale $\mW$ and modes for $H_S$ at the scale $M_\chi (1-z)$.  These two modes are only separated in energy, but have no angular hierarchy, while in addition there are collinear-soft modes at the same energy scale as $S$ but at a different angle.  This causes a potential issue when proving \Eq{eq:genrefactorize},
which no longer follows from a standard soft-collinear, hard-collinear, or hard-soft factorization. 

\begin{figure}[t!]
\begin{center}
\includegraphics[width=\columnwidth]{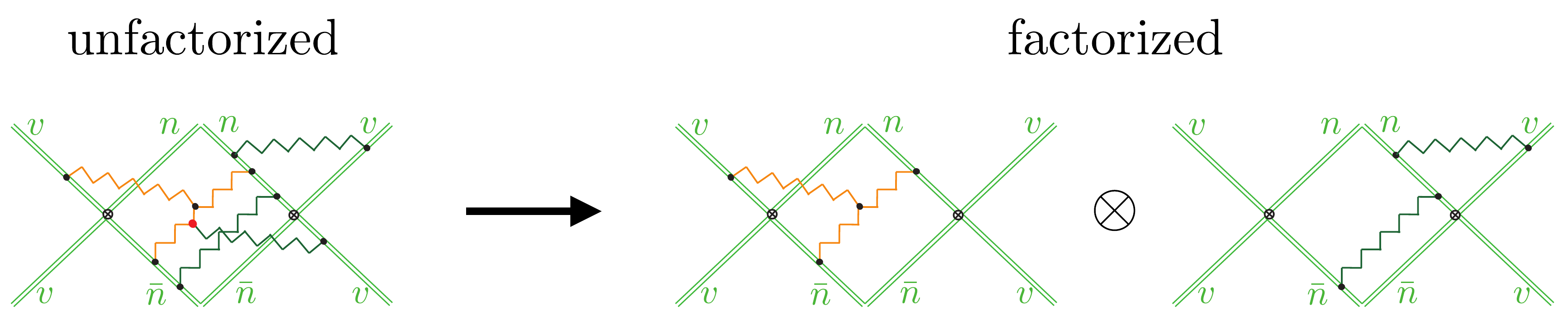}
\end{center}
\vspace{-15pt}
\caption{An illustration of the factorization of soft radiation at the scale $\mW$ (dark green) from those at the scale $M_\chi (1-z)$ (orange). To prove this factorization at NLL, one must show that subgraphs involving interactions between the radiation at the two scales, as illustrated by the red vertex in the unfactorized graph, do not contribute logarithms at this order.
}
\label{fig:soft_refac_general}
\end{figure}

More precisely, soft emissions at the scale $M_\chi (1-z)$ are much more energetic than those at the scale $\mW$, and therefore behave as eikonal sources for bosons at the scale $\mW$. Any boson radiated at the scale $M_\chi (1-z)$ can radiate many bosons at the scale $\mW$. An example of such a graph is shown by the red vertex in \Fig{fig:soft_refac_general}. From the perspective of the radiation at the scale $\mW$, it then appears as though additional Wilson lines are present, beyond those in the original soft function, and one might be worried that more complicated Wilson line operators are required to capture all effects to NLL accuracy. This would imply that the simple factorized formula of \Eq{eq:factorization} would need to be extended to go beyond LL order. It is known that this occurs in other cases, for example in the study of non-global logarithms (NGLs) \cite{Dasgupta:2001sh} and their factorization \cite{Larkoski:2015zka,Becher:2015hka,Becher:2016mmh,Larkoski:2016zzc}. However, we will see that this behavior does not occur in the case studied here, 
and more complicated soft functions are not required at NLL.

We will show the validity of \Eq{eq:genrefactorize} at NLL through an explicit calculation. The refactorization in \Eq{eq:genrefactorize} asserts that one can describe the radiation by two separate soft functions, one describing radiation at the scale $M_\chi (1-z)$ and one describing radiation at the scale $\mW$. A particular example of this factorization is shown in \Fig{fig:soft_refac_general}. To show that this is valid at NLL, one must show that there cannot be a logarithm arising from a graph where the bosons at the scale $\mW$ are color entangled with those at the scale $M_\chi (1-z)$. (In this section we use the word ``color'' to refer to gauge index structure in the electroweak theory.) With two emissions, this corresponds physically to one emission at the scale $M_\chi (1-z)$ and one at the scale $\mW$ that did not factorize, as illustrated by the red vertex in \Fig{fig:soft_refac_general}. More precisely, this is the non-abelian component of the graphs, since eikonal emissions factor for the sum of abelian diagrams. Since the soft function is inclusive over radiation at the scale $\mW$, one would naively expect that such logarithms do not exist, because of the cancellation between real and virtual corrections. However, due to the presence of electroweak charged initial and final state particles, the cancellation of IR divergences is not guaranteed by the Kinoshita-Lee-Nauenberg (KLN) theorem \cite{Bloch:1937pw,Kinoshita:1962ur,Lee:1964is}.  KLN violating effects, which arise due to a mismatch in color  structures between the real and virtual corrections leading to an incomplete cancellation \cite{Ciafaloni:1998xg,Ciafaloni:1999ub,Ciafaloni:2000df} could interfere with this argument. However, we can directly show that such logarithms do not exist by performing a two-loop calculation of the soft function in the strongly ordered limit, with one boson at the scale $\mW$ and one at the scale $M_\chi (1-z)$. The strongly ordered limit is sufficient to detect logarithms.

We now consider the calculation of the two loop soft function in the strongly ordered limit. We will separately consider two classes of diagrams: the triple gauge boson vertex and the independent emissions.  We will show that for each class of diagrams contributing to a given soft operator, most of the real and virtual diagrams cancel pairwise.  This is the case for all the triple gauge boson vertex diagrams as we discuss below.  For the independent emission diagrams, the non-singlet nature of the soft function implies that the color structure of the real and the corresponding virtual diagram is not always the same.
However, we still find that the non-abelian piece of the color structure cancels out, as is required to show the validity of the factorization in \Eq{eq:genrefactorize} at NLL.

To demonstrate the cancellation, we present explicit results for a particular soft operator, $S'_2$ (whose precise definition is given in \Eq{eq:SprimeDefs}), since it includes the most general structure involving all three types of Wilson lines in the $n$, $\bar n$, and $v$ directions. In our calculation, we will denote the two loop momenta by $k_1$ and $k_2$, with $k_1 \sim M_{\chi}(1-z)(1, 1, 1) $ being the harder and $k_2 \sim \mW(1, 1, 1)$ the softer boson.

\vspace{10pt}
\noindent \textbf{Triple Gauge Boson Vertex Diagrams:}

\begin{figure}
\hspace{20pt}\subfloat[]{\label{fig:soft_a}
\includegraphics[width=5.7cm]{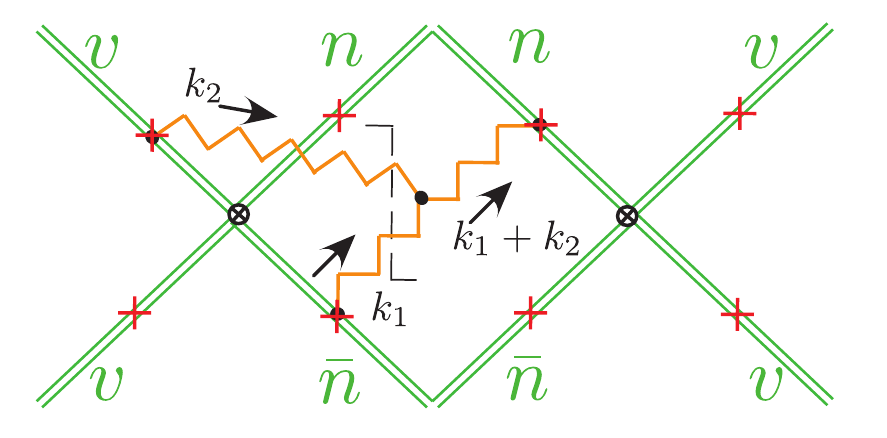}    
}\hspace{60pt}
\subfloat[]{\label{fig:soft_b}
\includegraphics[width=5.5cm]{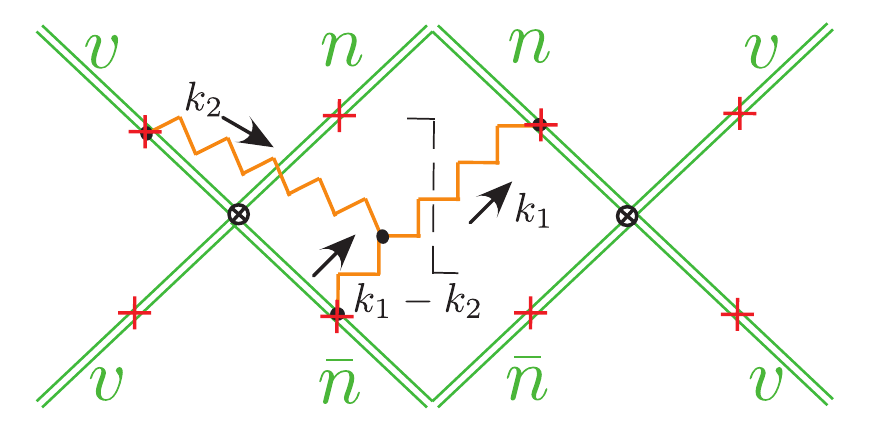}
}\hspace{20pt}
\caption[1]{Examples of real (a) and virtual (b) diagrams with a triple gauge boson vertex insertion contributing to the two loop soft function. Here the boson with momentum $k_1$ is at the scale $M_\chi (1-z)$, while $k_2$ is at the scale $\mW$.}
\label{fig:tripleNGLs} 
\end{figure}

We begin by considering diagrams with the triple gauge boson vertex. Representative real and virtual diagrams are shown in Fig.~\ref{fig:tripleNGLs}. These two diagrams are obtained simply by shifting the cut, and therefore their color structures are identical. In these diagrams, the higher energy particle has eikonalized from the perspective of the 
lower
energy particle, and therefore this is precisely the type of diagram that one could worry generates an operator with an additional Wilson line structure. However, since we are fully inclusive over the low energy radiation, and the color structures of the real and virtual diagrams are identical, we will see that the real and virtual graphs cancel, and no logarithm is generated.

The cancellation in the strongly ordered limit is easily checked. Defining the real integrand as $I_R$ and the virtual integrand as $I_V$, and taking the strongly ordered limit, we find 
\begin{align}
I_{R} &=\frac{F(k_1)\,2\,\pi\, \theta\big(k_2^0\big)\,\delta\big(k_2^2 -M_\chi^2\big)\, \delta\big(k_1^2\big)\, \delta\big( n\cdot q-n\cdot k_1\big)}{\big(\bar n \cdot k_2 + n \cdot k_2-i\hspace{0.4pt} \epsilon\big) \big(\bar n \cdot k_1+ i\hspace{0.4pt}\epsilon\big) \big(n \cdot k_1 +i\hspace{0.4pt} \epsilon\big)\big( k_1\cdot k_2 + i\hspace{0.4pt} \epsilon\big)}\,,  \nn\\[10pt]
I_{V} &= \frac{i\, F(k_1)\,\delta\big(k_1^2\big)\, \delta\big( n\cdot q-n\cdot k_1\big)}{\big(\bar n \cdot k_2 + n \cdot k_2-i \hspace{0.4pt}\epsilon\big) \big(\bar n \cdot k_1+ i\hspace{0.4pt}\epsilon\big) \big(n \cdot k_1 +i \hspace{0.4pt}\epsilon\big)\big( -k_1\cdot k_2 + i \hspace{0.4pt}\epsilon\big)\big(k_2^2-M_\chi^2+ i\hspace{0.4pt}\epsilon\big)} \,. 
\end{align}
Here $F(k_1)$ is a function of $k_1$ whose precise form is not relevant for the current argument. We can now perform the $n\cdot k_2$ integral in $I_V$ by contours. Since $k_1$ crosses the cut, we have $n\cdot k_1>0$. We can therefore close the contour in the upper half plane, and 
we only pick up
the residue from $1/\big(k_2^2-M_\chi^2+i \epsilon\big)$.  This is exactly equivalent to the on-shell condition for $k_2$ with an overall minus sign. Thus the real and virtual contributions cancel.  
Using this same approach, it can be verified that this cancellation happens for all triple gauge boson vertex diagrams at two loops.   Therefore, we see that for these graphs involving the triple gauge boson vertex, because the measurement function is inclusive over the low energy boson, no large logarithms are generated. We see explicitly that for these contributions, the presence of electroweak charged particles does not have an effect, since the color structure is manifestly identical between the real and virtual diagrams.

\vspace{10pt}
\noindent \textbf{Independent Emission Diagrams:}

We now consider the independent emission diagrams.
Most of the independent emission diagrams cancel between the real and virtual diagrams that are related by a shifted cut, in exactly the same way as was just demonstrated for the triple gauge boson vertex diagrams. However, shifting in the cut does not always produce the same color structure for the real and virtual diagrams. We will decompose the color structure into an abelian and non-abelian piece, and we will then show that the non-abelian piece always cancels, as is required for our factorization.

\begin{figure}
\hspace{20pt}\subfloat[]{\label{fig:soft_c}
\includegraphics[width=5.7cm]{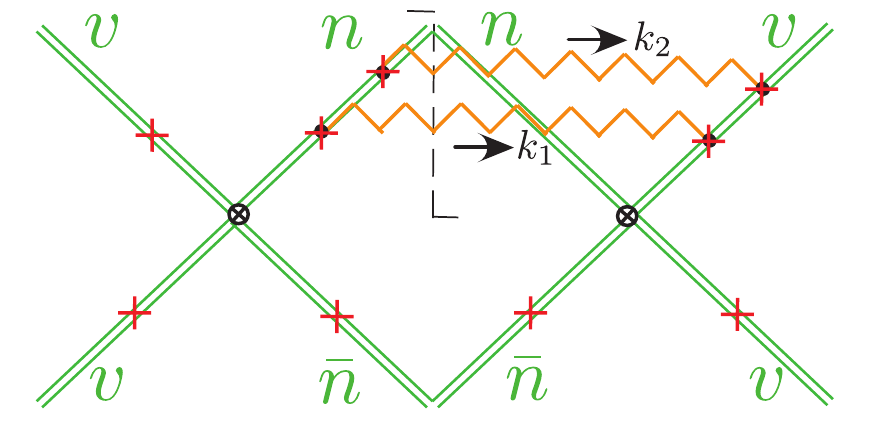}    
}\hspace{60pt}
\subfloat[]{\label{fig:soft_d}
\includegraphics[width=5.5cm]{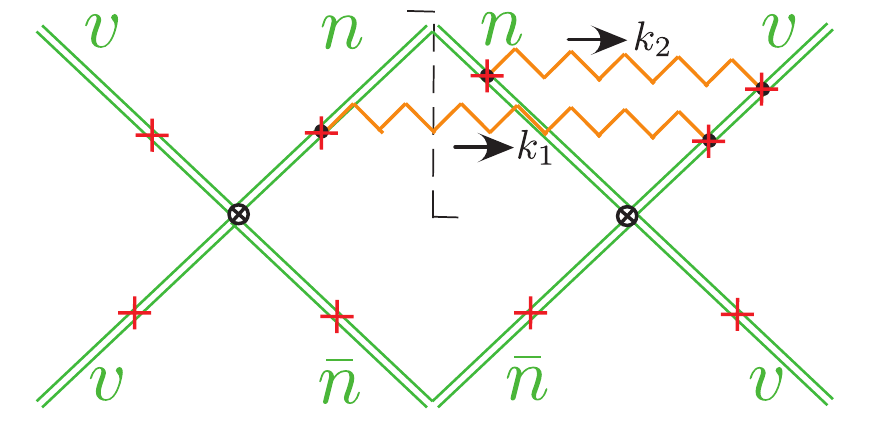}
}\hspace{20pt}
\caption{Examples of real (a) and virtual (b) diagram with independent emissions contributing to the two loop soft function. Here the particle with momentum $k_1$ is at the scale $M_\chi (1-z)$, while $k_2$ is at the scale $\mW$.}
\label{fig:indepNGLs} 
\end{figure}

We provide one illustrative example, which is shown in Fig.~\ref{fig:indepNGLs}.  Again, defining the real integral as $I_R$ and the virtual integral as $I_V$, we find
\begin{align}
I_{R} \!&= \!\left[\big(\tau^B \tau^A\big)_{3a}\delta^{bb'}\big(\tau^B \tau^A\big)_{a'3}\right]\!\frac{\delta\big(k_1^2\big)\,\delta\big(k_2^2-M_\chi^2\big)\, \delta\big(n\cdot q-n\cdot k_1\big)}{\big(n \cdot k_2 +i\hspace{0.4pt}\epsilon\big)\!\big(v \cdot k_2-i\hspace{0.4pt}\epsilon\big)\!\big(n \cdot k_1+i\hspace{0.4pt}\epsilon\big)\!\big(v \cdot k_1 -i\hspace{0.4pt}\epsilon\big)}\,,\\[8pt]
I_{V} \!&= \! \left[\big(\tau^A\big)_{3a}\delta^{bb'}\big(\tau^B \tau^A \tau^B\big)_{a'3}\right]\! \frac{-i\,\delta\big(k_1^2\big)\, \delta\big(n\cdot q-n\cdot k_1\big)}{\big(n \cdot k_2 -i\hspace{0.4pt}\epsilon\big)\!\big(v \cdot k_2-i\hspace{0.4pt}\epsilon\big)\!\big(n \cdot k_1-i\hspace{0.4pt}\epsilon\big)\!\big(v \cdot k_1 -i\hspace{0.4pt}\epsilon\big)\!\big(k_2^2-M_\chi^2+i\hspace{0.4pt}\epsilon\big)}. \nn
\end{align}
Here $\tau^A$ and $\tau^B$ are gauge factors from the $k_1$ and $k_2$ gauge boson-DM vertices respectively. 
To show that these two integrals cancel, we again perform a contour integral in $n \cdot k_2$. Closing the contour in the upper half plane, we 
only pick up
the residue $k_2^2-M_\chi^2+i\hspace{0.4pt}\epsilon$,  which implements the on-shell condition, similar to the triple gauge boson example. However, in this case the color structures are different, and therefore we do not have an exact cancellation.  We can investigate this further by rewriting the color structure for $I_R$
\bea
\big(\tau^B \tau^A\big)_{3a}\,\delta^{bb'}\,\big(\tau^B \tau^A\big)_{a'3} =\big(\tau^A \tau^B\big)_{3a}\,\delta^{bb'}\big(\tau^B \tau^A\big)_{a'3}+ i\,f^{BAC} \,\big(\tau^C\big)_{3a}\, \delta^{bb'}\,\big(\tau^B \tau^A\big)_{a'3} \,,
\eea
and for $I_V$
\bea
\big(\tau^A\big)_{3a}\,\delta^{bb'}\big(\tau^B \tau^A \tau^B\big)_{a'3} =  \big(\tau^A\big)_{3a}\,\delta^{bb'}\big(\tau^B \tau^B \tau^A\big)_{a'3}+i\,f^{ABC}\big(\tau^A\big)_{3a}\,\delta^{bb'}\big(\tau^B \tau^C\big)_{a'3}\,\,\, \nn\\[5pt]
=\big(\tau^A\big)_{3a}\,\delta^{bb'}\big(\big(\tau^B\big)^2  \tau^A\big)_{a'3}+i\,f^{BAC}\big(\tau^C\big)_{3a}\,\delta^{bb'}\big(\tau^B \tau^A\big)_{a'3}\,,
\eea
to show that it is possible in both cases to separate these terms into an abelian and a non-abelian part, with identical color structure for the non-abelian contributions.
Since the integrals have the same form with an opposite sign, these two expressions show that the non-abelian pieces cancel between the two diagrams, leaving behind an abelian term which describes the contribution to the soft function arising from independent emissions at the scale $\mW$ and $M_\chi (1-z)$. This contribution is already described by the factorization formula, which involves a soft function at each of these scales.

This argument holds for all strongly ordered two-loop contributions to the soft function, allowing us to prove that it factorizes into two independent soft functions, one describing radiation at the scale $\mW$, and one describing radiation at the scale $M_\chi (1-z)$.  Therefore, in particular we have proven that the refactorized formula in \Eq{eq:genrefactorize} is also valid at NLL order.

Given the above argument, we find that our factorization formula for the resummed photon spectrum in the endpoint region can be extended to NLL without modification
\begin{mdframed}[linewidth=1.5pt, roundcorner=10pt]
\begin{align}
\label{eq:factorization_NLL}
&\nn\\*[-30pt]
\hspace{-15pt}\left(\frac{\text{d} \hat{\sigma}}{\text{d}z}\right)^{\!\!\text{NLL}}\!=~&H\big(M_\chi, \mu\big)\, J_\gamma\big(\mW,\mu, \nu\big)\, J_\bn\big(\mW,\mu, \nu\big)\, S \big(\mW,\mu, \nu\big) \nn \\*[3pt]
&\times H_{J_\bn}\big(M_\chi, 1-z,\mu\big) \otimes H_{S}\big(M_\chi, 1-z,\mu\big) \otimes C_S\big(M_\chi, 1-z,\mW,\mu, \nu\big)  \,.\\[-10pt] \nn
\end{align}
\end{mdframed} 
Again we suppress the gauge indices $a'b'ab$ here.  It would be interesting to understand if this factorization continues to hold at higher logarithmic orders. However, since it is not required for this paper, we do not pursue this further.

\section{Renormalization Group Evolution at NLL}\label{sec:NLL}

Having demonstrated the validity of our factorization formula \Eq{eq:factorization_NLL} at NLL, deriving
the cross section at this level of accuracy requires computing the anomalous dimensions of each of the functions appearing in \Eq{eq:factorization} to $\mathcal{O}(\aW)$, and solving the associated renormalization group (RG) equations.  After briefly reviewing the technology utilized to achieve NLL accuracy in \Sec{sec:prelim}, in \Sec{sec:anom_dim_compile} we provide results for all the one-loop anomalous dimensions appearing in our factorization, as well as the required matching coefficients. In \Sec{sec:diag} we then solve the associated RG equations. Due to operator mixing, and the large number of functions appearing in our factorization formula, this turns out to be non-trivial. This section is more technical than the remainder of the paper, and the reader interested only in the final result can skip it on a first reading.

\subsection{RGE Preliminaries }\label{sec:prelim}

We will resum logarithms appearing in the cross section using renormalization group equations (RGEs) in both virtuality, $\mu$, and rapidity, $\nu$. We will always choose an appropriate integral transformation of the factorization formula so that the RG is multiplicative. In particular, we will work in Laplace space, with $s$ conjugate to $2\, M_\chi (1-z)\,e^{-\gamma_E}$, namely
\begin{align}\label{eq:LP_def}
f(s)=\text{\bf LP}[f(z)]= \int_0^\infty \text{d}(1-z) \, e^{- 2\,M_\chi\, (1-z)\, s\, e^{-\gamma_E}}f(z)\,,
\end{align}
and
\begin{align}\label{eq:LP_inv_def}
f(z)=\text{\bf LP}^{-1}[f(s)]=\frac{2\,M_\chi\, e^{-\gamma_E}}{2\,\pi\, i} \int\limits_{\gamma-i\,\infty}^{\gamma+i\,\infty} \text{d}s\, e^{2\,M_\chi\, (1-z)\, s\, e^{-\gamma_E}} f(s)\,.
\end{align}
As usual, the contour in the inverse Laplace transform is chosen such that $\gamma$ is to the right of all singularities in the complex plane.
The RGEs in virtuality will then take the form 
\begin{align}
\frac{\text{d}}{\text{d} \log\mu} F(s;\mu,\nu) =\gamma_\mu^F(s;\mu,\nu) \,F(s;\mu,\nu)\,.
\label{eq:muRG}
\end{align}
The anomalous dimension for the function $F$ can be shown to have the following structure to the order required for our analysis
\bea \label{eq:gammamuFform}
\gamma_\mu^F(s;\mu,\nu) 
  = \Gamma_F[\aW]\, \log \left(\frac{\mu^2}{\mu_F(s)^2}\right)
  + \tilde \Gamma_F[\aW]\, \log \left(\frac{\nu^2}{\tilde \mu_F(s)^2}\right) 
  + \gamma_F[\aW] \,.
\eea
Here $\Gamma_F$ and $\tilde \Gamma_F$ are proportional to the cusp anomalous dimension for the function $F$, which drives the double logarithmic evolution, and $\gamma_F[\aW]$ is the non-cusp anomalous dimension. To achieve NLL accuracy, the cusp anomalous dimension is needed to two loops, while the non-cusp anomalous dimension is needed to one loop.

It is known that at two-loops that the cusp anomalous dimension $\Gamma_F[\aW]$ is a multiple of the universal cusp anomalous dimension $\Gamma_\cusp$~\cite{Korchemsky:1987wg}, 
\begin{align}
\Gamma_F[\aW] &= c_F~ \Gamma_{\cusp}[\aW]\,,
 \qquad\quad
 & \tilde \Gamma_F[\aW] &= \tilde c_F~ \Gamma_{\cusp}[\aW]\,,
\end{align}
where $c_F$ and $\tilde c_F$ are constants depending on the function $F$, that do not have an expansion in $\aW$, \emph{i.e.}, it can be determined using a one-loop calculation (and should not be confused with the fundamental Casimir $C_F$).  We can expand the cusp anomalous dimension perturbatively as 
\begin{align}
\Gamma_{\cusp} [\aW]= \sum_{n=0}^{\infty} \Gamma_n \,\TaW^{n+1}\,,
\end{align}
where here, and throughout the rest of the text we use
\begin{align}
\TaW(\mu)=\frac{\aW(\mu)}{4\,\pi}\,,
\label{eq:TaW}
\end{align}
to simplify our notation. If the scale of $\aW$ is not made explicit, it should be taken to be evaluated at the natural scale of the function it appears in. 
The first two perturbative orders in the expansion of the cusp anomalous dimension, as required to achieve NLL accuracy, are well known~\cite{Korchemsky:1987wg} 
\begin{align}\label{eq:cusp}
\Gamma_0&=4\,, \nn \\
\Gamma_1&= C_A \left(\frac{268}{9}-\frac{4\,\pi^2}{3}\right)-\frac{80}{9}\,n_f\, T_F-\frac{16}{9}=4\left( \frac{70}{9}-\frac{2}{3}\,\pi^2  \right)\,.
\end{align}
Here $T_F=1/2$ is the SU($N$) gauge group index, $n_f=6$ is the number of fermions in the fundamental representation, and $C_A$ is the Casimir for the adjoint representation. Note that here we have defined $\Gamma_{\cusp}$ so as not to include an overall Casimir, which is included in $c_F$ and $\tilde c_F$. We will similarly expand the non-cusp anomalous dimension in $\TaW$ as
\bea
\gamma_F[\alpha_W] &=& \sum_{n=0}^{\infty} \gamma^F_n\, \TaW^{n+1} \,.
\eea

We will also need to RG evolve in rapidity~\cite{Becher:2011dz, Chiu:2011qc,Chiu:2012ir}, and to do so we will use the rapidity RG framework of~\cite{Chiu:2011qc,Chiu:2012ir}.  The rapidity RG equation takes the form
\begin{align}
\frac{\text{d}}{\text{d} \log\nu} F(s;\mu,\nu) =\gamma_\nu^F(s;\mu)\, F(s;\mu,\nu)\,,
\end{align}
where the anomalous dimension is given by 
\bea
\gamma_\nu^F(s;\mu) = \Gamma_\nu^F[\aW] \,\log \left(\frac{\mu^2}{\mu_F^{\prime\,2}}\right)+ \gamma_\nu^F[\aW] \,.
\eea
As for the $\mu$-anomalous dimension, $ \Gamma_\nu^F(\aW)$ is a multiple of the cusp anomalous dimension, $\Gamma_\cusp$. 
Therefore, as with the $\mu$-anomalous dimension, to achieve NLL accuracy, we need the one-loop non-cusp anomalous dimension, and the two-loop cusp anomalous dimension.
We will find that the one-loop non-cusp contributions $\gamma_\nu^F$ vanish.

At NLL, we will also need to take into account the running of the coupling, which is a single logarithmic effect. We define the perturbative expansion of the $\beta$ function as 
\begin{align} \label{eq:betafunction}
\beta[\aW] =
- 2 \,\aW \sum_{n=0}^\infty \beta_n\,\TaW^{n+1}\,,
\end{align}
where we will need 
\begin{align}
\beta_0 &=   \frac{11}{3}\, C_A -\frac{4}{3} \,n_f\, T_F-\frac{1}{3}\,T_F=\frac{19}{6} \,, \nn \\[5pt]
\beta_1 &= \frac{34}{3}\,C_A^2  - \left(\frac{20}{3}\,C_A\, + 4\, C_F\right)\, T_F\,n_f-\frac{13}{6}=-\frac{35}{6}\,, 
\end{align}
where $C_F=3/4$ is the quadratic Casimir for the fundamental representation.

\begin{figure}
\begin{center}
\includegraphics[width=0.55\columnwidth]{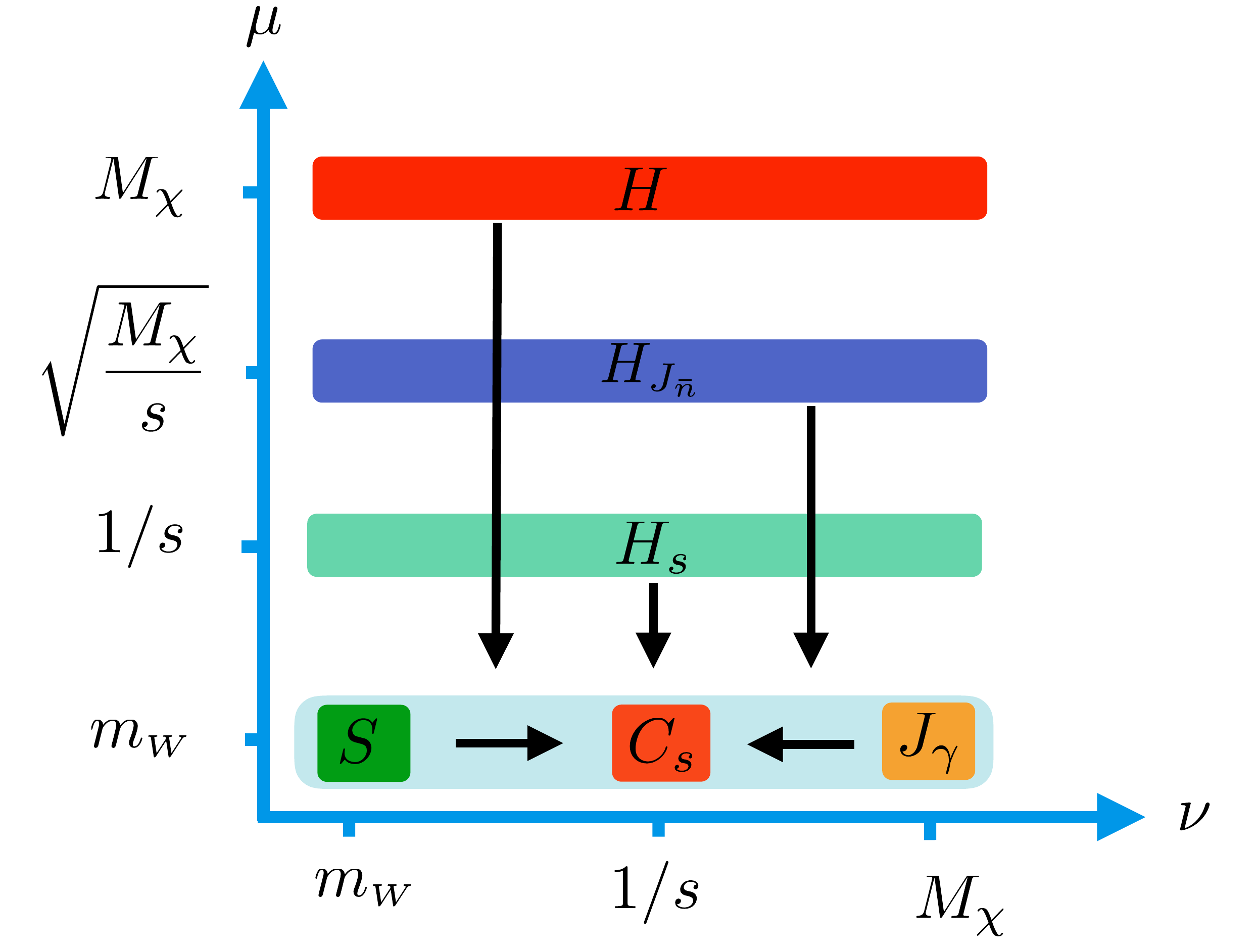}
\end{center}
\vspace{-15pt}
\caption{The RG path used to resum logarithms. The non-cusp term of the one-loop rapidity anomalous dimension vanishes, so that this path remains valid at NLL accuracy. } 
\label{fig:RG_path}
\end{figure}

The resummation of large logarithms is achieved by running all functions appearing in the factorization theorem to a single scale.  
Since the evolution in $\mu$ and $\nu$ commutes
\begin{align}
\left [ \frac{\text{d}}{\text{d}\log \mu},    \frac{\text{d}}{\text{d}\log \nu} \right ]=0\,,
\end{align}
one can choose an arbitrary path in the $(\mu, \nu)$ plane as long as large logarithms in the anomalous dimension are consistently avoided. In \Fig{fig:RG_path}  we show the path that was used in \cite{Baumgart:2017nsr} to simplify the RG evolution. Here all functions are run to the scale $\mu=\mW$. In principle, one must then run all the functions to a common $\nu$ value, which in the figure is shown as $\nu=1/s$. In practice, the $\nu$ running is trivial since the one-loop non-cusp rapidity anomalous dimension vanishes. Therefore, with this choice of path, it suffices to run the hard function and the hard matching coefficients for the jet and soft functions in $\mu$, greatly simplifying the RG structure. Furthermore, we can treat the combination $C_S\, S = \tilde S$ as unfactorized, see \Eq{eq:defStilde} below. Beyond NLL, this would no longer be possible, which would significantly complicate the RG evolution.

\subsection{Anomalous Dimensions and Matching Coeffecients}\label{sec:anom_dim_compile}

In this section, we give explicit results for all anomalous dimensions required to achieve NLL accuracy, along with the necessary matching coefficients. As discussed above, this requires the one-loop non-cusp anomalous dimensions.  For brevity, we only give the final results, noting that all the required integrals can be found in App.~A of~\cite{Baumgart:2017nsr}.  
Due to the large number of functions appearing in our factorization formula, instead of just giving the values of the cusp and non-cusp anomalous dimensions, we explicitly write the logarithms to show the natural scales appearing in the functions. Furthermore, to simplify the functions, we will write $\Gamma_\cusp[\aW]$ with the assumption that when working to NLL, this cusp anomalous dimension should be kept to second order in the coupling.

\subsubsection{Hard Function}\label{sec:hard}

We use a basis of hard scattering operators defined by \cite{Ovanesyan:2014fwa}
\begin{align}\label{eq:hardscattering}
 {\cal L}^{(0)}_\text{hard} &=\sum_{r=1}^2 C_r\big(M_\chi,\mu\big)\, \cO_r \nn \\
 &= \sum_{r=1}^2 \: C_r\big(M_\chi,\mu\big)\: \left( \chi_v^{aT} \,i\sigma_2\, \chi_v^b  \right) \left( Y_r^{abcd} \,\cB_{n\perp}^{ic}\, \cB_{\bar n \perp}^{jd}    \right) i\, \epsilon^{ijk} (n-\bar n)^k  \,.
\end{align}
Here $\chi_v$ are the non-relativistic heavy DM fields, which carry a label velocity $v$, as in heavy quark EFT~\cite{Neubert:1993mb,Manohar:2000dt}. We will take $v=(1,0,0,0)$. The $\cB_{n\perp}$ are collinear gauge invariant fields~\cite{Bauer:2000yr,Bauer:2001ct}, with labels $n=(1,0,0,1)$  and $\bn=(1,0,0,-1)$.  They are defined by
\begin{align} \label{eq:chiB}
\cB_{n\perp}^\mu(x)
= \frac{1}{g}\Bigl  [W_{n}^\dagger(x)\,i  D_{{n}\perp}^\mu W_{n}(x)\Bigr]
 \,.
\end{align}
Here $D_{n\perp}^{\mu}$ is the collinear gauge covariant derivative, and $W_n$ is a collinear Wilson line
\begin{align}
W_n(x) =\left[ \, \sum\limits_{\text{perms}} \exp \left(  -\frac{g}{\bar{n}\cdot \cP}\, \bar n \cdot A_n(x)  \right) \right]\,,
\end{align}
where $\cP^\mu$ is the label momentum operator, which when acting on a collinear field, returns its label momentum.  The ultrasoft Wilson lines, $Y$, appearing in the operator are determined by the BPS field redefinition \cite{Bauer:2001ct}
\begin{align} \label{eq:BPSfieldredefinition}
\cB^{a\mu}_{n\perp}\,\to\, Y_n^{ab}\, \cB^{b\mu}_{n\perp}\,,
\end{align}
which is performed in each collinear sector, and similarly for the non-relativistic DM particles. For a representation, $r$, we have
\begin{align}
Y^{(r)}_n(x)=\bold{P} \exp \left [ i\hspace{0.4pt}g \int\limits_{-\infty}^0 \text{d}s\, n\cdot A^a_{us}(x+s\,n)\,  T_{(r)}^{a}\right]\,,
\label{eq:YWilsonLine}
\end{align}
where $\bold P$ denotes path ordering, and similarly for the $\bn$ direction. For our particular basis of operators, the relevant Wilson line structures are
\begin{align}\label{eq:softwilsonY}
Y_1^{abcd}=\delta^{ab} \Big(Y_n^{ce} Y_{\bar n}^{de}\Big)\,, \qquad Y_2^{abcd} =\Big(Y_v^{ae}Y_n^{ce}\Big)\Big(Y_v^{bf} Y_{\bar n}^{df}\Big)\,.
\end{align}

At tree level, the Wilson coefficients are given by
\begin{align}
C_1(\mu)=-C_2(\mu)=-\pi\, \frac{\aW(\mu)}{M_\chi}\,.
\end{align}
To NLL accuracy, the anomalous dimensions of the Wilson coefficients are given by~\cite{Ovanesyan:2014fwa} 
\begin{align}
\hat{\gamma}^C(\mu) &= 2\, \gamma_{{\scriptscriptstyle W}_T}(\mu)\, \id+ \hat{\gamma}_S(\mu)\,.
\end{align}
Here the diagonal component
\begin{align}
\gamma_{\scriptscriptstyle W_T}^{\rm NLL} (\mu)&= C_A \,\Gamma_\cusp[\aW]\log \left( \frac{2\,M_{\chi}}{\mu} \right) - \TaW\, \beta_0\,, 
\end{align}
contains the cusp anomalous dimension and the beta function, and there is also an off-diagonal non-cusp component
\begin{align}
\hat{\gamma}_S^{\rm NLL}(\mu) &= 4\, \TaW\, \psi \begin{pmatrix} 2 & \,\,\,\,\,\,1 \\ 0 &\,\, -1 \end{pmatrix} - 8\,\TaW \begin{pmatrix} 1 & \,\,0 \\ 0 & \,\,1 \end{pmatrix}\,.
\end{align}
To simplify notation, we have defined
\begin{align}\label{eq:psi}
\psi = 1 - i\, \pi\,,
\end{align}
which will appear frequently in our results.

The hard function is defined as
\begin{align}
H_{ij}=C_i^* C_j\,.
\end{align}
We will use the simplified notation
\begin{align}
H_1= H_{11}\,,\hspace{0.5cm} 
H_2= H_{22}\,,\hspace{0.5cm}
H_3= H_{12}\,,\hspace{0.5cm}
H_4= H_{21}\,.
\end{align}
Note that a slightly different notation was used in~\cite{Baumgart:2017nsr}, where we defined $H_3 = H_{12} = H_{21}$. Beyond LL, it is necessary to treat $H_{12}$ and $H_{21}$ separately. For the same reason, when we redefine the soft operators below we will not follow the notation of~\cite{Baumgart:2017nsr}.\footnote{Due to this change of basis, the consistency conditions between the anomalous dimensions stated in Eqs.~(5.14) and (5.16) of~\cite{Baumgart:2017nsr} must also modified. In the basis used here, the consistency equations can be obtained from those in the LL basis by making the substitution $1/3\to 1/4$ on the left side of Eqs.~(5.14) and (5.16).}  The hard function in this new basis satisfies an RG equation with anomalous dimension matrix which is given by
\be
\hat{\gamma}_{\mu}^{H} (\mu)
= 
\left(
\begin{array}{cccc}
\vspace{2pt}
\gamma_{11}^{H}& 0 & \gamma_{13}^{H} & \gamma_{14}^{H}  \\[1.5pt]
0 & \gamma_{22}^{H} & 0& 0 \\[1.5pt]
0 & \gamma_{32}^{H} & \gamma_{33}^{H} & 0  \\[1.5pt]
0 & \gamma_{42}^{H} & 0 & \gamma_{44}^{H}  
\end{array}
\right)
\,.
\ee
The components of the anomalous dimension matrix are given by
\begin{align}\label{eq:hard_explicit}
\gamma_{11}^H &= \Gamma_H\,,\nn\\
\gamma_{13}^H & = \big(\gamma_{14}^H\big)^* = 4 \,\TaW\, \psi\,, \nn \\
\gamma_{22}^H & = \Gamma_H- 24\, \TaW\,, \nn \\
\gamma_{32}^H & = \big(\gamma_{42}^H\big)^*=4 \,\TaW \psi^*\,, \nn \\
\gamma_{33}^H & = \big(\gamma_{44}^H\big)^*= \Gamma_H - 12 \,\TaW\psi \,,
\end{align}
where
\begin{align}
\Gamma_H=-4\,\Gamma_\cusp[\aW]\log \left( \frac{\mu^2}{\big( 2\,M_{\chi} \big)^2} \right) - 4\, \TaW\, \beta_0\,,
\label{eq:defGammaH}
\end{align}
where here and below we have explicitly used $C_A=2$ to simplify the formulas. At LL order this matrix is diagonal, while at NLL one encounters non-trivial operator mixing.

\subsubsection{Photon Jet Function}\label{sec:photon_jet}

The photon jet function $J_\gamma$ is defined as 
\begin{align}
J_\gamma(\mu,\nu,E_\gamma, \mW) =  \Big\langle 0\Big| \cB_{n \perp}^{c}(0)\Big| \gamma \Big\rangle \Big\langle \gamma \Big| \cB_{n \perp}^c(0) \Big|0 \Big\rangle\,.
\end{align}
Its $\mu$- and $\nu$-anomalous dimensions are given by
\begin{align}
\gamma_{\mu}^{n} (E_\gamma;\mu, \nu) &=  4\,\Gamma_\cusp [\aW] \log \left(\frac{\nu}{2\,E_{\gamma}}\right)+2\, \TaW\, \beta_0\,,\nn\\
\gamma_{\nu}^{n} (\mW;\mu) &=    4\,\Gamma_\cusp [\aW]  \log\left(\frac{\mu}{\mW}\right)\,.
\end{align}

With our choice of resummation path, we do not need to RG evolve the photon jet function, and therefore only need its boundary value. However, to ensure a consistent definition of NLL accuracy in Laplace and cumulative space, it is well known that in Laplace space, one must also keep the $\cO(\aW)$ RG generated logs in the boundary terms (see \emph{e.g.}~\cite{Almeida:2014uva} for a detailed discussion). We will do this throughout the forthcoming sections without further comment. For the photon jet function, we have
\begin{align}
J_{\gamma}\big(\mu^0_\gamma, \nu\big) = &- 2\, \sW^2\big(\mu_{\gamma}^0\big) \left[ 1 + 4\, \Gamma_0\, \TaW\big(\mu_{\gamma}^0\big)\, \log \left( \frac{\mu_{\gamma}^0}{\mW} \right) \log \left( \frac{\nu}{2\, E_{\gamma}} \right) \right] \nn\\[5pt]
= &- 2 \,\sW^2\big(\mu_{\gamma}^0\big)\, C_{\gamma}\,.
\end{align}
Here $\sW = \sin \thetaW$ is the sine of the weak mixing angle and the canonical value for $\mu_{\gamma}^0$ is $\mW$.

\subsubsection{Recoiling Jet Function}\label{sec:recoil_jet}

The recoiling jet function is defined before refactorization by
\begin{align}
  J'_{\bar n}\big(M_\chi, 1-z, \mW\big) &= \Big\langle 0\Big| \cB_{\bar n\perp}^{d}\!(0)\,\delta\big((1-z) -\widehat{\cM}_c\big)\,\delta\big(M_{\chi} - \mathcal{P}^-/2\big)\,\delta^2\big(\vec{\mathcal{P}}_{\perp}\big) \,\cB_{\bar n\perp}^d\!(0)\Big|0 \Big\rangle\,,
\end{align}
where
\begin{align}
\widehat\cM_c\, |X_c\rangle= \frac{1}{4\,M_\chi^2} \left( \sum\limits_{\,\,i\in X_c}p_i^\mu \right)^2 \big|X_c\big\rangle\,,
\end{align}
is the collinear measurement function, which returns the value of $(1-z)$ when acting on a collinear state. The recoiling jet function is refactorized into a low scale function $J_\bn(\mW,\mu, \nu)$ and a high scale matching coefficient $H_{J_\bn}\big(M_\chi, 1-z,\mu\big)$ as
\begin{align}\label{eq:jet_matching}
J'_\bn\big(M_\chi, 1-z, \mW, \mu,\nu\big) = H_{J_\bn}\big(M_\chi, 1-z, \mu\big)\, J_\bn\big(\mW, \mu, \nu\big) \left[ 1+ \cO\left( \frac{\mW}{M_\chi \sqrt{1-z}} \right) \right ]\,.
\end{align}
The low scale function is defined as
\begin{align}
 J_{\bar n}(\mW, \mu,\nu) &=  \Big\langle 0\Big| \cB_{\bar n\perp}^{d}\!(0)\,\delta\big(M_{\chi} - \mathcal{P}^-/2\big)\,\delta^2\big(\vec{\mathcal{P}}_{\perp}\big)\, \cB_{\bar n\perp}^d\!(0)\Big|0 \Big\rangle\,,
\end{align}
and is understood to be evaluated in the broken theory. \Eq{eq:jet_matching} then defines the hard matching coefficient $H_{J_\bn}\big(M_\chi, 1-z, \mu\big)$.

To NLL accuracy, the low scale function has neither a $\mu$- or $\nu$-anomalous dimension
\begin{align}
\gamma_{\mu}^{J_\bn}=\gamma_{\nu}^{J_\bn}=0\,.
\end{align}
The hard matching coefficient 
only has
a $\mu$-anomalous dimension, which at NLL order is  
\bea
\gamma_{\mu}^{H_{J_\bn}}(s;\mu) = 4\, \Gamma_\cusp [\aW] \log \left(\frac{\mu^2 \,s}{2\,M_{\chi}} \right) + 2\, \TaW\, \beta_0\,.
\label{eq:hardjetmuanomdim}
\eea
Recall that $s$ is the Laplace space variable conjugate to $2\,M_\chi\, (1-z)\, e^{-\gamma_E}$, as defined in \Eq{eq:LP_def}. The boundary values for the matching coefficient and the jet function are given at tree level by 
\begin{align}
H_{J_\bn}&=\frac{1}{2\,M_{\chi}} \,,\qquad J_{\bn}=1\,.
\end{align}

\subsubsection{Soft Function}\label{sec:soft}

Before refactorization, we have the soft functions $S'$  
\begin{align}
S_1' (z_S)\equiv  S_{11}^{\prime} &= \bigg \langle 0 \, \bigg | \left(Y_n^{3k}Y_{\bar n}^{dk}\right)^{\!\dagger}\!\!\,\delta\Big( (1-z_S)-\widehat{\cM}_s \Big)  \left(Y_n^{3j}Y_{\bar n}^{dj}\right)\! \bigg| \, 0 \bigg \rangle\, \delta^{a'b'} \delta^{ab} \,, \nn \\
S_2'(z_S) \equiv  S_{22}^{\prime} &= \bigg \langle 0 \, \bigg |  \left(Y_n^{3f'}Y_{\bar n}^{dg'}Y_v^{a'f'}Y_v^{b'g'}\right)^{\!\dagger}\!\!\,\delta\Big( (1-z_S) -\widehat{\cM}_s\Big)  \left(Y_n^{3f}Y_{\bar n}^{dg}Y_v^{af}Y_v^{bg}\right)\! \bigg| \, 0 \bigg \rangle \,, \nn \\
S_3' (z_S)\equiv  S_{12}^{\prime} &= \bigg \langle 0 \, \bigg |  \left(Y_n^{3k}Y_{\bar n}^{dk}\right)^{\!\dagger}\!\!\,\delta\Big( (1-z_S)-\widehat{\cM}_s\Big)  \left(Y_n^{3g}Y_{\bar n}^{df}Y_v^{ag}Y_v^{bf}\right)\!  \bigg| \, 0 \bigg \rangle\, \delta^{a'b'}\,, \nn\\
S_4' (z_S)\equiv S_{21}^{\prime}&=  \bigg \langle 0 \, \bigg |  \left(Y_n^{3f'}Y_{\bar n}^{dg'}Y_v^{a'f'}Y_v^{b'g'}\right)^{\!\dagger}\!\!\,\delta\Big( (1-z_S) -\widehat{\cM}_s\Big)  \left(Y_n^{3k}Y_{\bar n}^{dk}\right)\!  \bigg| \, 0 \bigg \rangle\,  \delta^{ab}\,,
\label{eq:SprimeDefs}
\end{align}
where all functions $S'$ should be read as carrying gauge index structure $a'b'ab$. These definitions involve the measurement function
\begin{align}
\widehat\cM_S\,\big  |X_S\big\rangle&=\frac{1}{2\,M_\chi}  \sum\limits_{\,\,i\in X_s} \bar{n}\cdot p_i\, \big |X_S\big\rangle\,,
\end{align}
which returns the value of $(1-z)$  when acting on a soft state. See \cite{Baumgart:2017nsr} for more details.
Here, and in all subsequent expressions, we keep the time ordering and dependence on $x=0$ implicit. Again note that the definition of the operators differs slightly from that used in our LL calculation \cite{Baumgart:2017nsr}, since we distinguish $S_{12}^{\prime}$ and $S_{21}^{\prime}$ at NLL. At NLL the non-vanishing $\mu$-anomalous dimensions are given by
\begin{align}
\gamma_{\mu,11}^{S'} = &-4\,\Gamma_\cusp[\aW] \log \big( \nu\, s\big)\,, \nn\\
\gamma_{\mu,22}^{S'} = &-4 \,\Gamma_\cusp[\aW] \log \big( \nu \,s\big) + 24\, \TaW\,,\nn \\
\gamma_{\mu,23}^{S'} = & \big( \gamma_{\mu,24}^{S'} \big)^* = -4\, \TaW \psi^*\,, \nn\\
\gamma_{\mu,31}^{S'} = & \big( \gamma_{\mu,41}^{S'} \big)^* = -4 \,\TaW \psi\,, \nn\\
\gamma_{\mu,33}^{S'} = & \big( \gamma_{\mu,44}^{S'} \big)^* = -4\, \Gamma_\cusp[\aW] \log \left( \nu s\right) + 12 \,\TaW \psi\,.
\end{align}
The $\nu$-anomalous dimension is diagonal in gauge index space, and is given by
\be
\hat \gamma_\nu^{\tilde S}=  -4 \,\Gamma_\cusp[\aW] \log \left(\frac{\mu}{\mW}\right) \id\,.
\ee

We write the refactorization of the high scale soft function into a hard matching coefficient, a collinear-soft function, and a low scale soft function, as
\bea\label{eq:HS_refactorize}
 S_{i}^{\prime\,aba'b'}= H_{S, ikl}\, (C_{S,k}\,S_l)^{aba'b'} = H_{S, ij}\, \tilde S_{j}^{aba'b'}\,.
 \label{eq:defStilde}
\eea
As discussed above, we are able to choose a path such that we do not need to separately run the $C_S$ and $S$ functions. Therefore, as in \cite{Baumgart:2017nsr} we only give the anomalous dimension for the combined function $\tilde S$, as well as for the matching coefficient $H_S$. This significantly simplifies the structure. However, even with this simplification, we will find that there are 5 relevant refactorized soft functions, so that $j$ runs from 1-5, and $i$ runs from 1-4 in \Eq{eq:HS_refactorize}. This implies that there are $20$ hard-soft functions. Fortunately, we will find that $10$ of them vanish at this order, simplifying the structure of the RGE.

For the low scale soft functions, we have 
\begin{align}
\tilde S_{1}(z_S) &\equiv\tilde S_{11} = \Big\langle 0\Big|\! \left(X_n^{3f'}V_{n}^{dg'}\right)^{\dagger}\delta \Big((1-z_S)-\widehat M_{c_S}\Big)\left(X_n^{3f}V_{n}^{dg}\right)\! \Big|0\Big \rangle \,\delta^{f'g'}\delta^{a'b'}\delta^{fg}\delta^{ab}\,, \nn\\
\tilde S_{2}(z_S) &\equiv \tilde S_{22} = 
 \Big\langle 0\Big|\! \left[ \left( X_n^{ce} V_{n}^{Ae} \right)^{\dagger} \delta \Big((1-z_S)-\widehat M_{c_S}\Big) \left( X_n^{c'g'} V_{n}^{A'g'} \right) \right]\!\!\left[ S_n^{3c} S_n^{3c'}S_v^{a'A'}S_v^{aA}\right] \!  \Big|  0  \Big\rangle \delta^{bb'}\,, \nn \\
\tilde S_{3}(z_S) &\equiv \tilde S_{12} = 
 \Big\langle 0\Big|\! \left[ \left( X_n^{ce} V_{n}^{B'e} \right)^{\dagger} \delta \Big((1-z_S)-\widehat M_{c_S}\Big) \left( X_n^{c'g'} V_{n}^{A'g'} \right) \right]\!\!\left[ S_n^{3c} S_n^{3c'}S_v^{a'A'}S_v^{b'B'}\right]\!  \Big|  0  \Big\rangle  \delta^{ab} 
\,,\nn\\
\tilde S_{4}(z_S) &\equiv \tilde S_{21} = 
 \Big\langle 0\Big|\! \left[ \left( X_n^{ce} V_{n}^{B'e} \right)^{\dagger} \delta \Big((1-z_S)-\widehat M_{c_S}\Big) \left( X_n^{c'g'} V_{n}^{A'g'} \right) \right]\!\!\left[ S_n^{3c} S_n^{3c'}S_v^{aA'}S_v^{bB'}\right] \! \Big|  0  \Big\rangle  \delta^{a'b'} \,,\nn\\
\tilde S_{5}(z_S) &= \Big\langle 0\Big| \left(X_n^{3f'}V_{n}^{df'}\right)^{\dagger}\delta \Big((1-z_S)-\widehat M_{c_S}\Big) \left(X_n^{3f}V_{n}^{df}\right) \Big|0 \Big\rangle\, \delta^{aa'}\delta^{bb'}\,.
\end{align}
For simplicity, we have not written the free gauge indices on the left hand side of these equations. Here $X_n$ and $V_n$ are collinear soft Wilson lines
\begin{align}
X^{(r)}_n(x)=\bold{P} \exp \left [ i\hspace{0.4pt}g \int\limits_{-\infty}^0 \text{d}s\, n\cdot A^a_{cs}(x+s\,n) \, T_{(r)}^{a}\right]\,,
\end{align}
and
\begin{align}
V^{(r)}_n(x)=\bold{P} \exp \left [ i\hspace{0.4pt}g \int\limits_{-\infty}^0 \text{d}s\, \bar n\cdot A^a_{cs}(x+s\,\bar n)\,  T_{(r)}^{a}\right]\,.
\end{align}
We have also defined the collinear soft measurement function,
\begin{align}\label{eq:CSmeasurement}
\widehat{\cM}_{c_S}\, \big| X_{c_S} \big\rangle = \frac{1}{2\,M_\chi}  \sum\limits_{\,\,i\in X_{c_S}} \bn\cdot p_i \, \big| X_{c_S} \big\rangle \, ,
\end{align}
which returns the value of $(1-z)$  on a collinear soft state, see \cite{Baumgart:2017nsr} for more details.
This then defines the matching coefficients $H_{S,ij}$. Again, we reiterate that as compared to \cite{Baumgart:2017nsr} we have slightly modified our basis to incorporate the complete set of gauge index structures that are generated at NLL.

The soft function $\tilde S$ satisfies a $\nu$ RGE 
\begin{align}\label{eq:StildeNuRGE}
\frac{\text{d}}{\text{d} \log\nu} \tilde S(s) =\hat \gamma_\nu^{\tilde S}(\mu)\, \tilde S(s)\,,
\end{align}
where the matrix is diagonal
\begin{align}\label{eq:StildeNuRGEresult}
\hat \gamma_\nu^{\tilde S}(\mu)=  -4 \,\Gamma_\cusp[\aW] \log \left(\frac{\mu}{\mW}\right) \id\,.
\end{align}
Note that the non-cusp term of the $\nu$-anomalous dimension vanishes at one-loop.

The soft function $\tilde S$ also satisfies a $\mu$ RGE
\begin{align}\label{eq:StildeMuRGE}
\frac{\text{d}}{\text{d} \log\mu} \tilde S(s) =\hat \gamma_\mu^{\tilde S}(s;\mu, \nu)\, \tilde S(s)\,.
\end{align}
The $\mu$-anomalous dimension for $\tilde S$ has a non-trivial mixing structure, 
\be
\hat{\gamma}_{\mu}^{\tilde{S}} (s;\mu, \nu)
= 
\left(
\begin{array}{ccccc}
\gamma_{\mu,11}^{\tilde{S}} & 0 & 0 & 0 & 0 \\
0 & \gamma_{\mu,22}^{\tilde{S}} & \gamma^{\tilde{S}}_{\mu,23} & \gamma^{\tilde{S}}_{\mu,24} & \gamma^{\tilde{S}}_{\mu,25} \\[1.7pt]
\gamma_{\mu,31}^{\tilde{S}} & 0 & \gamma_{\mu,33}^{\tilde{S}} & 0 & 0 \\[1.7pt]
\gamma_{\mu,41}^{\tilde{S}} & 0 & 0 & \gamma_{\mu,44}^{\tilde{S}} & 0 \\[1.7pt]
0 & 0 & 0 & 0 & \gamma_{\mu,55}^{\tilde{S}}
\end{array}
\right)\,.
\ee
The non-vanishing terms of this matrix are given by
\begin{align}
\gamma_{\mu,11}^{\tilde{S}} &= - 4\,\Gamma_\cusp [\aW]\, \log \big(\nu\, s \big)\,, \nn \\
\gamma_{\mu,22}^{\tilde{S}} &= - 4\,\Gamma_\cusp [\aW]\, \log \big(\nu\, s \big) + 6\, \Gamma_\cusp [\aW]\, \log \big( \mu \,s \big) + 12\, \TaW\,,  \nn \\
\gamma_{\mu,23}^{\tilde{S}} &= \Big( \gamma_{\mu,24}^{\tilde{S}} \Big)^* = - 4\, i \,\pi\, \TaW\,,  \nn\\
\gamma_{\mu,25}^{\tilde{S}} &= - 2\, \Gamma_\cusp [\aW]\,\log \big( \mu\, s \big) - 4\, \TaW\,, \nn \\
\gamma_{\mu,31}^{\tilde{S}} &= \Big( \gamma_{\mu,41}^{\tilde{S}} \Big)^* = - 2\,\Gamma_\cusp [\aW] \,\log \big( \mu\, s \big) - 4\, \TaW\psi\,,  \nn\\
\gamma_{\mu,33}^{\tilde{S}} &= \Big( \gamma_{\mu,44}^{\tilde{S}} \Big)^* = - 4\,\Gamma_\cusp [\aW] \,\log \big(\nu\, s \big) + 6\,\Gamma_\cusp [\aW] \log \big( \mu\, s \big) + 12\, \TaW\psi\,,  \nn\\
\gamma_{\mu,55}^{\tilde{S}} &= - 4\,\Gamma_\cusp [\aW]\,\log \big(\nu\, s \big) \,.
\end{align}

The matching coefficient for the soft function $H_S$ only has a $\mu$ RGE, which again exhibits non-trivial operator mixing. Making the indices explicit, we have
\begin{align}
{\text{d} \over \text{d} \log\mu} {H_{S,ij}}(s)=\gamma_{\mu,ijkl}^{H_S}\, H_{S,kl}(s)\,.
\end{align}
Recall that here $i$ runs from 1-4, and $j$ runs from 1-5. We therefore have a system of 20 coupled differential equations. Fortunately, to NLL accuracy 10 equations drop out, due to the fact that
\begin{align}
H_{S,12}(\mu) &= H_{S,13}(\mu) = H_{S,14}(\mu) = H_{S,15}(\mu) = 0\,, \nn \\
H_{S,32}(\mu) &= H_{S,34}(\mu) = H_{S,35}(\mu) = 0\,, \nn \\
H_{S,42}(\mu) &= H_{S,43}(\mu) = H_{S,45}(\mu) = 0\,.
\end{align}
Therefore, we must solve a $10\times10$ matrix RGE.
The RGE for the remaining hard matching coefficients are given by
\begin{align}\label{eq:hard_soft_undiag}
\frac{\df}{\df\log \mu} H_{S,11} =&\,\, 0\,,\nn \\
\frac{\df}{\df\log \mu} H_{S,31} =& - 4\, \TaW\, \psi\, H_{S,11} + 12\, \TaW \,\psi \,H_{S,31} + \left( 2\, \Gamma_\cusp[\aW]\, \log \big( \mu\, s \big) + 4\, \TaW\, \psi \right) H_{S,33}\,,\nn \\
\frac{\df}{\df\log \mu} H_{S,33} =&  - 6\,\Gamma_\cusp[\aW]\, \log \big( \mu \,s \big) \, H_{S,33} \,, \nn\\
\frac{\df}{\df\log \mu} H_{S,41} =& - 4 \,\TaW\, \psi^*\, H_{S,11} + 12 \,\TaW\, \psi^*\, H_{S,41} + \left( 2\, \Gamma_\cusp[\aW]\, \log \big( \mu\, s \big) + 4\, \TaW \psi^* \right) H_{S,44}\,,\nn \\
\frac{\df}{\df\log \mu} H_{S,44} =&  - 6\, \Gamma_\cusp[\aW] \,\log \big( \mu \,s \big)  H_{S,44}\,, \nn\\
\frac{\df}{\df\log \mu} H_{S,21} =&\,\, 24 \,\TaW H_{S,21} - 4\, \TaW\, \psi^*\, H_{S,31} - 4\, \TaW\, \psi\, H_{S,41}  \nn\\
&\hspace{-25pt}+\! \left( 2\, \Gamma_\cusp[\aW] \log \big( \mu\, s \big) + 4\, \TaW \psi \right) H_{S,23}+\! \left( 2\, \Gamma_\cusp[\aW]\, \log \big( \mu\, s \big) + 4\, \TaW \,\psi^* \right) H_{S,24}\,,\nn \\
\frac{\df}{\df\log \mu} H_{S,22} =&  \left(- 6\,\Gamma_\cusp[\aW] \,\log \big( \mu\, s \big) + 12\, \TaW \right) H_{S,22}\,, \nn\\
\frac{\df}{\df\log \mu} H_{S,23} =& \left( - 6\, \Gamma_\cusp[\aW] \,\log \big( \mu\, s \big) + 12\, \TaW \,\psi^* \right) H_{S,23} - 4\, \TaW \,\psi^*\, H_{S,33} + 4 \,i \,\pi\, \TaW \,H_{S,22} \,,\nn \\
\frac{\df}{\df\log \mu} H_{S,24} =& \left( - 6\, \Gamma_\cusp[\aW]\, \log \big( \mu \,s \big) + 12 \,\TaW\, \psi\, \right) H_{S,24} - 4\, \TaW\, \psi\, H_{S,44} - 4\, i\, \pi\, \TaW \,H_{S,22} \,, \nn\\
\frac{\df}{\df\log \mu} H_{S,25} =&\,\, 24\, \TaW\, H_{S,25} + \left( 2 \,\Gamma_\cusp[\aW]\, \log \big( \mu \,s \big) + 4\, \TaW \right) H_{S,22} \,,
\end{align}
where $\psi$ was defined in \Eq{eq:psi}.
These anomalous dimensions can be written in the form of a matrix equation as
\begin{align}
\frac{\df}{\df\log \mu} H_S(s)=\Gamma_\cusp[\aW]\, \log\big(\mu\, s\big)\, \hat \Gamma_{H_S} \,H_S(s) + \hat \gamma_{H_S}\, H_S(s) \,.
\end{align}
Due to the size of the matrices, we do not explicitly give them here, although they can be directly read off of \Eq{eq:hard_soft_undiag}. This provides the complete set of anomalous dimensions required to achieve NLL accuracy. We have checked that they satisfy all RG consistency constraints.

\subsection{Solution to NLL Evolution Equations}\label{sec:diag}

Armed with all the necessary anomalous dimensions, we next turn to solving the RG equations at NLL.
Due to our choice of resummation path, as was discussed in \Sec{sec:prelim}, we must $\mu$-evolve all functions from their natural scale to the scale $\mW$. Recall that the homogeneous $\mu$-evolution equation takes the form 
\begin{align}
\frac{\text{d}}{\text{d} \log\mu} F(\mu,\nu) =\gamma_\mu^F(\mu,\nu)\, F(\mu,\nu)\,,
\end{align}
where $\gamma_\mu^F(\mu,\nu)$ takes the form given in \eq{gammamuFform}.  For the purpose of the solution the terms not involving an explicit $\log(\mu^2)$ can be grouped together, so for simplicity we use $\gamma_F[\alpha_W]$ below, instead of writing out $\gamma_F[\alpha_W]+\tilde \Gamma_F[\alpha_W] \log(\nu^2/\tilde \mu_F(s)^2)$. 
Here we have suppressed all kinematic arguments for simplicity.
At NLL we need to include the effects of the running coupling,  so it is useful to change variables from $\mu$ to $\aW$
\begin{align}
	\frac{\text{d}\mu}{\mu}= \frac{\text{d}\aW}{\beta[\aW]}\,.
\end{align}
The solution to the RGE is then given by\footnote{For detailed discussions of the solution of RGEs of this structure, see \emph{e.g.}~\cite{Korchemsky:1993uz,Neubert:2005nt,Becher:2006mr,Fleming:2007xt,Ellis:2010rwa,Almeida:2014uva}.}
\bea
F(\mu) =  e^{K_F(\mu,\mu_0)}\, F[\mu_0]\, \left(\frac{\mu_0^2}{\mu_F^2}\right)^{\omega_F(\mu,\mu_0)}\,, 
\label{NLL}
\eea
where the evolution kernels are 
\begin{align}
\omega_F (\mu, \mu_0) = \int\limits_{\aW(\mu_0)}^{\aW(\mu)}   \frac{\text{d}\aW}{\beta[\aW]}\,  \Gamma_F[\aW]\,,
\end{align}
and 
\begin{align}
K_F(\mu, \mu_0) = \int\limits_{\aW(\mu_0)}^{\aW(\mu)} \hspace{-3pt}  \frac{\text{d}\aW}{\beta[\aW]}  \,\gamma_F[\aW] + 2 \hspace{-8pt} \int\limits_{\aW(\mu_0)}^{\aW(\mu)}   \frac{\text{d}\aW}{\beta[\aW]} \, \Gamma_F[\aW] \hspace{-8pt} \int\limits_{\aW(\mu_0)}^{\aW}  \hspace{-3pt} \frac{\text{d}\aW'}{\beta[\aW']}\,.
 \end{align}
To NLL accuracy, we have
\begin{align}
K_F(\mu, \mu_0) &= \frac{c_F\,\Gamma_0}{2\, \beta_0^2} \left(\frac{1}{\TaW(\mu_0)} \left(\log r +\frac{1}{r} -1\right)+\left(\frac{\Gamma_1}{\Gamma_0}-\frac{\beta_1}{\beta_0}\right)(r-1-\log r)-\frac{\beta_1}{2\,\beta_0} \log^2 r\right)\nn \\[5pt]
&\hspace{15pt} -\frac{\gamma_0}{2\,\beta_0}\log r\,, \nn\\[10pt]
\omega_F(\mu, \mu_0) &= -\frac{c_F\,\Gamma_0}{2\,\beta_0} \left(\log r + \TaW(\mu_0) \left(\frac{\Gamma_1}{\Gamma_0}-\frac{\beta_1}{\beta_0}\right)(r-1)\right)\,,
\end{align}
where recall $\Gamma_0$ and $\gamma_0$ control the cusp and non-cusp parts of the evolution and
\begin{align}
 \frac{1}{r} = \frac{\TaW(\mu_0)}{\TaW(\mu)} = 1+ 2\, \TaW(\mu_0)\,\beta_0\, \log \left(\frac{\mu}{\mu_0}\right)\,.
\end{align}
Using this evolution kernel, we can now run each of the functions to the scale $\mW$. We will separately consider the jet function, hard function, and hard-soft matching coefficients. For the hard and hard-soft functions, due to the complicated mixing structure, we will have to first diagonalize the system of equations before using the kernels given in this section.

\subsubsection{Recoiling Jet Function Evolution}\label{sec:jet_evol}

The jet function must be evolved from its initial natural scale $\mu_J^0=2\, M_{\chi}\sqrt{1-z}$ to the common scale $\mW$.   In Laplace space, the natural scale is
\begin{align}\label{eq:jet_natural_scale}
\mu_J(s) = \sqrt{\frac{2\,M_{\chi}}{s}}\,.
\end{align}
From \Eq{eq:hardjetmuanomdim}, we can see that for the hard jet evolution we have $c_J = 4$ and $\gamma_0 = 2\, \beta_0$. This allows us to write the evolution kernel as
\begin{align} 
U_J\big(\mW; \mu_J^0, \mu_J(s)\big) & = \Big( \big( V_J - 1 \big) \Theta_J + 1 \Big) \left( \frac{(\mu_J^0)^2}{(\mu_J(s))^2} \right)^{\omega_J}\,,
\label{eq:UJ}
\end{align}
which we have written in terms of
\begin{align}
V_J &= \exp \left\{ \frac{2\, \Gamma_0}{\beta_0^2} \left[ \frac{1}{\TaW(\mu_J^0)} \left( \log r_J + \frac{1}{r_J} - 1 \right) + \left( \frac{\Gamma_1}{\Gamma_0} - \frac{\beta_1}{\beta_0} \right) \left( r_J - 1 - \log r_J \right) - \frac{\beta_1}{2\,\beta_0} \log^2 r_J \right] \right. \nn \\[2pt]
&\hspace{45pt}\left.- \log r_J \vphantom{\frac{2 \Gamma_0}{\beta_0^2}} \right\}\,, \nn \\[5pt]
\omega_J &=- \frac{2\,\Gamma_0}{\beta_0} \left[ \log r_J + \TaW\big(\mu_J^0\big) \left( \frac{\Gamma_1}{\Gamma_0} - \frac{\beta_1}{\beta_0} \right) (r_J-1) \right] \Theta_J\,, \nn \\[5pt]
r_J &=  \frac{\TaW(\mW)}{\TaW\big(\mu_J^0\big)} = \left[ 1 + 2\, \TaW\big(\mu_J^0\big)\, \beta_0\, \log \left( \frac{\mW}{\mu_J^0} \right) \right]^{-1}\,, \nn \\[5pt]
\Theta_J &=  \Theta \Big( 2\,M_{\chi}\, \sqrt{1-z} - \mW \Big)\,,
\end{align}
where $\Theta$ is the Heaviside step function. The step function enforces that the mass of radiation in the jet is greater than $\mW$.

In addition to the RG evolution, we also need the appropriate initial value of the hard jet function at the initial scale $\mu_J^0$. We remind the reader again that to ensure a consistent NLL accuracy in Laplace and cumulative space, in Laplace space one must also keep the $\cO(\alpha_s)$ RG generated logs in the boundary terms. For $H_{J_{\bar{n}}}$, we therefore have 
\begin{align}
\hspace{-10pt}H_{J_{\bar{n}}}\big(\mu_J^0\big) = \frac{1}{2\,M_{\chi}} \left[ 1 + \TaW\, \Gamma_0\, \log^2 \left( \frac{(\mu_J^0)^2}{(\mu_J(s))^2} \right) + \TaW\, \beta_0\, \log \left( \frac{(\mu_J^0)^2}{(\mu_J(s))^2} \right) \right] = \frac{1}{2\,M_{\chi}}\,C_J\,,\hspace{-5pt}
\end{align}
where $C_J$ contains the additional logs.  Combining these results, we can then write down the hard-jet function evolved to the common scale $\mW$ as
\begin{align}
H_{J_{\bar{n}}}(\mW) = \frac{1}{2\,M_{\chi}}\,C_J\, U_J\,.
\end{align}

\subsubsection{Hard Function Evolution}\label{sec:diag_hard}

We now consider the RG equation for the hard function. The hard function satisfies an evolution equation with non-trivial mixing at NLL
\bea
\frac{\text{d}}{\text{d} \log \mu} 
\left(
\begin{array}{c} H_1 \\ H_2 \\ H_3 \\ H_4 
\end{array}\right)
= 
\left(
\begin{array}{cccc}
\vspace{2pt}
\gamma^H_{11} & 0 & \gamma^H_{13} & \gamma^H_{14} \\[1.5pt]
0 & \gamma^H_{22} & 0 & 0 \\[1.5pt]
0 & \gamma^H_{32} & \gamma^H_{33} & 0 \\[1.5pt]
0 & \gamma^H_{42} & 0 & \gamma^H_{44}
\end{array}\right)
\left(
\begin{array}{c} H_1 \\ H_2 \\ H_3 \\ H_4 
\end{array}\right)\,,
\eea
where the explicit form of the entries were given in \Eq{eq:hard_explicit}. We can write this as a diagonal evolution involving the $\Gamma_\cusp$ and the beta function plus an off-diagonal non-cusp contribution
\begin{align}
\hspace{-10pt}\frac{\text{d}}{\text{d} \log \mu} \vec H
= \Gamma_H\, \vec H 
+  4\, \TaW
\left(
\begin{array}{cccc}
0 & 0 & \psi & \psi^* \\
0 & - 6 & 0 & 0 \\
0 & \psi^* & - 3\, \psi & 0 \\
0 & \psi & 0 & - 3\, \psi^*
\end{array}\right)
\vec H\,.
\end{align}
Here $\Gamma_H$ is as defined in \Eq{eq:defGammaH} and $\psi$ is as defined in \Eq{eq:psi}. The remaining non-cusp anomalous dimension can now be diagonalized
\bea
\left(
\begin{array}{cccc}
0 & 0 & \psi & \psi^* \\
0 & - 6 & 0 & 0 \\
0 & \psi^* & - 3 \,\psi & 0 \\
0 & \psi & 0 & - 3\, \psi^*
\end{array}\right)
&= V^{-1}
\left(
\begin{array}{cccc}
-3\, \psi & 0 & 0 & \hphantom{-}0 \\
0 & -3\, \psi^* & 0 & \hphantom{-}0 \\
0 & 0 & -6 & \hphantom{-}0 \\
0 & 0 & 0 & \hphantom{-}0
\end{array}\right)
V\,, 
\eea
where the matrix $V$ defines the change of basis
\be
\left(
\begin{array}{c} H_A \\ H_B \\ H_C \\ H_D 
\end{array}\right)
 = 
 \left(
\begin{array}{rrcc}
0 & 1/3 & \hspace{3pt}1 & 0 \\
0 & 1/3 & \hspace{3pt}0 & 1 \\
0 & -1/3 &\hspace{3pt} 0 & 0 \\
1 & 1/9 & \hspace{3pt}1/3 & 1/3
\end{array}\right)
\left(
\begin{array}{c} H_1 \\ H_2 \\ H_3 \\ H_4 
\end{array}\right)\,.
\ee
In this basis, 
writing $\Gamma_H$ explicity, 
we have the diagonal equations
\begin{align}
\frac{\text{d}}{\text{d} \log \mu} H_A &=  \left[-4\,\Gamma_\cusp[\aW]\log \left( \frac{\mu^2}{\big( 2\,M_{\chi} \big)^2} \right)  -4\, \TaW\, \beta_0  -12\, \TaW\, \psi \right] H_A\,, \nn \\
\frac{\text{d}}{\text{d} \log \mu} H_B &=  \left[-4\,\Gamma_\cusp[\aW]\log \left( \frac{\mu^2}{\big( 2\,M_{\chi} \big)^2} \right)  -4\, \TaW\, \beta_0  -12\, \TaW\, \psi^* \right] H_B\,, \nn\\
\frac{\text{d}}{\text{d} \log \mu} H_C &=  \left[-4\,\Gamma_\cusp[\aW]\log \left( \frac{\mu^2}{\big( 2\,M_{\chi} \big)^2} \right)  -4\, \TaW\, \beta_0  -24\, \TaW \right] H_C\,,\nn \\
\frac{\text{d}}{\text{d} \log \mu} H_D &=  \left[-4\,\Gamma_\cusp[\aW]\log \left( \frac{\mu^2}{\big( 2\,M_{\chi} \big)^2} \right)  -4\, \TaW\, \beta_0 \right] H_D\,,
\end{align}
which we can now easily evolve from the natural scale $\mu_H^0$ (which has the canonical value $2\, M_{\chi}$) down to the scale $\mW$. Writing the non-cusp piece of the anomalous dimension as $-4\,\beta_0+a$, we can now define a hard evolution kernel  
\begin{align}
e^{K_H} \left( \frac{\big(\mu_H^0\big)^2}{\big( 2\, M_{\chi} \big)^2} \right)^{\omega_H} = &\,\,r_H^2 \left( \frac{\big(\mu_H^0\big)^2}{\big( 2\, M_{\chi} \big)^2} \right)^{\omega_H} \exp \left[ -\frac{2\, \Gamma_0}{\beta_0^2} \left[ \frac{1}{\TaW\big(\mu_H^0\big)} \left( \log r_H + \frac{1}{r_H} - 1 \right) \right. \right.\nn\\[5pt]
&\hspace{2cm}\left. \left.+ \left( \frac{\Gamma_1}{\Gamma_0} - \frac{\beta_1}{\beta_0} \right) \left( r_H - 1 - \log r_H \right) - \frac{\beta_1}{2\,\beta_0} \log^2 r_H \right] \right]  \times r_H^{-\frac{a}{2\,\beta_0}} \nn \\[5pt]
\equiv &\,\, U_H \, r_H^{-\frac{a}{2\beta_0}}\,,
\end{align}
where we have also defined
\begin{align}
\omega_H = &\frac{2\,\Gamma_0}{\beta_0} \left[ \log r_H + \TaW\big(\mu_H^0\big) \left( \frac{\Gamma_1}{\Gamma_0} - \frac{\beta_1}{\beta_0} \right) (r_H-1) \right] \,, \nn \\[5pt]
r_H = &\frac{\TaW(\mW)}{\TaW\big(\mu_H^0\big)} = \left[ 1 + 2\, \TaW\big(\mu_H^0\big)\, \beta_0\, \log \left( \frac{\mW}{\mu_H^0} \right) \right]^{-1}\,.
\end{align}
Note that at the canonical scale  the $\omega_H$ contribution will vanish, but it is important to retain in order to estimate the impact of scale variation.
We can then write
\begin{align}
H_A(\mW) &= r_H^{6\psi/\beta_0} U_H\, H_A(\mu_H^0)\,, \nn\\
H_B(\mW) &= r_H^{6\psi^*/\beta_0} U_H\, H_B(\mu_H^0)\,,\nn \\
H_C(\mW) &= r_H^{12/\beta_0} U_H\, H_C(\mu_H^0)\,,\nn \\
H_D(\mW) &= U_H\, H_D(\mu_H^0)\,.
\end{align}
Before inverting, we need the boundary values of the hard functions at their natural scale, which are given by
\begin{align}
H_1\big(\mu_H^0\big) &= H_2\big(\mu_H^0\big) = - H_3\big(\mu_H^0\big) = - H_4\big(\mu_H^0\big)\nn\\[5pt]
&= \frac{\pi^2\, \aW^2\big(\mu_H^0\big)}{M_{\chi}^2} \left[ 1 - \TaW\big(\mu_H^0\big)\, \Gamma_0\, \log^2 \left( \frac{\big(\mu_H^0\big)^2}{\big( 2\, M_{\chi} \big)^2} \right) \right] \nn\\[5pt]
&= \frac{\pi^2\, \aW^2\big(\mu_H^0\big)}{M_{\chi}^2} C_H\,,
\end{align}
where we used the results for the one loop cusp contribution to the hard scale matching coefficients from~\cite{Ovanesyan:2016vkk,Beneke:2018ssm}. Substituting these in we find
\begin{align}
H_1(\mW) &= \frac{\pi^2\, \aW^2\big(\mu_H^0\big)}{9\,M_{\chi}^2}\, U_H\,C_H \left( 2 + r_H^{6\, \psi/\beta_0} \right) \left( 2 + r_H^{6 \,\psi^*/\beta_0} \right)\,,\nn \\
H_2(\mW) &= \frac{\pi^2\, \aW^2\big(\mu_H^0\big)}{M_{\chi}^2}\, U_H\,C_H\, r_H^{12/\beta_0}\,, \nn\\
H_3(\mW) &= - \frac{\pi^2\, \aW^2\big(\mu_H^0\big)}{3\,M_{\chi}^2}\, U_H\,C_H\, r_H^{12/\beta_0} \left( 1 + 2\, r_H^{-6\, \psi^*/\beta_0} \right)\,,\nn \\
H_4(\mW) &= - \frac{\pi^2\, \aW^2\big(\mu_H^0\big)}{3\,M_{\chi}^2}\, U_H\,C_H\, r_H^{12/\beta_0} \left( 1 + 2\, r_H^{-6\, \psi/\beta_0} \right)\,.
\end{align}
Note that although $\psi$ appears in these results, the final result for the cross section will be purely real as it must -- $\psi$ will lead to the appearance of the sine or cosine of a phase. 
This occurs already for $H_1(\mW)$, 
\begin{align}
H_1(\mW) &= \frac{\pi^2\, \aW^2\big(\mu_H^0\big)}{9\,M_{\chi}^2}\, U_H\,C_H \left( 4 + 2\, r_H^{6\, \psi/\beta_0} + 2\, r_H^{6\, \psi^*/\beta_0} + r_H^{12/\beta_0} \right) \nn\\[5pt]
&= \frac{\pi^2\,\aW^2\big(\mu_H^0\big)}{9\,M_{\chi}^2}\, U_H\,C_H \left( 4 + 4\, r_H^{6/\beta_0} \cos \bigg( \frac{6\, \pi}{\beta_0} \, \log r_H \bigg) + r_H^{12/\beta_0} \right)\,,
\end{align}
and we see that a cosine of a phase appears in the resummed result. We will explain the physical origin of these phases in \Sec{sec:glauber}.

\subsubsection{Soft Matching Coefficient Evolution}\label{sec:diag_hardsoft}

The soft matching coefficient $H_S$  satisfies an RG equation involving a $10\times10$ mixing matrix. To diagonalize the system, we perform the following invertible change of basis
\be
\left(
\begin{array}{c}
H_{S,A} \\ H_{S,B} \\ H_{S,C} \\ H_{S,D} \\ H_{S,E} \\ H_{S,F} \\ H_{S,G} \\ H_{S,H} \\ H_{S,I} \\ H_{S,J}
\end{array}\right)
=
\left(
\begin{array}{rrrrrrrrrr}
1 & 0 & 0 & 0 & 0 & \hphantom{-}0 & \hphantom{-}0 & \hphantom{-}0 & \hphantom{-}0 & \hphantom{-}0 \\
-\frac{1}{3} & 1 & \frac{1}{3} & 0 & 0 & 0 & 0 & 0 & 0 & 0 \\
0 & 0 & 1 & 0 & 0 & 0 & 0 & 0 & 0 & 0 \\
-\frac{1}{3} & 0 & 0 & 1 & \frac{1}{3} & 0 & 0 & 0 & 0 & 0 \\
0 & 0 & 0 & 0 & 1 & 0 & 0 & 0 & 0 & 0 \\
\frac{1}{9} & - \frac{1}{3} & - \frac{1}{9} & - \frac{1}{3} & -\frac{1}{9} & 1 & 0 & \frac{1}{3} & \frac{1}{3} & 0 \\
0 & 0 & 0 & 0 & 0 & 0 & 1 & 0 & 0 & 0 \\
0 & 0 & - \frac{1}{3} & 0 & 0 & 0 & \frac{1}{3} & 1 & 0 & 0 \\
0 & 0 & 0 & 0 & - \frac{1}{3} & 0 & \frac{1}{3} & 0 & 1 & 0 \\
0 & 0 & 0 & 0 & 0 & 0 & \frac{1}{3} & 0 & 0 & 1
\end{array}\right)
\left(
\begin{array}{c}
H_{S,11} \\ H_{S,31} \\ H_{S,33} \\ H_{S,41} \\ H_{S,44} \\ H_{S,21} \\ H_{S,22} \\ H_{S,23} \\ H_{S,24} \\ H_{S,25}
\end{array}\right)\,.
\ee
After performing this change of basis, we obtain the following set of decoupled equations
\begin{alignat}{2}
\frac{\text{d}}{\text{d} \log \mu} H_{S,A} &= 0\,, \quad
&&\frac{\text{d}}{\text{d} \log \mu} H_{S,B} = 12\, \TaW\, \psi H_{S,B}\,,\nn \\
\frac{\text{d}}{\text{d} \log \mu} H_{S,C} &= - 6\, \Gamma_{\rm cusp}[\aW] \log\big(\mu\, s\big) H_{S,C}\,,\quad
&&\frac{\text{d}}{\text{d} \log \mu} H_{S,D} = 12\, \TaW\, \psi^* H_{S,D}\,,\nn \\
\frac{\text{d}}{\text{d} \log \mu} H_{S,E} &= - 6\, \Gamma_{\rm cusp}[\aW] \log\big(\mu\, s\big)  H_{S,E}\,, \quad
&&\frac{\text{d}}{\text{d} \log \mu} H_{S,F} = 24\, \TaW\, H_{S,F}\,, \nn\\
\mathrlap{\hspace{-56pt}\frac{\text{d}}{\text{d} \log \mu} H_{S,G} = 3\, \Big( 4\, \TaW - 2\,\Gamma_{\rm cusp}[\aW]\log\big(\mu\, s\big)  \Big) H_{S,G}\,,}\nn\\
\mathrlap{\hspace{-57pt}\frac{\text{d}}{\text{d} \log \mu} H_{S,H} = 3\, \Big( 4\, \TaW\, \psi^*  - 2\,\Gamma_{\rm cusp}[\aW]\log\big(\mu\, s\big)  \Big) H_{S,H}\,,}\nn\\
\frac{\text{d}}{\text{d} \log \mu} H_{S,I} &= 3\, \Big( 4\, \TaW\, \psi - 2\,\Gamma_{\rm cusp}[\aW]\log\big(\mu\, s\big)  \Big) H_{S,I}\,, \quad
&&\frac{\text{d}}{\text{d} \log \mu} H_{S,J} = 24\, \TaW\, H_{S,J}\,. 
\end{alignat}
We can now solve these equations to evolve the functions from the high scale $\mu_S^0$ (with canonical value $2\, M_{\chi}(1-z)$) down to $\mW$. To do so, we define the soft evolution kernel 
\begin{align}
U_S\big(\mW, \mu_S^0; \mu_S(s)\big) & = \Big( \left( V_S - 1 \right) \Theta_S + 1 \Big) \left( \frac{\mu_S^0}{\mu_S(s)} \right)^{2\,\omega_S}\,,
\end{align}
which is given in terms of
\begin{align}
V_S &= \exp \left\{ - \frac{3\,\Gamma_0}{2\,\beta_0^2} \left[ \frac{1}{\TaW\big(\mu_S^0\big)} \left( \log r_S + \frac{1}{r_S} - 1 \right) + \left( \frac{\Gamma_1}{\Gamma_0} - \frac{\beta_1}{\beta_0} \right) \left( r_S - 1 - \log r_S \right) - \frac{\beta_1}{2\,\beta_0} \log^2 r_S \right] \right\}\,, \nn\\[5pt]
\omega_S &= \frac{3\,\Gamma_0}{2\,\beta_0} \left[ \log r_S + \TaW\big(\mu_S^0\big) \left( \frac{\Gamma_1}{\Gamma_0} - \frac{\beta_1}{\beta_0} \right) (r_S-1) \right] \Theta_S\,, \nn \\[5pt]
r_S &= \frac{\TaW(\mW)}{\TaW(\mu_S^0)} = \left[ 1 + 2\, \TaW\big(\mu_S^0\big)\, \beta_0 \,\log \left( \frac{\mW}{\mu_S^0} \right) \right]^{-1}\,, \nn \\[5pt]
\Theta_S &= \Theta \Big( 2\, M_{\chi}\,(1-z) - \mW \Big)\,,
\end{align} 
where we have written the results in terms of the natural Laplace variable scale,\footnote{Note that we use $S$ to denote soft and $s$ is the Laplace space variable.}
\begin{align}
\mu_S(s) &= \frac{1}{s}\,.
\end{align}
The $\Theta_S$ functions prevent the soft scale from going below $\mW$.
Inverting to our original basis, and using the boundary values
\begin{alignat}{2}
H_{S,11}\big(\mu_S^0\big) &= 1\,,\qquad
&&H_{S,33}\big(\mu_S^0\big) = H_{S,44}\big(\mu_S^0\big) = 1 - 12\, \TaW\, L_S^2(s)\,,\nn \\
H_{S,31}\big(\mu_S^0\big) &= H_{S,41}\big(\mu_S^0\big) = 4\, \TaW\, L^2_S(s)\,,\qquad
&&H_{S,22}\big(\mu_S^0\big) = 1 - 12\, \TaW\, L^2_S(s) + 12\, \TaW\, L_S(s)\,, \nn\\
H_{S,23}\big(\mu_S^0\big) &= H_{S,24}\big(\mu_S^0\big) = - 4\, \TaW\, L_S(s)\,,\qquad
&&H_{S,25}\big(\mu_S^0\big) = 4\, \TaW\, L^2_S(s) + 4\, \TaW\, L_S(s)\,,
\end{alignat}
where 
\begin{align}
L_S(s) = \log \left(\frac{\mu_S^0}{\mu_S(s)}\right)\,,
\end{align} 
and all other boundary coefficients are zero,
we obtain
\begin{align}
H_{S,11}(\mW) &= 1\,, \nn\\
H_{S,31}(\mW) &= H_{S,41}(\mW) = \frac{1}{3} - \frac{1}{3}\, U_S\, \lambda_S^b(s)\,, \nn\\
H_{S,33}(\mW) &= H_{S,44}(\mW) = U_S\, \lambda_S^b(s)\,,\nn \\
H_{S,21}(\mW) &= \frac{1}{9} \left( 1 - 2\,U_S\,\lambda_S^b(s) \right) - \frac{1}{9}\, r_S^{-12/\beta_0}\, \lambda_S^c(s) + \frac{2}{9}\, r_S^{-6/\beta_0}\, U_S\, \lambda_S^a(s)\,, \nn \\
H_{S,22}(\mW) &= r_S^{-6/\beta_0}\, U_S \, \lambda_S^a(s) \,,\nn \\
H_{S,23}(\mW) &= H_{S,24}(\mW) = - \frac{1}{3}\, r_S^{-6/\beta_0}\, U_S \, \lambda_S^a(s) + \frac{1}{3}\, U_S\, \lambda_S^b(s)\,,\nn \\
H_{S,25}(\mW) &= - \frac{1}{3}\, r_S^{-6/\beta_0}\, U_S \, \lambda_S^a(s) + \frac{1}{3}\, r_S^{-12/\beta_0}\, \lambda_S^c(s)\,.
\end{align}
Here to simplify the expressions we have defined
\begin{align}
\lambda_S^a(s) &= 1 - 12\, \TaW\, L_S(s) (L_S(s) - 1)\,, \nn\\
\lambda_S^b(s) &= 1 - 12\, \TaW\, L^2_S(s)\,,\nn \\
\lambda_S^c(s) &= 1 + 24\, \TaW\, L_S(s)\,.
\label{eq:DsEsFs}
\end{align}
Recall that the remaining ten $H_{S,ij}(\mW)$ functions not listed here are zero. This solution resums all logarithms appearing in the hard-soft function to NLL accuracy.

\section{Analytic Resummed Cross Section at NLL Accuracy}\label{sec:answer}
We now have all the relevant pieces to derive analytic cumulative and differential endpoint spectra for wino annihilation at NLL accuracy.
The NLL cross section is given by
\begin{align} \label{eq:NLLpre}
\hspace{-3pt}\left(\frac{\df \sigma}{\df z}\right)^{\text{NLL}}\! &=   \frac{2\,\pi\, \aW^2\big(\mu_H^0\big) \sW^2\big(\mu_{\gamma}^0\big)}{9\, M_{\chi}\, v}\, C_{\gamma}\,C_H\,U_H\, \nn \\
&\hspace{10pt}\text{\bf LP}^{-1} \bigg \{ U_J\bigg(\hspace{5pt} \big|s_{00}\big|^2  \Big[ 
4\, \Lambda^d(s) + 2 \,r_{HS}^{12/\beta_0} \Lambda^c(s) \Big] 
+ \big|s_{0\pm}\big|^2 \bigg[ 
8\, \Lambda^d(s) + r_{HS}^{12/\beta_0} \Lambda^c(s) \bigg] \nn \\
&\hspace{73pt}+ \sqrt{2} \operatorname{Re}\!\Big[s_{00}\, s_{0\pm}^*\Big]\bigg[ 
8 \,\Lambda^d(s) - 2\,r_{HS}^{12/\beta_0} \Lambda^c(s) \bigg] \bigg) \nn \\
&\hspace{45pt}+ U_J\, U_S \, r_H^{6/\beta_0}\bigg(\hspace{5pt} \big|s_{00}\big|^2  \Big[ 2 \,r_{HS}^{6/\beta_0} \Lambda^a(s) - 8\, c_H\, \Lambda^b(s) \Big]  \nn \\
&\hspace{113pt}+ \big|s_{0\pm}\big|^2 \bigg[ r_{HS}^{6/\beta_0}\, \Lambda^a(s)
+8\, c_H\, \Lambda^b(s) \bigg]  \nn\\
&\hspace{113pt}+ \sqrt{2} \operatorname{Re}\!\Big[s_{00}\, s_{0\pm}^*\Big]\bigg[ - 2\,r_{HS}^{6/\beta_0}\, \Lambda^a(s)-4\, c_H\, \Lambda^b(s) \bigg] \nn\\
&\hspace{113pt}+ \sqrt{2} \operatorname{Im}\!\Big[s_{00}\, s_{0\pm}^*\Big]\bigg[ -12\, s_H\, \Lambda^b(s) \bigg] \bigg) \bigg \}\,,
\end{align}
where $\text{\bf LP}^{-1}$ denotes the inverse Laplace transform as defined in \Eq{eq:LP_inv_def}. 
In addition to previously defined shorthand notation, we have also defined the phases
\begin{align}
c_H = \cos \left( \frac{6\, \pi}{\beta_0} \log r_H\right)\,,\;\;
s_H = \sin \left( \frac{6\, \pi}{\beta_0} \log r_H\right)\,,
\end{align}
as well as
\be
r_{HS} = \frac{r_H}{r_S}\,.
\ee
Finally the $\Lambda^{a-d}$ expressions result from expanding the product $\lambda^{a-c}_S \times C_J$ and keeping only the terms relevant at NLL order.  Note that $\Lambda^d = C_J$, where we have chosen to redefine it to make the notation consistent. In detail, we have
\begin{align}\label{eq:CapitalLambda}
\Lambda^a &= 1 + \TaW\, \Gamma_0\, L^2_J(s) + \TaW\, \beta_0\, L_J(s) - 12\, \TaW\, L_S(s) \big(L_S(s) - 1\big)\,, \nn\\
\Lambda^b &= 1 + \TaW\, \Gamma_0\, L^2_J(s) + \TaW\, \beta_0\, L_J(s) - 12\, \TaW\, L^2_S(s)\,,\nn \\
\Lambda^c &= 1 + \TaW\, \Gamma_0\, L^2_J(s) + \TaW\, \beta_0\, L_J(s) + 24\, \TaW\, L_S(s)\,,\nn \\
\Lambda^d &= 1 + \TaW\, \Gamma_0\, L^2_J(s) + \TaW\, \beta_0\, L_J(s)\,,
\end{align}
where in analogy with $L_S(s)$ we have 
\be
L_J(s) = \log \left( \frac{\big(\mu_J^0\big)^2}{\big(\mu_J(s)\big)^2} \right)\,.
\ee
The complexity of this result is both due to the multiple gauge index structures in the hard and hard-soft functions, and their contractions with the Sommerfeld factors.

To perform the inverse Laplace transform analytically, we set scales in cumulative space. The natural scales of the functions in Laplace space are therefore taken to be formally independent of the Laplace space variable $s$. The only required transform between Laplace space and cumulative space is 
\begin{align}\label{eq:LaplaceSub}
\big(s \,\mu_0\big)^q \big(s \, \mu'_0\big)^{q'} \to \frac{1}{2\,M_{\chi}} \left(\frac{e^{\gamma_E} \mu_0}{t}   \right)^q \left(\frac{e^{\gamma_E} \mu'_0}{t}   \right)^{q'} \frac{1}{\Gamma[1-q-q']}\,.
\end{align}
To deal with logs of the form\footnote{These arguments of the logarithms are chosen due to the square root entering the jet scale, see \Eq{eq:jet_natural_scale}. } 
\be
\log \left( \frac{\big(\mu_J^0\big)^2}{\big(\mu_J(s)\big)^2} \right)\quad \text{and} \quad\log \left( \frac{\mu_S^0}{\mu_S(s)} \right)\,,
\ee
which appear in the $\Lambda^i(s)$ expressions, we use that in Laplace space all of these terms appear multiplied by an expression of the form $s^q$. We can therefore rewrite these logs in terms of derivatives using 
\be
\partial_q^n s^q =  s^q \log^n s\,.
\ee
Once rewritten in this manner, the logs no longer depend on $s$ and the only objects we need to Laplace transform are of the form given in \Eq{eq:LaplaceSub}. 
In detail, the substitutions made are
\begin{align}\label{eq:derivsub}
L_J(s) &\to \partial_{\omega_J}\,, \qquad
L_S(s) \to \partial_{2\, \omega_S}\,.
\end{align}
The derivatives can then be evaluated analytically once the Laplace transform has been performed. These differential operators generate logarithms and polygamma functions. For detailed results see App.~\ref{app:explicitxsec}. 

Using \eq{NLLpre} and these steps we
can then derive the following expression for the cumulative spectrum, as defined in \Eq{eq:cumulative_def_intro}:
\\
\\
\noindent\begin{minipage}{\linewidth}
\begin{mdframed}[linewidth=1.5pt, roundcorner=10pt]
\begin{align}\label{eq:final_cumulant}
\sigma^{\text{NLL}}(\zcut)=\,& \frac{\pi\, \aW^2\big(2\, M_{\chi}\,) \sW^2\big(\mW\big)}{9\, M_{\chi}^2\, v} \,U_H \,\big((V_J-1)\, \Theta_J+1\big)\nn \\
&\Bigg \{\bigg(\hspace{5pt} \big|s_{00}\big|^2  \Big[ 
4\,\Lambda^d + 2\, r_{HS}^{12/\beta_0} \Lambda^c \Big] 
+ \big|s_{0\pm}\big|^2 \bigg[ 
8\,\Lambda^d + r_{HS}^{12/\beta_0} \Lambda^c \bigg] \nn \\
&\hspace{21pt}+ \sqrt{2}\, \operatorname{Re}\!\Big[s_{00}\, s_{0\pm}^*\Big]\bigg[ 
8\,\Lambda^d - 2\,r_{HS}^{12/\beta_0} \Lambda^c \bigg] \bigg) \frac{e^{\gamma_E\,\omega_J}}{\Gamma\big(1-\omega_J\big)} \nn \\
&\hspace{8pt}+ \big((V_S-1)\, \Theta_S+1\big) \, r_H^{6/\beta_0} \nn \\
&\hspace{20pt}\bigg(\hspace{5pt} |s_{00}|^2  \Big[ 
2 \,r_{HS}^{6/\beta_0} \Lambda^a - 8\, c_H\, \Lambda^b \Big] 
+ \big|s_{0\pm}\big|^2 \bigg[ r_{HS}^{6/\beta_0}\, \Lambda^a
+8\, c_H\, \Lambda^b \bigg]  \nn\\
&\hspace{35pt}+ \sqrt{2}\, \operatorname{Re}\!\Big[s_{00}\, s_{0\pm}^*\Big]\bigg[ 
- 2\,r_{HS}^{6/\beta_0}\, \Lambda^a -4\, c_H\, \Lambda^b \bigg] \nn \\
&\hspace{35pt}+ \sqrt{2} \operatorname{Im}\!\Big[s_{00}\, s_{0\pm}^*\Big]\bigg[ -12\, s_H\, \Lambda^b \bigg] \bigg) 
\frac{e^{\gamma_E(\omega_J+2\,\omega_S)}}{\Gamma\big(1-\omega_J-2\,\omega_S\big)}
\Bigg \}\,.
\end{align}
\end{mdframed}
\end{minipage}\vspace{0.3cm}
Here, for simplicity, we have only given the result evaluated with all scales at their canonical values. In addition we no longer show an $s$ argument on the $\Lambda^i$ expressions, as all  logarithms have now been replaced by derivatives according to \Eq{eq:derivsub}. Results for general scales, as are required to generate scale variations, are given in App.~\ref{app:explicitxsec}.

The differential spectrum in $z$ is generated by differentiating the result in \Eq{eq:final_cumulant} by $z$. We find: 
\\
\\
\noindent\begin{minipage}{\linewidth}
\begin{mdframed}[linewidth=1.5pt, roundcorner=10pt]
\begin{align}\label{eq:final_spectrum}
\left(\frac{\text{d}\sigma}{\text{d}z}\right)^{\text{NLL}} =\,& \frac{\pi\, \aW^2\big(2\, M_{\chi}\big) \sW^2\big(\mW\big)}{9\, M_{\chi}^2\, v\,(1-z)}\, U_H\, \big((V_J-1)\, \Theta_J+1\big)\nn \\
&\Bigg \{\bigg(\hspace{5pt} \big|s_{00}\big|^2  \Big[ 
4\,\Lambda^d + 2\, r_{HS}^{12/\beta_0} \Lambda^c \Big] 
+ \big|s_{0\pm}\big|^2 \bigg[ 
8\,\Lambda^d + r_{HS}^{12/\beta_0} \Lambda^c \bigg] \nn \\
&\hspace{21pt}+ \sqrt{2}\, \operatorname{Re}\!\Big[s_{00}\, s_{0\pm}^*\Big]\bigg[ 
8\,\Lambda^d - 2\,r_{HS}^{12/\beta_0} \Lambda^c \bigg] \bigg) \frac{e^{\gamma_E\,\omega_J}}{\Gamma\big(-\omega_J\big)} \nn \\
&\hspace{8pt}+ \big((V_S-1) \,\Theta_S+1\big) \, r_H^{6/\beta_0} \nn \\
&\hspace{20pt}\bigg(\hspace{5pt} |s_{00}|^2  \Big[ 
2 \,r_{HS}^{6/\beta_0} \Lambda^a - 8\, c_H\, \Lambda^b \Big] 
+ \big|s_{0\pm}\big|^2 \bigg[ r_{HS}^{6/\beta_0}\, \Lambda^a
+8\, c_H\, \Lambda^b \bigg]  \nn\\
&\hspace{35pt}+ \sqrt{2}\, \operatorname{Re}\!\Big[s_{00}\, s_{0\pm}^*\Big]\bigg[ 
- 2\,r_{HS}^{6/\beta_0}\, \Lambda^a -4\, c_H\, \Lambda^b \bigg] \nn \\
&\hspace{35pt}+ \sqrt{2} \operatorname{Im}\!\Big[s_{00}\, s_{0\pm}^*\Big]\bigg[ -12\, s_H\, \Lambda^b \bigg] \bigg) 
\frac{e^{\gamma_E(\omega_J+2\,\omega_S)}}{\Gamma\big(-\omega_J-2\,\omega_S\big)}
\Bigg \}\nn \\
&+ \sigma_{\rm exc}^{\rm NLL} \delta(1-z)\,.
\end{align}
\end{mdframed}\end{minipage}\vspace{0.3cm}
Here $\sigma_{\rm exc}^{\rm NLL}$ is the NLL exclusive cross section, which is given by
\begin{align}
\!\!\!\sigma_{\rm exc}^{\rm NLL} = &\,\frac{\pi\, \aW^2\big(2\,M_{\chi}\big) \sW^2\big(\mW\big)}{9\, M_{\chi}^2\, v}\, U_H \nn \\[3pt]
&\times \bigg\{ \left[ 4 + 4\, r_H^{12/\beta_0} - 8\, r_H^{6/\beta_0}  c_H \right] \big|s_{00}\big|^2
+ \left[ 8 + 2\, r_H^{12/\beta_0} + 8\, r_H^{6/\beta_0}  c_H \right] \big|s_{0\pm}\big|^2 \nn \\
&\hspace{13pt}+ \sqrt{2} \left[ 8 - 4\, r_H^{12/\beta_0} - 4\, r_H^{6/\beta_0}  c_H  \right] \operatorname{Re}\!\Big[s_{00}\, s_{0\pm}^*\Big] 
- 12 \sqrt{2}\, r_H^{6/\beta_0}  s_H \operatorname{Im}\!\Big[s_{00}\, s_{0\pm}^*\Big] \bigg\}\,.
\end{align}
Our result for $\sigma_{\rm exc}^{\rm NLL}$ reproduces the original NLL result for the wino from~\cite{Ovanesyan:2014fwa}.  The analogous line result for scalar DM at NLL was computed in~\cite{Bauer:2014ula}. In App.~\ref{app:intbrem} we verify that our resummed result expanded to NLO, exactly reproduces existing fixed order wino calculations. We also briefly comment on how internal bremsstrahlung processes, which have received considerable interest in the literature, are reproduced in our framework.

This provides a closed form NLL result which simultaneously resums all logarithms of $(1-z)$ and $\mW/M_\chi$  and correctly incorporates Sommerfeld enhancement effects. 
\eqs{final_cumulant}{final_spectrum} are the main results of this paper.
Note that in \Eq{eq:final_spectrum} the $\Theta$ functions cut off the $1/(1-z)$ power law singularities. Although this result is considerably more complicated than the corresponding LL expression presented in \cite{Baumgart:2017nsr}, it simply dresses the same $1/(1-z)$ power law growth with additional logarithms. This structure persists to all logarithmic orders.

\subsection{Non-Vanishing Electroweak Glauber Phase}\label{sec:glauber}

Here we briefly comment on an interesting effect appearing in our final resummed result that occurs when electroweak charged external states are present. This effect is not specific to the case we are considering here, and indeed it also appears in fully exclusive calculations with charged electroweak states in the initial or final states (for the heavy dark matter annihilation case, this includes \cite{Ovanesyan:2014fwa,Bauer:2014ula,Beneke:2018ssm}, and in the collider context from an EFT perspective~\cite{Chiu:2007yn,Chiu:2008vv,Chiu:2007dg,Manohar:2014vxa,Manohar:2018kfx}). However, recent advances in the understanding of the treatment of Glauber gauge bosons in SCET~\cite{Rothstein:2016bsq} give a clean interpretation of these terms.  Since this connection has not previously been emphasized, we briefly deviate from our main goal to explain it here. 

Our final result for the cross section involves the following phases, 
\bea\label{eq:phase_Glaub_section}
c_H = \cos \left( \frac{6\, \pi}{\beta_0} \log r_H\right)\,,\qquad
s_H = \sin \left( \frac{6\, \pi}{\beta_0} \log r_H\right)\,.
\eea
These first appear at NLL in both the hard function $H$ and the soft function $S$. These phases arise because the $\psi=1-i\,\pi$ and $\psi^*=1+i\,\pi$ do not fully cancel in the cross section. This lack of cancellation has a physical origin when understood in terms of Glauber gauge boson contributions.  In our factorization, the soft function describes soft modes with a homogeneous scaling
\begin{align}
\big(n\!\cdot\! p, \bn \!\cdot\! p, p_{\perp}\big)_\text{soft} \sim Q \,\Big(\la^2,\la^2,\la^2\Big)\,,
\end{align}
corresponding to on-shell degrees of freedom in the EFT.
However, when evaluating the virtual soft integrals, we also integrate over regions of phase space where 
\begin{align}
\big(n\!\cdot\! p, \bn \!\cdot\! p, p_{\perp}\big)_\text{Glauber} \sim Q \,\Big(\la^a,\la^b,\la\Big)\,,
\end{align}
with $a+b<2$.
This is the scaling for the so-called Glauber region, which corresponds to off-shell exchanges that are instantaneous  in both lightcone times. 

\begin{figure}
\centerline{\scalebox{.4}{\includegraphics{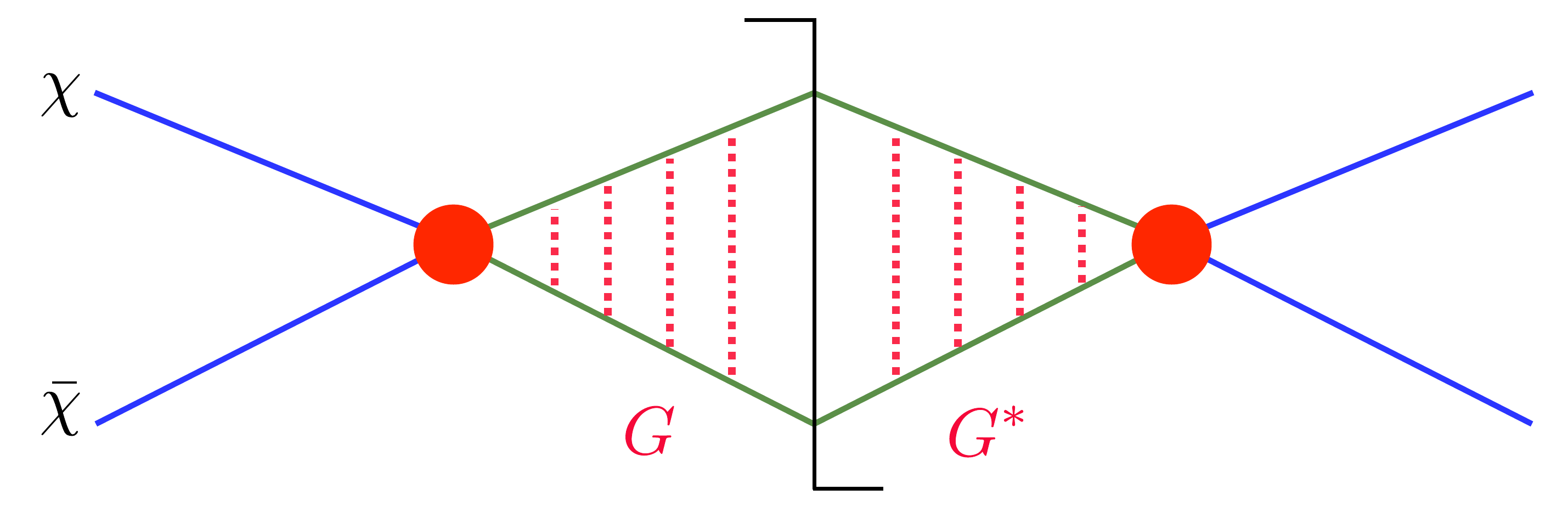}}}
\vspace{-10pt}
\caption[1]{This figure illustrates Glauber bursts $G$. For identical gauge index structures, the Glauber contributions to  the squared amplitude multiply their conjugates and cancel. In the presence of electroweak charged final and initial states, different gauge index structures are present on either side of the cut.  There can then be a miscancellation of the Glauber contributions, leaving behind the residual phase that contributes to our NLL result.}
\label{fig:Glaubers} 
\end{figure}

In our calculation, we do not treat the Glaubers as being separate from the soft modes.   The fact that they are captured by the soft function in a hard scattering is due to their ``cheshire'' nature~\cite{Rothstein:2016bsq}. Once properly defined and regulated, Glaubers do not cross the cut, nor do they connect the initial to the final state. Example diagrams are shown in \Fig{fig:Glaubers}, illustrating the final state Glauber bursts $G$.  These off-shell Glauber contributions to the virtual corrections give rise to the $i\,\pi$ appearing in the one-loop amplitude, \Eq{eq:phase_Glaub_section}. 
These $i\, \pi$ contributions at the level of the amplitude are ubiquitous.  What is interesting here is that a phase contribution survives at the cross section level, since in many cases Glauber effects cancel for inclusive cross sections. 

To understand why we are left with a phase in the cross section, we consider calculating the Glauber contribution to the $S_{12}$ soft function (the soft function arising from the interference of the two distinct gauge index structures). Computing the two relevant diagrams
\begin{align}
\fd{6.3cm}{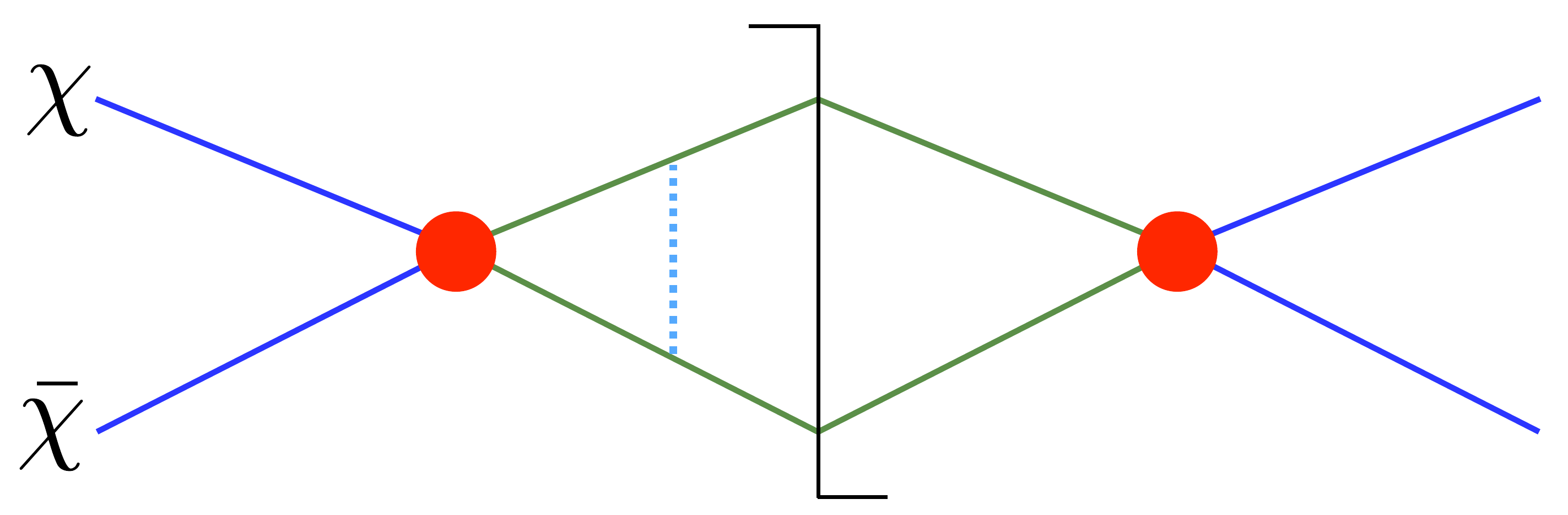}+\fd{6.3cm}{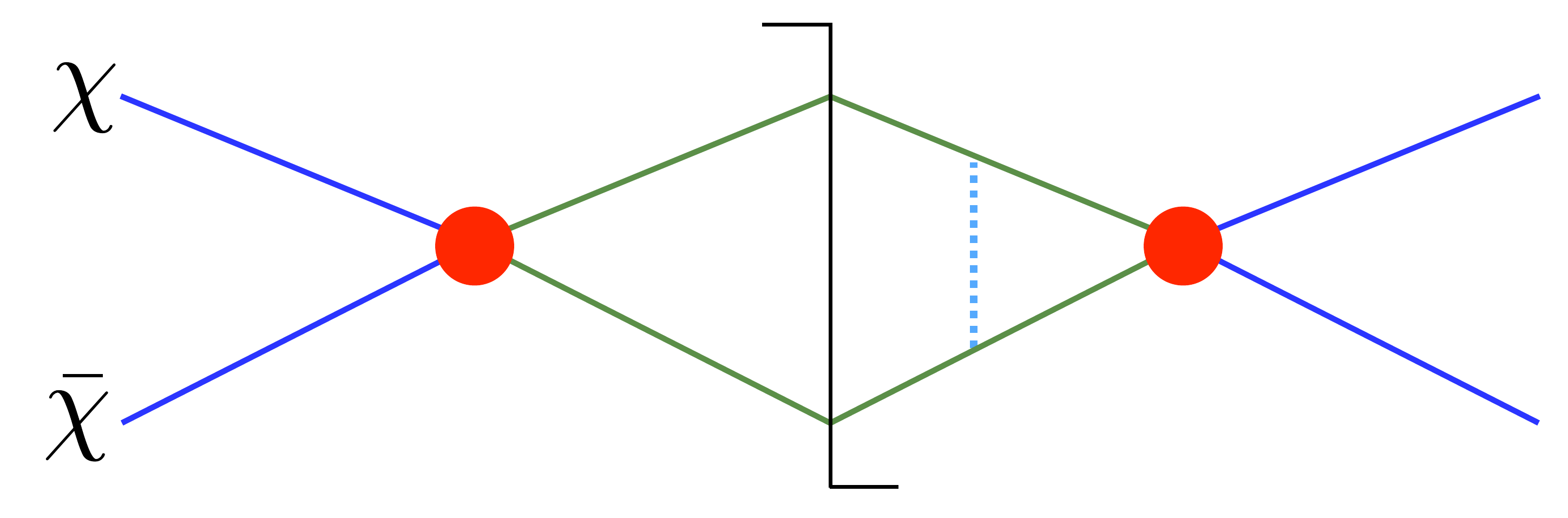}\,,
\end{align}
which have Glaubers on either side of the cut,
we find
\begin{align}\label{eq:Glaub_s12}
S_{12}^{G, a'b'ab}= \delta^{a'b'} \Big(\delta^{ab}-3\,\delta^{a3}\delta^{b3}  \Big)(i\,\pi) \frac{\aW}{\pi}\log \left(\frac{\mu}{\mW}\right)\,,
\end{align}
which is non-zero. Here the superscript $G$ denotes the Glauber contribution. The two diagrams give different gauge index structures, which sum to the result shown here. \eq{Glaub_s12} gives the $i\,\pi$ terms that are included in our soft function $S$, which are related to the ones in $H$ by renormalization group consistency.

Note that if the external states were electroweak singlets, electroweak charge conservation would imply that the two diagrams would have the same gauge index structure.  Then the two diagrams would be exactly conjugate, and the imaginary Glauber contribution vanishes, $\psi+\psi^*=\text{Re}[\psi]$.  In \Eq{eq:Glaub_s12}, this can be seen by contracting the result with a gauge index singlet final state
\begin{align}
\delta^{ab}S_{12}^{G, a'b'ab}= \delta^{ab}\delta^{a'b'} \Big(\delta^{ab}-3\,\delta^{a3}\delta^{b3}  \Big)(i\,\pi) \frac{\aW}{\pi}\log \left(\frac{\mu}{\mW}\right)=0\,.
\end{align}
However, for wino annihilation the electroweak charges of the initial and final states are non-singlet, and Glauber phases contribute to the cross section.  Furthermore, the Glauber $i\, \pi$ multiplies a logarithm, which once resummed yields the phases in~\Eq{eq:phase_Glaub_section}.  This can be viewed as a manifestation of the KLN violation in the Glauber sector, and is a completely general phenomenon when one has multiple electroweak charged initial or final states. This cancellation (or lack thereof) extends to multiple Glauber exchanges. It would be interesting to investigate the properties of the (non-) cancellation of electroweak Glaubers further. This is beyond the scope of this work since 
for our current application these phases appear in the hard-soft matching coefficient, and are correctly captured within our framework.

\section{Numerical Results and Uncertainties}\label{sec:results}

\begin{figure}[h!]
\begin{center}
\subfloat[]{
\includegraphics[width=6.9cm]{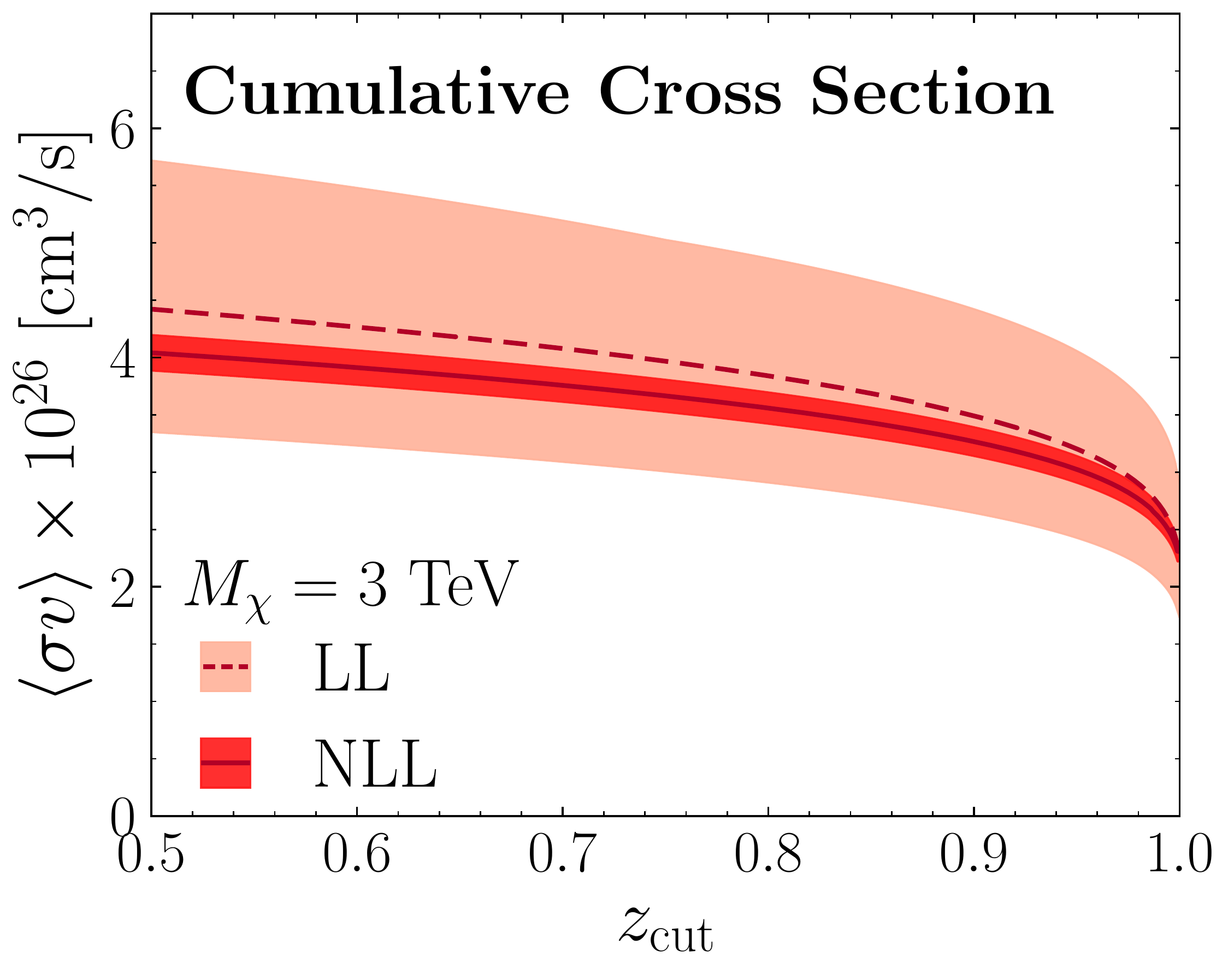}    
}\hspace{5pt}
\subfloat[]{
\includegraphics[width=6.9cm]{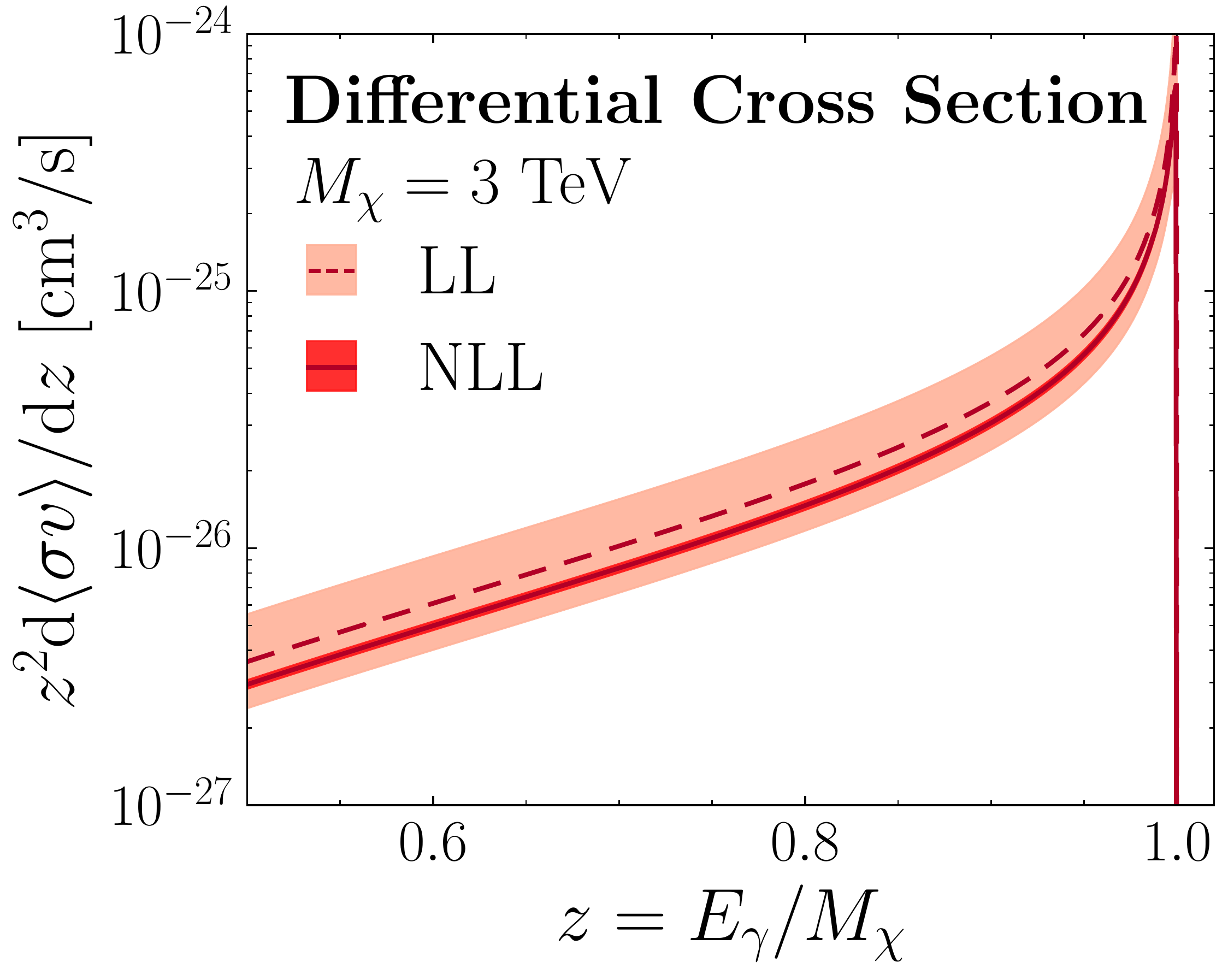}
}
\\
\subfloat[]{
\includegraphics[width=6.9cm]{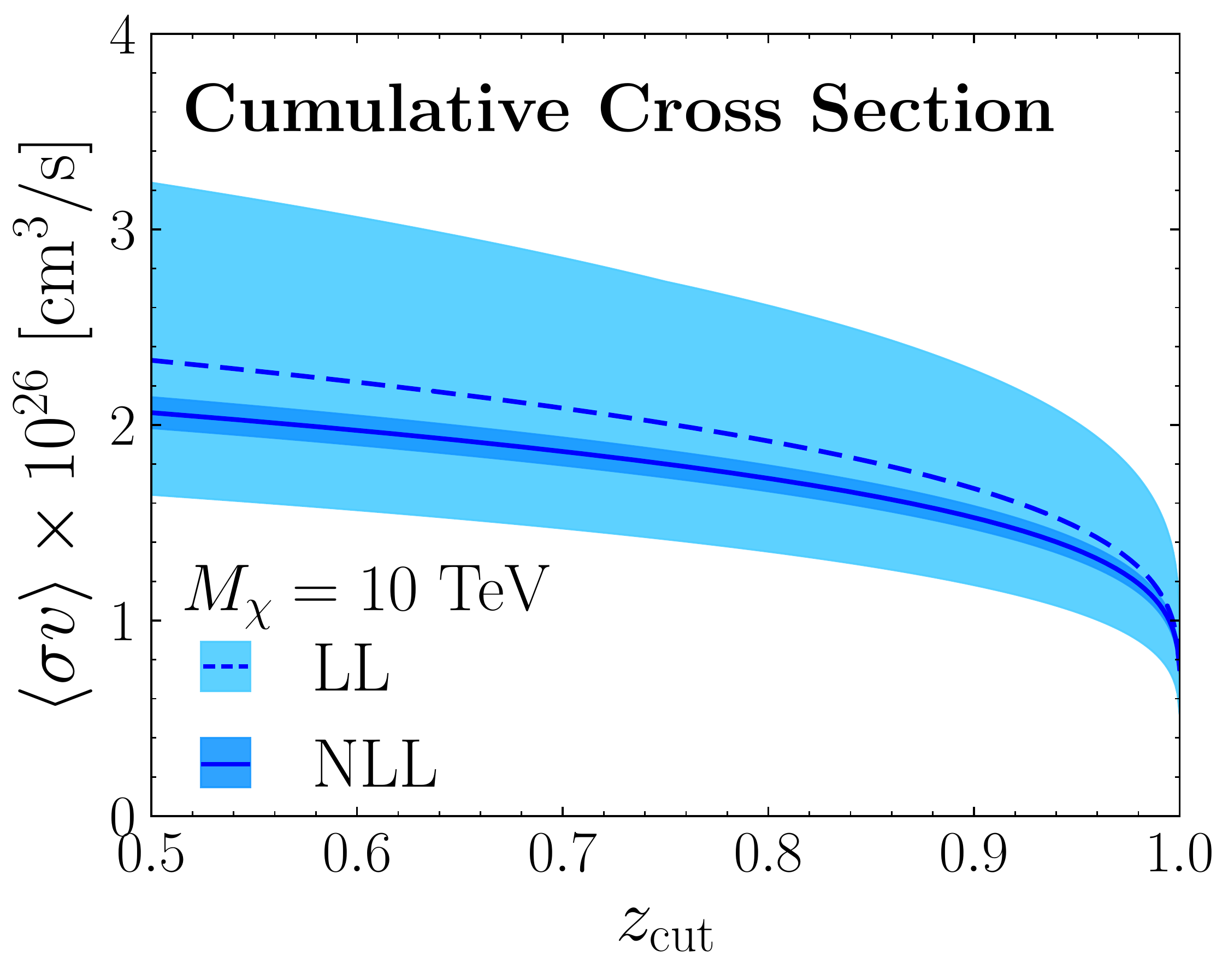}    
}\hspace{5pt}
\subfloat[]{
\includegraphics[width=6.9cm]{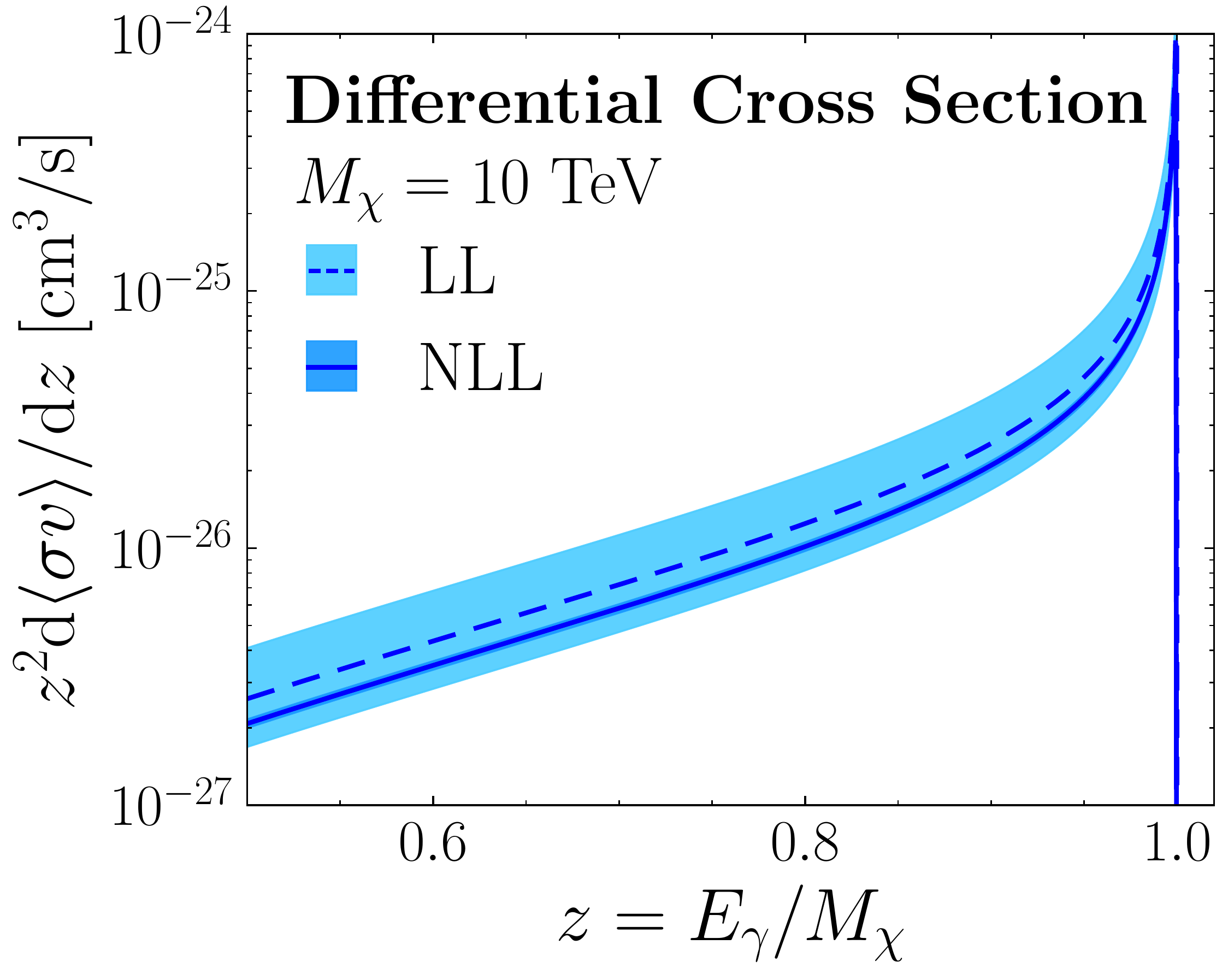}
}\\
\subfloat[]{
\includegraphics[width=6.9cm]{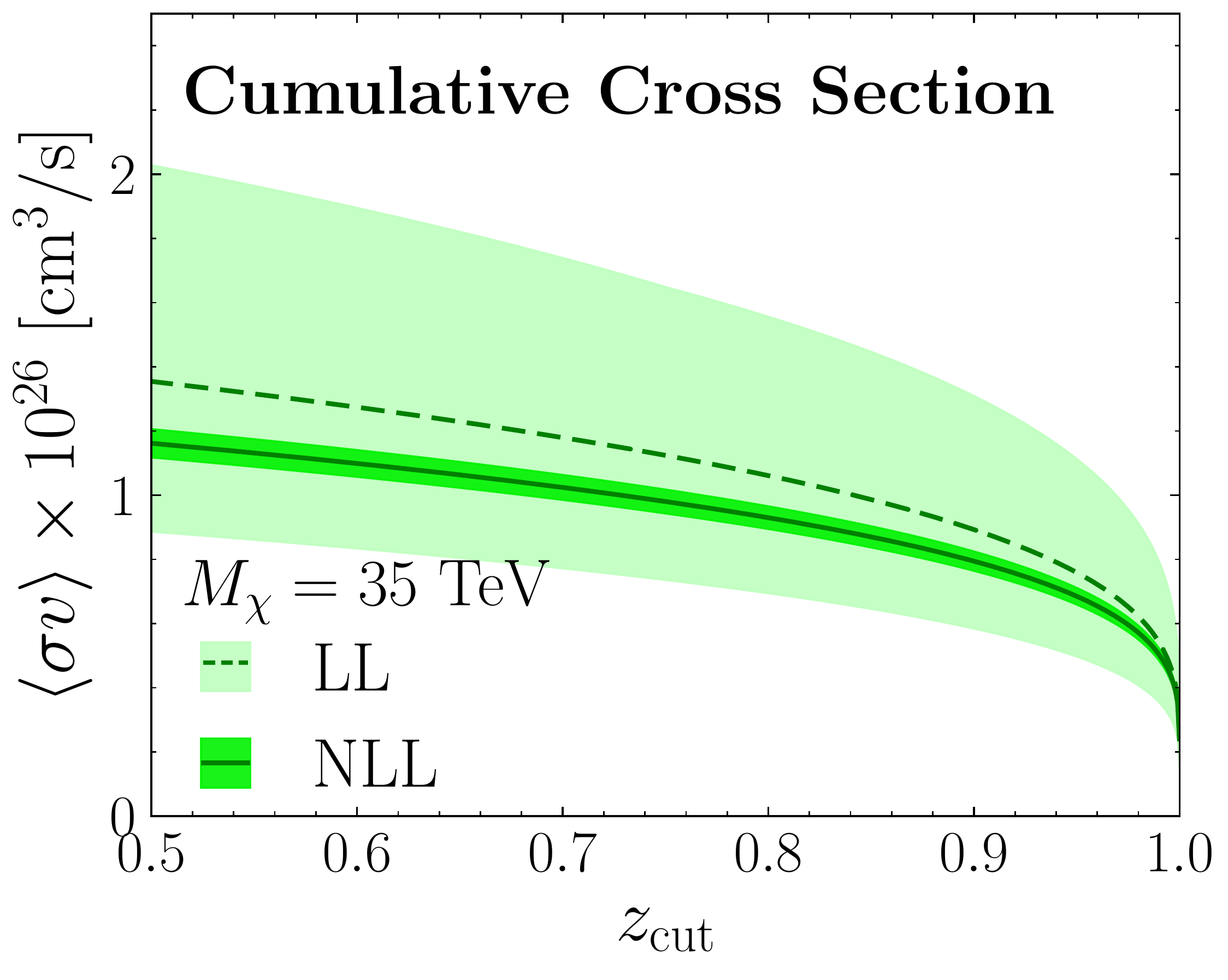}    
}\hspace{5pt}
\subfloat[]{
\includegraphics[width=6.9cm]{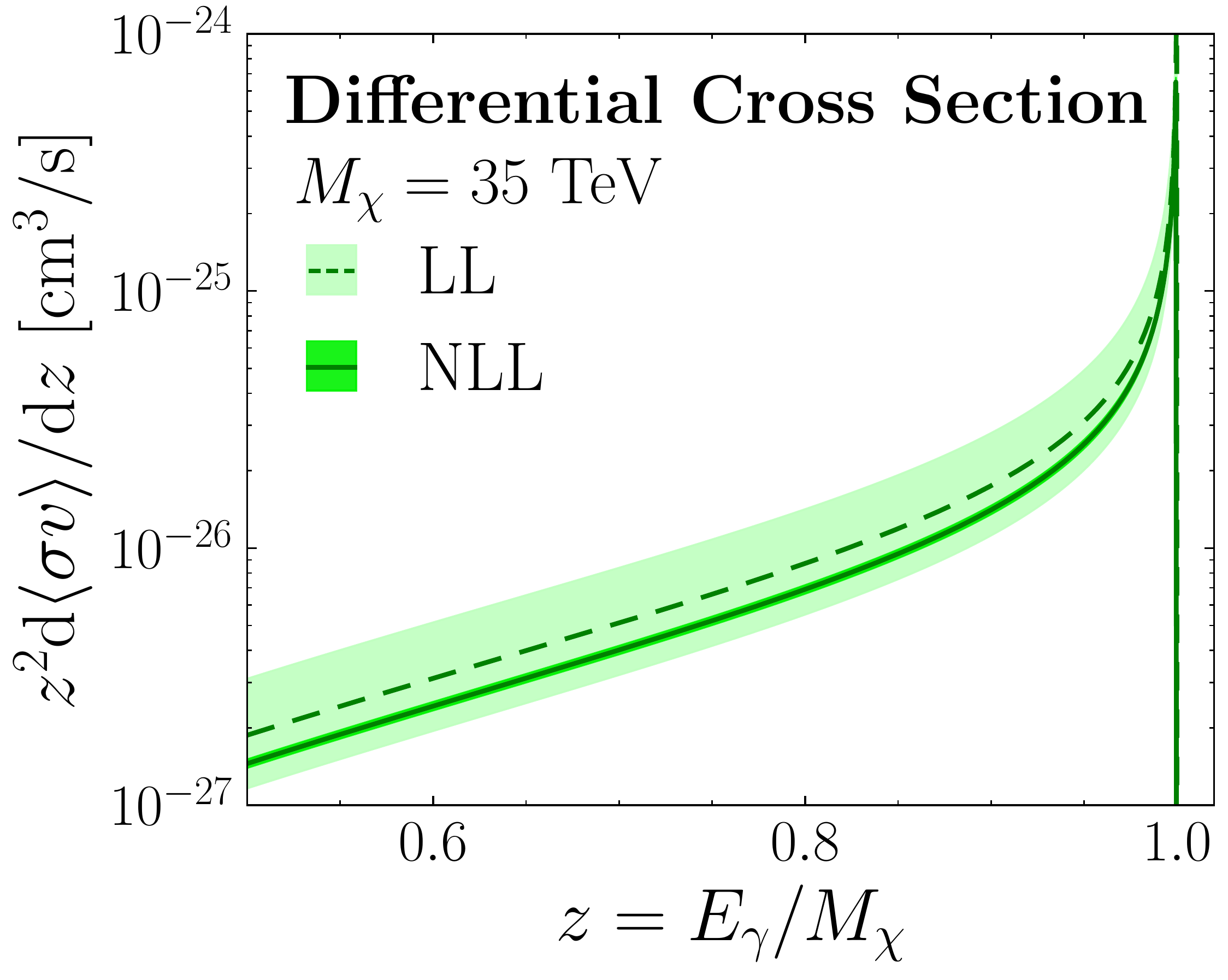}
}
\end{center}
\vspace{-10pt}
\caption{The cumulative (left column) and $z^2$ weighted differential (right column) spectrum for wino annihilation in the endpoint region. Spectra are shown for three different wino masses, $3$, $10$ and $35$ TeV, at both LL and NLL accuracy. Theoretical uncertainties obtained by scale variation are shown by the shaded bands -- the uncertainty bands are hardly visible at NLL at high masses.
\vspace{-30pt}
}
\label{fig:diff}
\end{figure}

In this section, we present numerical results using our NLL formula. In particular, we focus on the reduction in the theoretical uncertainty as compared with LL. The uncertainty bands are generated by scale variations, which probes higher order logarithms. Due to the structure of the $\mu$- and $\nu$-anomalous dimensions, as described in \Sec{sec:NLL}, we are able to choose a path up to NLL accuracy that does not require rapidity evolution (see \Fig{fig:RG_path}). Since no logarithms at NLL are generated by $\nu$-evolution, when performing the scale variation at LL, all logarithms at NLL can be probed by $\mu$-scale variations. The LL uncertainty bands are generated by varying the $\mu$-scales about their central value by a factor of $2$.

The NLL error bands probe next-to-next-to-leading logarithmic (NNLL) logarithms.  Note that at NNLL one has a non-trivial rapidity evolution 
between the collinear-soft and soft functions, so that they could not be combined into a single function as was done here. To capture this in our uncertainty estimate, our NLL uncertainty band requires both a $\mu$- and $\nu$-variation: explicitly we move both scales 
independently up and down by a factor of $2$ for the photon jet function, and 
take the maximal band. This uncertainty is then added in quadrature with those from the other $\mu$-scale variations. We believe that this is a reasonable estimate of the scale uncertainties.  As a reference, we will find that the 
scale uncertainties for our result are comparable to those for the fully exclusive NLL wino calculation in~\cite{Ovanesyan:2014fwa}, which also found 5\% perturbative uncertainty. When the $\cO(\alpha_W)$ fixed order corrections were added in~\cite{Ovanesyan:2016vkk} to obtain NLL' accuracy, the perturbative uncertainties shrank to $\sim 1\%$, and the central value was near the boundary of the $\sim 5\%$ NLL uncertainty bands. We expect the $\cO(\alpha_W)$ fixed order corrections to have a comparable effect in our calculation (some of these corrections, such as those to the hard function, are even identical).  This further supports that our estimate of the higher order perturbative error is reasonable.

\begin{figure}
\begin{center}
\includegraphics[width=0.65\columnwidth]{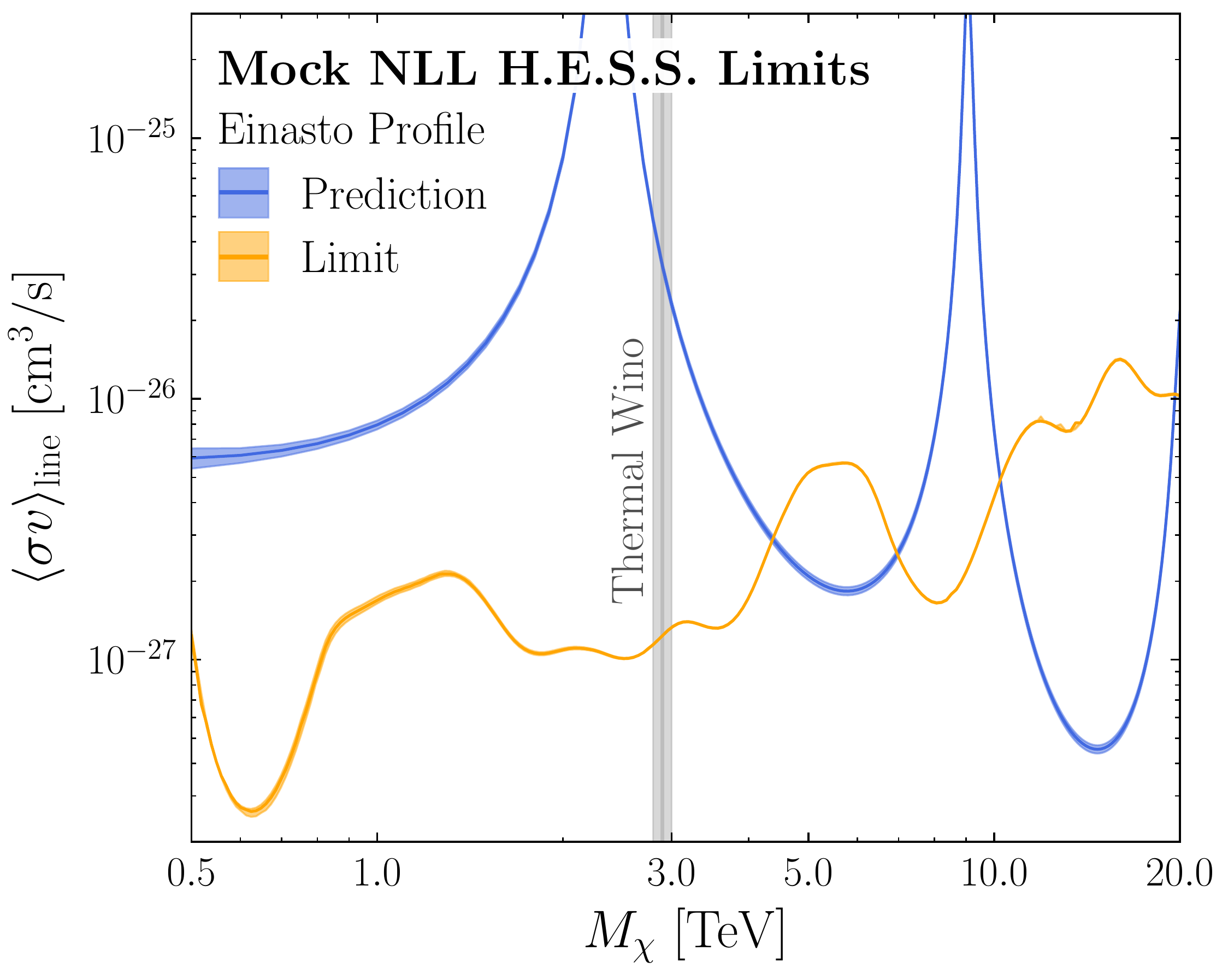}
\end{center}
\vspace{-15pt}
\caption{A mock limit on wino annihilation derived by reinterpreting the 2013 H.E.S.S. line search~\cite{Abramowski:2013ax}, see~\cite{Baumgart:2017nsr} for details of the procedure.  The comparison is made in terms of the line annihilation cross section $\langle \sigma v\rangle_\text{line} = \langle \sigma v \rangle_{\gamma \gamma} + \frac{1}{2} \langle \sigma v \rangle_{\gamma Z}$ (see text for details), as a function of the wino mass. Plotted here are the prediction (blue) and the mock limit (orange); the parameter space above the limit line is excluded assuming the Einasto DM profile. In both cases the bands represent the theoretical uncertainty associated with the NLL calculation.  The overall normalization error is captured by the prediction band.  For the mock limit, the uncertainty originates both from the variation in the shape of the endpoint spectrum and its normalization relative to the line. Finally, the thermal wino prediction $M_\chi = 2.9 \pm 0.1 \text{ TeV}$ is also shown. }
\label{fig:mockLimit}
\end{figure}

In \Fig{fig:diff} we show the cumulative (left column) and differential (right column) spectrum for wino annihilation for $M_\chi=3,10,\text{and } 35$ TeV (top, middle, and bottom rows respectively). This range of values was chosen since on the high end H.E.S.S. is currently probing DM masses up to $70$ TeV \cite{Abdallah:2018qtu}, while on the low end our EFT expansion breaks down as we approach $1$ TeV.  For $M_\chi \lesssim 1 \text{ TeV}$, one should smoothly match our EFT onto an EFT where $\mW\sim M_\chi (1-z)$. This would be particularly interesting to consider for the Higgsino, which we leave to future work. We emphasize that the value $3$ TeV (or more precisely $2.9$ TeV) is particularly motivated as this is the mass that corresponds to a thermal relic wino.  In all cases, we find that moving from LL to NLL yields a relatively small change in the central value.   What is particularly noteworthy is that NLL demonstrates a large reduction in the theoretical uncertainty as compared to LL.  With the NLL results in hand, the spectrum for heavy wino annihilation near the endpoint is now under excellent theoretical control.

The impact of this calculation is that it can be used to quantitatively explore constraints on the parameter space for winos from indirect detection.  To this end, we provide Fig.~\ref{fig:mockLimit}, which shows the results of the mock analysis developed in~\cite{Baumgart:2017nsr}, updated using our NLL prediction -- more discussion on the details of the analysis is provided in the next section.  For concreteness, the comparison between the mock exclusion and the theory prediction is made in terms of the line annihilation cross section
$\langle \sigma v\rangle_\text{line} = \langle \sigma v \rangle_{\gamma \gamma} + \frac{1}{2} \langle \sigma v \rangle_{\gamma Z}$.  We note that in order to convert from the endpoint cross section $\sigma^\text{NLL}$ computed here to $\sigma_\text{line}$, one must evaluate \Eq{eq:final_spectrum} in the limit $z\rightarrow 1$, and be careful to keep track of the fact that here we are computing the rate for $\gamma + X$, which introduces a factor of 2 in the conversion since both of the $\gamma$'s from $\sigma_{\gamma\gamma}$ contribute, \emph{i.e.},
\begin{align}
\lim\limits_{\zcut \rightarrow 1}\,\int\limits_{\zcut}^1\text{d}z \left(\frac{\text{d}\sigma}{\text{d}z}\right)^\text{NLL} = \sigma_{\rm exc}^{\rm NLL} =2\,\left(\sigma_{\gamma \gamma} + \frac{1}{2} \sigma_{\gamma Z}\right)=2\,\sigma_\text{line}\,\,,
\end{align}
and the approximation made here treats the kinematics for $\gamma\,\gamma$ and $\gamma\,Z$ identically, see~\cite{Baumgart:2017nsr, Rinchiuso:2018ajn} for additional discussion of this convention.  Note that these mock limits only include the contribution from the line and endpoint spectrum; the justification to neglect the contribution from continuum production resulting from wino annihilation to $W^+\,W^-$ at  lower masses was provided in~\cite{Baumgart:2017nsr}.  While we caution that a genuine analysis of the 2013 H.E.S.S. data should be done to provide an actual limit, we see that our mock limit shows that the thermal wino with mass $M_\chi = 2.9\text{ TeV}$ is excluded by a factor of $\sim25$.  We also emphasize that this assumes an Einasto DM profile, so that one way to avoid this seemingly stringent bound on wino annihilation is to core the profile, see \emph{e.g.}~\cite{Cohen:2013ama, Rinchiuso:2018ajn} for a discussion.  Importantly, the theory error band shown for the prediction in this plot is now under excellent control, justifying the need for our NLL calculation. For a more careful exploration of the implications of the NLL endpoint spectrum in the context of H.E.S.S. forecast limits, see~\cite{Rinchiuso:2018ajn}. 

 \begin{figure}
\begin{center}
\includegraphics[width=0.65\columnwidth]{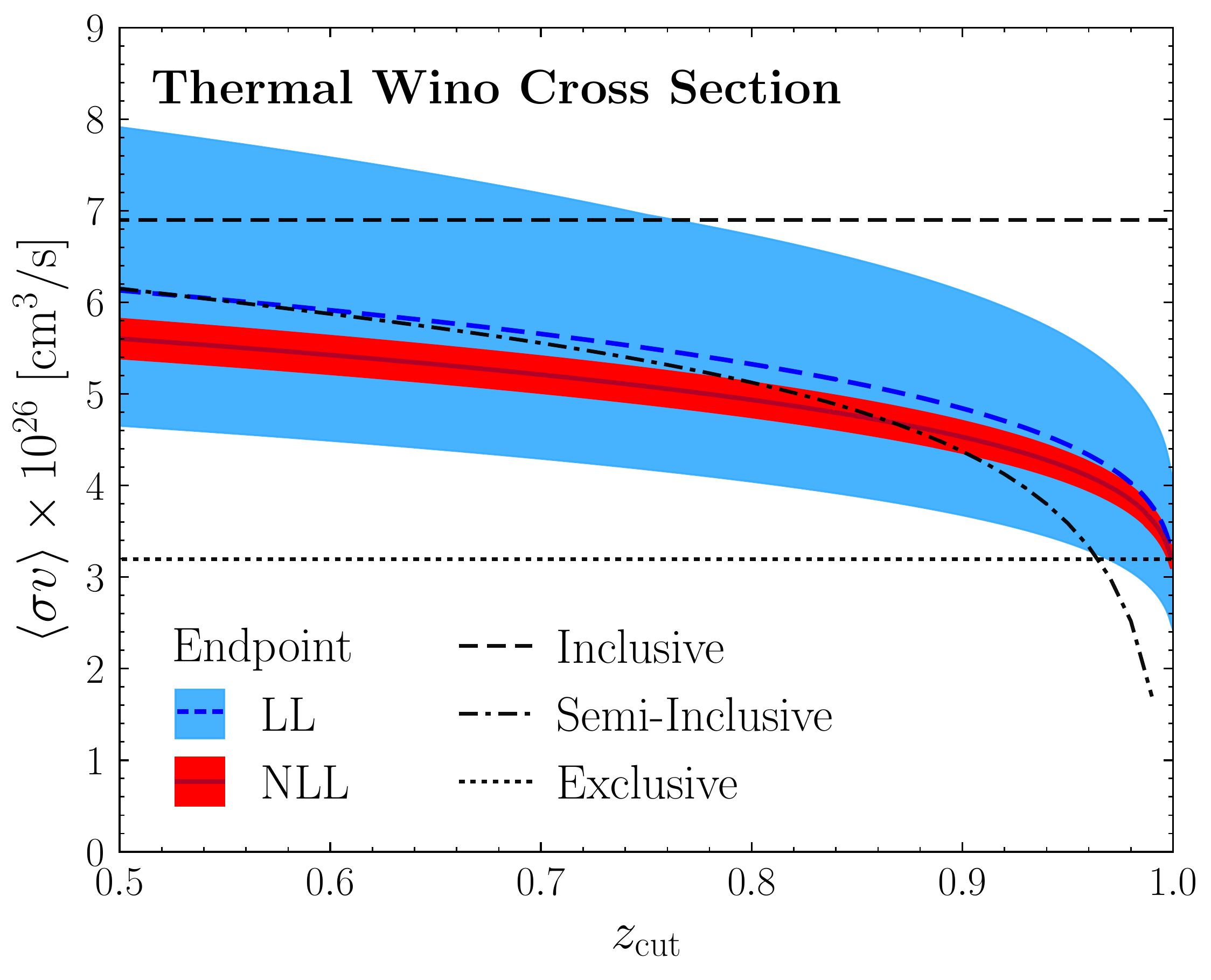}
\end{center}
\vspace{-15pt}
\caption{A comparison of our LL and NLL calculations with inclusive, exclusive and semi-inclusive predictions from the literature. The best agreement is found with the semi-inclusive calculation. The disagreement as $\zcut\to 1$ is due to unresummed logarithms of $(1-\zcut)$ in the semi-inclusive calculation which are correctly captured using our formalism.
}
\label{fig:intro_fig_later}
\end{figure}

Finally, in \Fig{fig:intro_fig_later} we show a comparison of our NLL cross section with several calculations that exist in the literature. In particular, we compare with the fully exclusive (line) calculation at NLL \cite{Ovanesyan:2014fwa,Ovanesyan:2016vkk}, the inclusive calculation at LL \cite{Baumgart:2014vma,Baumgart:2014saa}, and the semi-inclusive calculation at LL$^\prime$ \cite{Baumgart:2015bpa}. With the reduced NLL uncertainties, we see that for $\zcut\sim 0.8$-$0.9$, our prediction differs significantly from the exclusive and inclusive predictions, being approximately intermediate between the two, which individually each sum large $\log(M_\chi/\mW)$ logarithms at NLL order. As expected, the semi-inclusive provides a better approximation, agreeing with the shape and norm of the LL endpoint result away from $\zcut\to 1$. However, this calculation does not resum logarithms of $1-\zcut$, which become important as $\zcut\to 1$. The effect of these logarithms, which are properly captured in our result, is clearly seen by the fact that the semi-inclusive result rapidly diverges from our LL and NLL endpoint results as $\zcut\to 1$.

\section{Endpoint Spectrum Versus Fixed Bin Approach for Experiments} \label{sec:compare}

Obtaining a reliable theoretical interpretation of indirect detection line searches requires correctly incorporating experimental constraints into the underlying theoretical setup. This consideration has resulted in the appearance of a number of different approaches: fully inclusive~\cite{Baumgart:2014vma, Baumgart:2014saa}, fully exclusive~\cite{Bauer:2014ula, Ovanesyan:2014fwa, Ovanesyan:2016vkk}, semi-inclusive~\cite{Baumgart:2015bpa}, non-zero fixed bin width~\cite{Beneke:2018ssm}, and endpoint spectrum~\cite{Baumgart:2017nsr}.  Clearly, what is observed by the experiment is the true photon spectrum convolved with the appropriate instrument response functions. In our endpoint approach, this can be correctly treated since we have computed the full shape of the photon spectrum. With the NLL calculation presented here, we have a theoretically robust result with an estimated 5\% residual perturbative uncertainty.  

In this section we investigate how well the full result could be approximated by assuming that the resolution effects at an experiment such as H.E.S.S. are captured by integrating the photon spectrum over a bin with some effective width (with the bin maximum being $z=1$). This is important since it allows for an understanding of which approximations are valid for correctly describing the experimental setup. 
One motivation is that in certain cases determining the rate in a single bin near the endpoint is easier than achieving the full spectral shape. If this approach were demonstrated to be a good approximation, it could also pave the way for more calculational efficiency.

To address this question we compare the experimental constraints on the wino obtained using our full spectrum to those derived by rescaling constraints on a gamma-ray line. One might hope that for an appropriate choice of bin size, the true constraint would trace the line limits up to a simple rescaling. However, we will demonstrate that there does not appear to be any simple way to choose such a bin a priori.

In the interest of thoroughness, we will provide the results for two different analyses (for more details, see~\cite{Baumgart:2017nsr} and~\cite{Rinchiuso:2018ajn} respectively) 
\begin{enumerate}
\item {\bf Mock Analysis:} A simplified re-analysis of a 2013 H.E.S.S. result~\cite{Abramowski:2013ax}, following the methodology we developed previously~\cite{Baumgart:2017nsr}. We perform a $\chi^2$ analysis on the data, floating a seven-parameter background model in addition to the signal. The functional form of the background model was chosen~\cite{Abramowski:2013ax} to provide a good description of the data in the region of interest. We confirm that we approximately reproduce the quoted constraints on a pure line, and then apply the same analysis taking our full NLL spectrum as the signal. We estimate the energy resolution for this dataset based on interpolating the values given for photon energies of 0.5 and 10 TeV~\cite{Abramowski:2013ax}.  \label{item:MockAnalysis}
\item {\bf Forecast Analysis:} A detailed forecast of H.E.S.S. sensitivity, based on Monte Carlo simulations of the expected background, the instrument response functions, and the analysis pipeline~\cite{Rinchiuso:2018ajn}. We perform a likelihood analysis that incorporates both spatial and spectral information, for both signal and background. In this section we show results (reproduced from \cite{Rinchiuso:2018ajn}) assuming an Einasto profile for the DM density, although this method is generalizable to alternate DM density profiles. We compute the sensitivity to a pure line and compare that to the prediction of our NLL spectrum. This forecast includes cuts on the data that allow an effective energy resolution of $10\%$ independent of energy. 
\end{enumerate}

\begin{figure}
\begin{center}
\subfloat[]{\label{fig:bin_b}
\hspace{-3pt}\includegraphics[width=7.4cm]{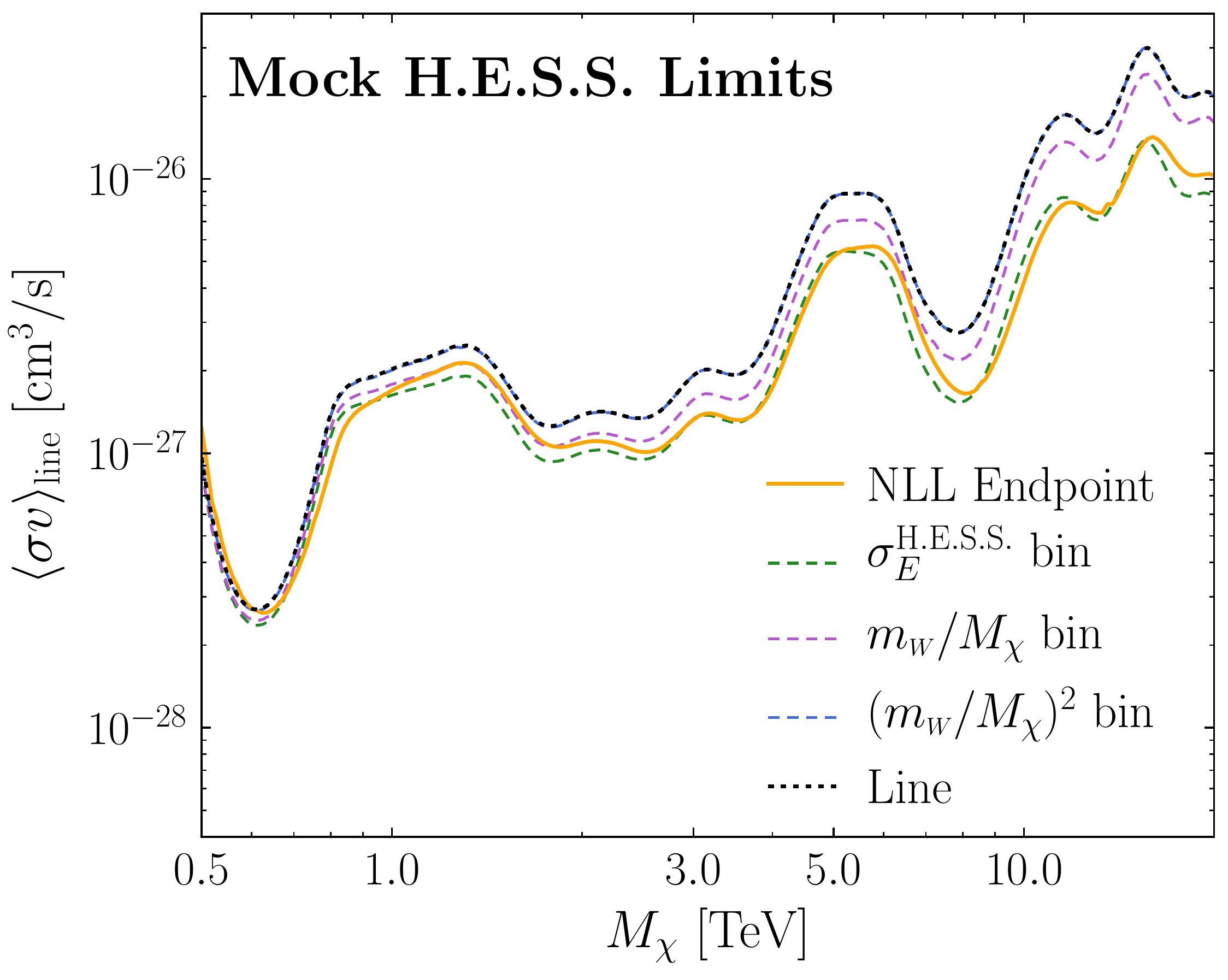}
}\hspace{5pt}
\subfloat[]{\label{fig:bin_b_H.E.S.S.}
\hspace{-5pt}\includegraphics[width=7.4cm]{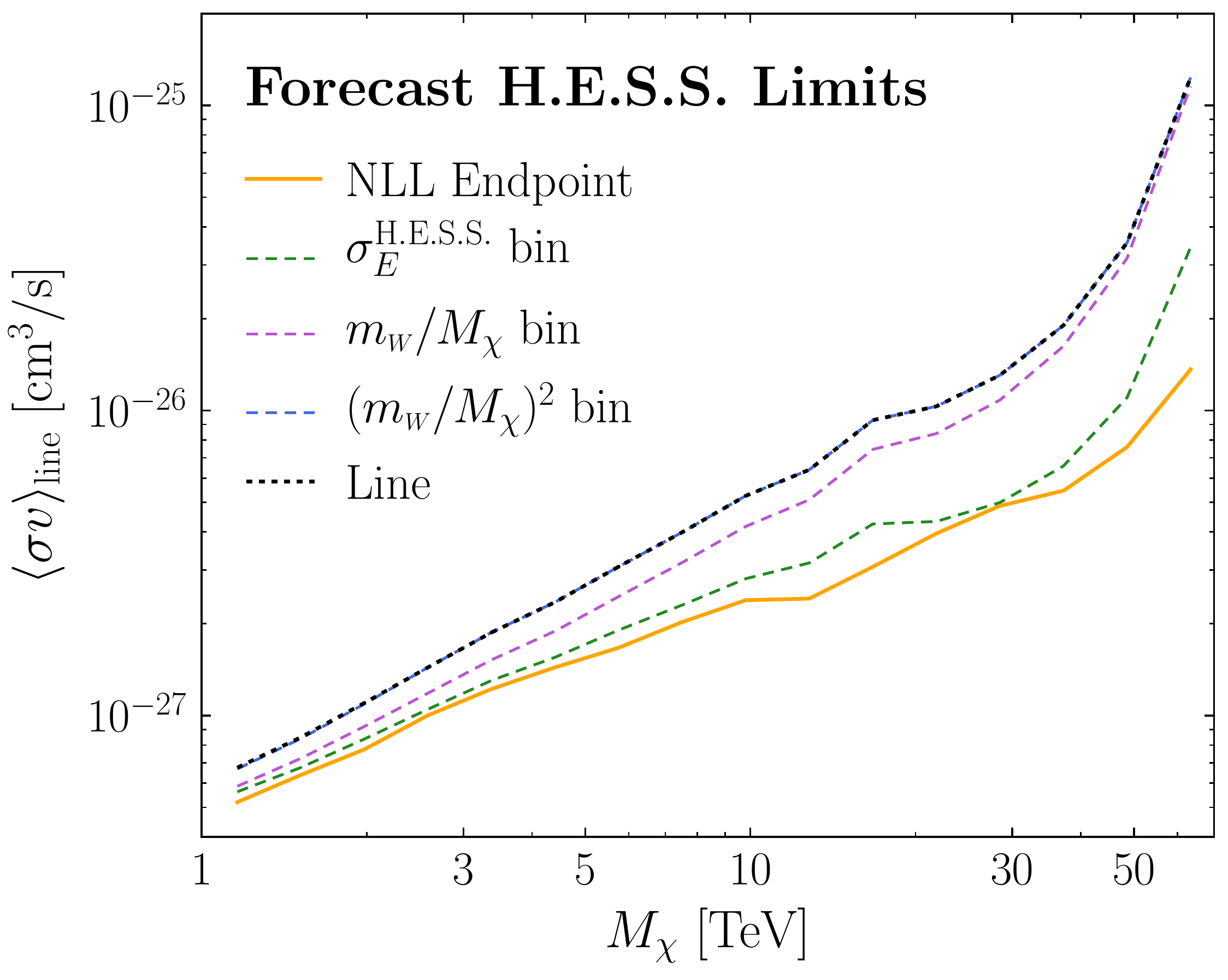}
}\\
\subfloat[]{\label{fig:bin_a}
\hspace{8pt}\includegraphics[width=7.0cm]{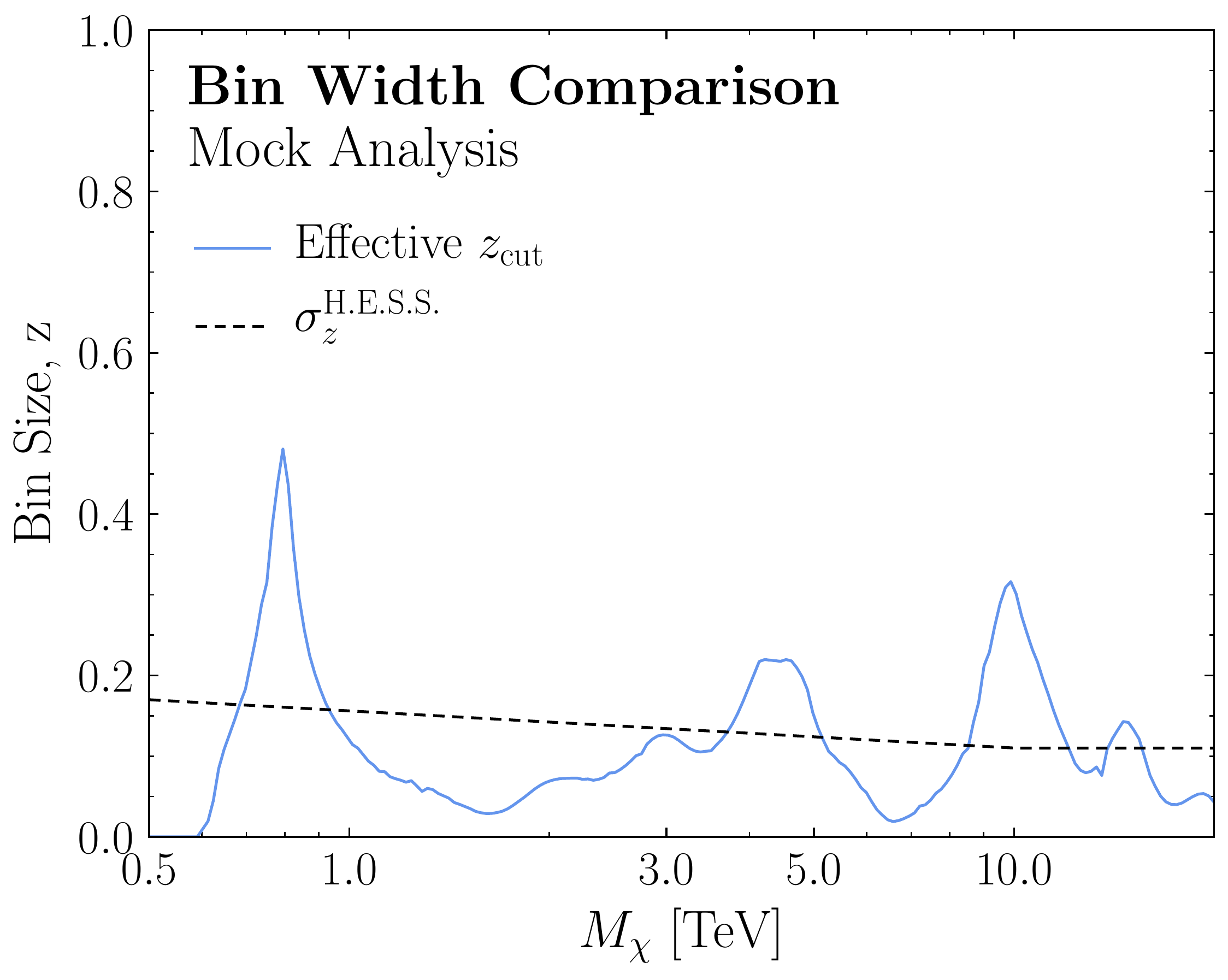}    
}
\subfloat[]{\label{fig:bin_a_H.E.S.S.}
\hspace{14pt}\includegraphics[width=7.0cm]{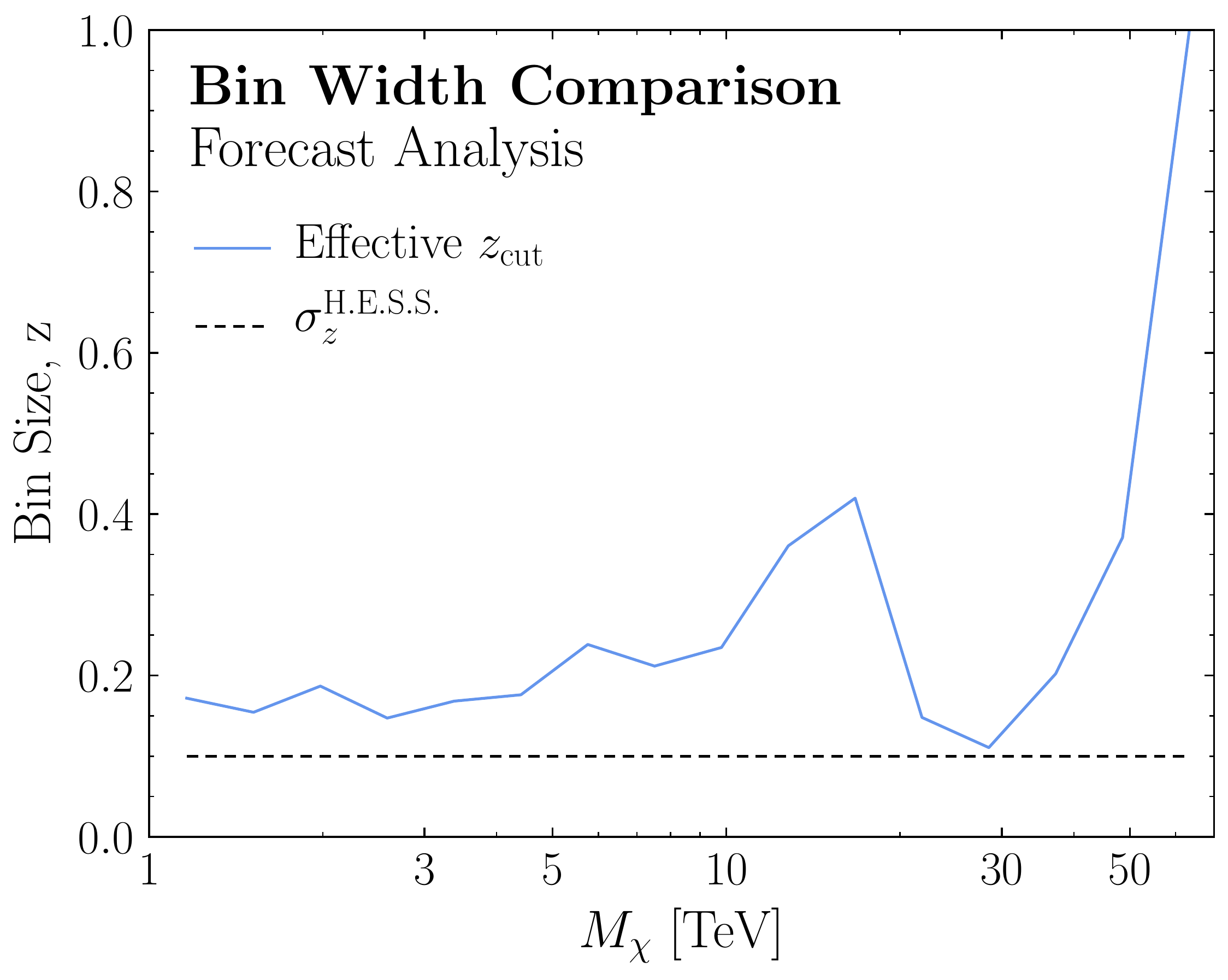}    
}
\end{center}
\vspace{-10pt}
\caption{(top)  Estimated limits obtained using a variety of different assumptions on the treatment of the spectrum for the (a) mock analysis and (b) forecast analysis.  The NLL endpoint curve relies on the computed spectrum, while all other curves rescale a line-only signal shape.  Note that the curves corresponding to ``Line'' and ``$(\mW/M_\chi)^2$'' lie on top of each other.
(bottom) The effective bin size required to reproduce the endpoint analysis, compared with the true H.E.S.S. resolution for the (c) mock analysis and (d) forecast analysis.  We see that in both approaches, the bin width required to map the line-only signal onto the limit derived for the full spectral shape is a non-trivial function of the DM mass.
}
\label{fig:bin_sizing}
\end{figure}

We begin by understanding the effect the various approximations would have on H.E.S.S. limits, in \Fig{fig:bin_b} (\Fig{fig:bin_b_H.E.S.S.}) we plot the mock (forecast) limits for $\langle \sigma v\rangle_\text{line}$, considering a number of different ways of modeling the experimental analysis given the theory calculation.  First, in solid orange we show the limit derived using the NLL prediction, which incorporates the full spectral shape. The black dotted line shows the limit assuming the pure line analysis using the resummed exclusive prediction for the annihilation rate. As was shown in~\cite{Baumgart:2017nsr}, the inclusion of the photon distribution near the endpoint almost always enhances the limits. Next we consider a number of different approximations for effective bin widths. In dashed green, we show what happens if instead of using the shape of the distribution, we take our NLL prediction for the spectrum and just assume a bin of width equal to the H.E.S.S. resolution (as appropriate for the calculation in question). Although this is not a terrible approximation, it is well outside our estimate of the theoretical uncertainty at NLL -- this emphasizes the need for a correct treatment of the spectrum when working at this accuracy.  Finally, we also consider the limits derived when taking a bin width of $\mW/M_\chi$ and $(\mW/M_\chi)^2$. These scales are natural from the EFT perspective, as they tie the bin width to the other physical scales in the problem, simplifying the calculation~\cite{Baumgart:2014saa,Baumgart:2014vma,Beneke:2018ssm}.  The appropriate EFT for a bin width $\mW/M_\chi$ uses the scaling $\mW\sim M_\chi (1-z)$, while the appropriate EFT for a bin width $(\mW/M_\chi)^2$ uses the scaling $\mW\sim M_\chi \sqrt{1-z}$. Relative to our EFT, these include additional power corrections. However, at the level of accuracy that we work, namely NLL, we do not include the matrix elements in the low energy theory,  only logarithms from evolution, and therefore we should correctly reproduce the large logarithms in these bins for the $\mW\sim M_\chi (1-z)$ case, which also bounds the $\mW\sim M_\chi \sqrt{1-z}$ case. Therefore, for simplicity, we do not perform dedicated analyses for these bins, and instead simply take our EFT result evaluated with bin widths of these values. We believe that the qualitative conclusions that we draw should be robust. Results for a resolution of $(\mW/M_\chi)^2$ are shown by the dashed blue curves, and for realistic H.E.S.S. resolutions we see that this bin size is too small (with our approach the dashed blue curves fully overlap the line curves, but will presumably differ somewhat in a full treatment of $\mW\sim M_\chi \sqrt{1-z}$ as in~\cite{Beneke:2018ssm}).  Finally we see that a EFT setup which utilizes a $\mW/M_\chi$ bin width gives the dashed purple curves, which lie in-between the line and endpoint results for $M_\chi\gtrsim 3\,{\rm TeV}$. These EFTs are perhaps more relevant for a DM mass of around $1$ TeV, \emph{e.g.} in the case of the Higgsino. However, the region of DM masses they can describe is likely to be rather limited since they are forced to have a resolution that scales like $1/M_\chi$. This is distinct from the behavior of the H.E.S.S. resolution, which is approximately flat with increasing DM mass.

Another aspect of \Fig{fig:bin_b} and \Fig{fig:bin_b_H.E.S.S.} that deserves comment is the fact that the mock limit and the forecast limit performances are similar for the important mass range near $M_\chi \sim 3 \text{ TeV}$, even though the forecast limit assumes a larger data set. This behavior occurs for two reasons. First, the forecast limit utilizes a more sophisticated background determination method, with less flexibility to incorrectly absorb a signal into the background model, and less sensitivity to fluctuations in the background determination. Consequently, estimating the relative strength of the expected limits in the two analyses is non-trivial. Additionally, the mock analysis is based on observed data, not expected data, and the 2013 dataset from which the mock limit was derived placed a stronger-than-expected upper limit on the flux near 3 TeV \cite{Abdallah:2018qtu}, likely due to a downward fluctuation in the background.

Another way of testing the fixed-bin approach is to explore the extent to which the H.E.S.S. endpoint analysis can be reproduced by using a single endpoint bin with an effective size, and compare that size with the H.E.S.S. resolution. To this end, we provide results computed by rescaling the limit on (or sensitivity estimate for) the line cross section, proportionally to the number of photons in the endpoint bin, and then determine what bin size is needed to reproduce the constraint (or sensitivity estimate) obtained by the endpoint analysis involving the full spectrum.

In \Fig{fig:bin_a}, we plot the effective bin size required to reproduce the constraint obtained by the endpoint analysis in the first calculation above, compared with the true H.E.S.S. resolution for the 2013 dataset. We see that the effective bin size varies non-trivially as a function of the DM mass, which could not be predicted without a full calculation of the shape of the distribution. Indeed, at low masses the limit with the full spectrum is weaker than the limit from the line only.  This is likely due to the non-trivial interplay between the signal and the flexible background model used in the mock analysis (based on that of Ref.~\cite{Abramowski:2013ax}). These results emphasize the need to convolve the NLL spectrum computed here with the experimental line shape and full background fit.  To further demonstrate this point, \Fig{fig:bin_a_H.E.S.S.} shows the same result for the sensitivity estimate from the second calculation described above, using state-of-the-art background modeling methods. The requisite effective bin size is somewhat more stable with mass in this case; sensitivity estimates do not include the statistical fluctuations inherent to real data, and the background modeling also has less freedom in this analysis. However, the bin size needed to match the predicted level of sensitivity again does not match the nominal H.E.S.S. energy resolution over the vast majority of DM masses, and shows non-trivial variation with DM mass.

One might also ask whether further improvements in the accuracy of the endpoint spectrum would allow corresponding improvements in the experimental constraints. For the H.E.S.S. experiment, however, the theoretical uncertainties at NLL are currently subdominant to the experimental systematic uncertainties.  In more detail, the Galactic Center (GC) region is a very crowded environment in very-high-energy (VHE) gamma rays. The determination of the residual background, mostly coming from misidentified CR hadrons, is challenging in the GC due to the presence of gamma-ray sources and regions of diffuse gamma-ray emission. A proven robust approach to probe signals from a cusped DM density distribution in the GC region consists of masking regions of the sky with known VHE emission, and then making use of the reflected background method where background and signal are measured in the same observational and instrumental conditions allowing for a precise determination of the background level \cite{Abdallah:2016ygi, Abdallah:2018qtu}. However, systematic uncertainties arise from the imperfect knowledge of the energy scale and energy resolution of the instrument. In addition, the Night Sky Background (NSB) rate -- corresponding to the unavoidable optical photon light emitted by bright stars in the field of view -- varies significantly over degree scales in the GC region. This variation induces a systematic uncertainty in the background determination when using the reflected background method. Propagating this uncertainty into the DM constraints implies a systematic uncertainty in the limit on the line annihilation cross section ranging from a few percent up to 60\%, depending on the DM mass \cite{Abdallah:2018qtu}, which dominates over the $\mathcal{O}(5\%)$ NLL theoretical uncertainties.

Future studies will make use of precise Monte Carlo simulations for the expected residual background determination in the GC, in the same observational and instrumental conditions as for the signal measurement. This could alleviate the level of systematic uncertainties substantially; for example, the inhomogeneous NSB could be accurately simulated in each pixel, allowing a careful subtraction of this component. The main remaining systematic uncertainty might then become the uncertainties in the energy scale and energy resolution of H.E.S.S.; a systematic uncertainty in the energy scale of 10\% shifts the limits by up to 15\%. The energy resolution uncertainties are a smaller effect; while the energy resolution is weakly dependent on the observational conditions, and this can impact the limit, a deterioration of a factor of two in the energy resolution only induces
a decrease of 25\% in the expected limit \cite{Abdallah:2018qtu}. Greater precision in the theoretical calculation thus might eventually become valuable from an experimental perspective, but would require multiple current sources of systematic uncertainty at the $10\%$ and higher level to be reduced below 5\% by improved analyses.

The results of this analysis demonstrate that to have a reliable theoretical interpretation of the experimental results requires a calculation of the full shape of the photon spectrum. Approximations using effective bin widths are simply not reliable at the level of accuracy of an NLL calculation.  In the event that multiple sources of experimental systematic error can be reduced in the future, extending the calculation presented here to NLL+NLO accuracy (or NNLL) could become necessary.

\section{Conclusions}\label{sec:conc}

In this paper, we have extended the calculation of the hard photon spectrum for wino annihilation in the endpoint region, as is relevant for indirect detection experiments, to NLL accuracy. This calculation was performed using an EFT framework developed in~\cite{Baumgart:2017nsr}, which facilitates the factorization of distinct physical effects. In particular, our result includes both the resummation of Sudakov logarithms and the Sommerfeld enhancement.
The theoretical uncertainties of our calculation are of the order of $5\%$, which is a significant reduction as compared with our earlier LL prediction.   In particular, the theory uncertainties are now sufficiently under control so as to make a subdominant contribution to the total uncertainty relevant for experimental exploration. 

In the course of our calculation we encountered a number of interesting effects associated with electroweak radiation and the presence of electroweak charged initial and final states.  For example, we found that the non-electroweak-gauge-singlet nature of the incoming and outgoing states led to a non-trivial remaining Glauber phase in the final cross section. We think this would be interesting to explore further in a more general context than that considered here. 

The H.E.S.S.~telescope has collected a large dataset of photons from the Galactic Center region, with an energy resolution of $\sim 10\%$, permitting sensitive searches for spectral features. Using both a mock H.E.S.S.~analysis and a detailed forecasting framework, we studied the importance of using the full photon spectrum computed in this paper as compared with a single endpoint bin approximation when computing experimental limits. We find that the mapping to an effective bin width is a highly non-trivial function of the DM mass. This emphasizes the importance of having theoretical control over the shape of the distribution in the endpoint region, and not simply the photon count in an endpoint bin, for deriving accurate limits from experimental data. 

With an understanding of our factorization formula at NLL accuracy, it is now straightforward to calculate the spectrum at this accuracy for other heavy WIMP candidates, such as the pure Higgsino, the mixed bino-wino-Higgsino, or the minimal DM quintuplet~\cite{Cirelli:2005uq, Cirelli:2007xd, Cirelli:2008id, Cirelli:2009uv, Cirelli:2015bda,Mitridate:2017izz}. This will allow for the robust theoretical interpretation of indirect detection constraints for these compelling DM candidates from the wealth of data at current and future experiments.

\begin{acknowledgments}

We are grateful to Mikhail Solon for his collaboration on the leading log result. We thank Torsten Bringmann, Camilo Garcia-Cely, and Graham Kribs for useful discussions. Several Feynman diagrams were drawn using~\cite{Ellis:2016jkw}.
MB is supported by the U.S. Department of Energy, under grant number DE-SC-0000232627.
TC is supported by the U.S. Department of Energy, under grant number DE-SC0018191 and DE-SC0011640. 
IM is supported by the U.S. Department of Energy, under grant number DE-AC02-05CH11231.
NLR and TRS are supported by the U.S. Department of Energy, under grant numbers DE-SC00012567 and DE-SC0013999.
NLR is further supported by the Miller Institute for Basic Research in Science at the University of California, Berkeley.
IWS is supported by the Office of Nuclear Physics of the U.S. Department of Energy under the Grant No. DE-SCD011090 and by the Simons Foundation through the Investigator grant 327942.
VV is supported by the Office of Nuclear Physics of the U.S. Department of Energy under the Grant No.  Contract DE-AC52-06NA25396 and through the LANL LDRD Program.

\end{acknowledgments}

\appendix
\addcontentsline{toc}{section}{\protect\numberline{}Appendices}%
\addtocontents{toc}{\protect\setcounter{tocdepth}{1}}
\section*{Appendix}
\section{Cross Section with Scale Dependence}
\label{app:explicitxsec}

In the main text we provided expressions for the cumulative cross section \Eq{eq:final_cumulant} and the differential cross section \Eq{eq:final_spectrum}.  For brevity, the scales have been set to their canonical values in these formulas, which causes many logarithms to vanish and reduces the complexity of the expressions dramatically.  This is all one needs to compute the central value of the NLL prediction.  However, since our goal is to also demonstrate that the uncertainties due to scale variation are well under control, it is important to provide the explicit results leaving the hard scale $\mu_H^0$, photon jet scale $\mu_\gamma^0$, recoiling jet scale $\mu_J^0$, and soft scale $\mu_S^0$ as parameters that can be varied. Note that since our EFT is no longer meaningful at scales below $\mW$, if the argument of any function of scale falls below $\mW$, we simply freeze that function to its value at $\mW$ in our numerical evaluations. For example, this is relevant for the $r_S$, $r_J$, and $r_{HS}$ functions defined below. 

We begin by defining a variety of functions that will be used to build both the cumulative and differential cross sections.  We will use the notation $\TaW=\aW/(4\,\pi)$, as defined in \Eq{eq:TaW} above, and will write the results in terms of the cusp anomalous dimension and the $\beta$ function, as defined in \Eqs{eq:cusp}{eq:betafunction}.
The {\bf hard} function is expressed in terms of (see Sec.~\ref{sec:diag_hard}) 
\begin{align}
U_H = &\,r_H^2 \left( \frac{\big(\mu_H^0\big)^2}{\big( 2\, M_{\chi} \big)^2} \right)^{\omega_H} \exp \left[ -\frac{2\, \Gamma_0}{\beta_0^2} \left[ \frac{1}{\TaW\big(\mu_H^0\big)} \left( \log r_H + \frac{1}{r_H} - 1 \right) \right. \right.\nn\\[5pt]
&\hspace{2cm}\left. \left.+ \left( \frac{\Gamma_1}{\Gamma_0} - \frac{\beta_1}{\beta_0} \right) \left( r_H - 1 - \log r_H \right) - \frac{\beta_1}{2\,\beta_0} \log^2 r_H \right] \right]\,,\\[5pt]
C_H =&\,  1 - \TaW \big(\mu_H^0\big)\, \Gamma_0\, \log^2 \left( \frac{\big(\mu_H^0\big)^2}{\big( 2\, M_{\chi} \big)^2} \right) \,,
\end{align}
where
\begin{align}
\omega_H = &\frac{2\,\Gamma_0}{\beta_0} \left[ \log r_H + \TaW\big(\mu_H^0\big) \left( \frac{\Gamma_1}{\Gamma_0} - \frac{\beta_1}{\beta_0} \right) (r_H-1) \right] \,, \nn \\[5pt]
r_H = &\frac{\TaW(\mW)}{\TaW\big(\mu_H^0\big)} = \left[ 1 + 2\, \TaW\big(\mu_H^0\big)\, \beta_0\, \log \left( \frac{\mW}{\mu_H^0} \right) \right]^{-1}\,.
\end{align}
We will also use the notation
\begin{align}
c_H = \cos \left( \frac{6\, \pi}{\beta_0} \log r_H\right)\,,\;\;
s_H = \sin \left( \frac{6\, \pi}{\beta_0} \log r_H\right)\,.
\end{align}
The {\bf photon jet} function is given by (see Sec.~\ref{sec:photon_jet})
\begin{align}
C_{\gamma} = 1 + 4\, \Gamma_0\, \TaW(\mu_{\gamma}^0)\, \log \left( \frac{\mu_{\gamma}^0}{\mW} \right) \log \left( \frac{\nu}{2\, E_{\gamma}} \right)\,.
\end{align}
The {\bf recoiling jet} evolution is built using (see Sec.~\ref{sec:jet_evol})
\begin{align}
V_J &= \exp \left\{ \frac{2\, \Gamma_0}{\beta_0^2} \left[ \frac{1}{\TaW(\mu_J^0)} \left( \log r_J + \frac{1}{r_J} - 1 \right) + \left( \frac{\Gamma_1}{\Gamma_0} - \frac{\beta_1}{\beta_0} \right) \left( r_J - 1 - \log r_J \right) - \frac{\beta_1}{2\,\beta_0} \log^2 r_J \right] \right. \nn \\[2pt]
&\hspace{45pt}\left.- \log r_J \vphantom{\frac{2 \Gamma_0}{\beta_0^2}} \right\}\,, \nn \\[5pt]
\omega_J &=- \frac{2\,\Gamma_0}{\beta_0} \left[ \log r_J + \TaW\big(\mu_J^0\big) \left( \frac{\Gamma_1}{\Gamma_0} - \frac{\beta_1}{\beta_0} \right) (r_J-1) \right] \Theta_J\,, \nn \\[5pt]
r_J &=  \frac{\TaW(\mW)}{\TaW\big(\mu_J^0\big)} = \left[ 1 + 2\, \TaW\big(\mu_J^0\big)\, \beta_0\, \log \left( \frac{\mW}{\mu_J^0} \right) \right]^{-1}\,, \nn \\[5pt]
\Theta_J &=  \Theta \Big( \mu_J^0 - \mW \Big)\,.
\end{align}
The {\bf soft} matching coefficient evolution is built using (see Sec.~\ref{sec:diag_hardsoft})
\begin{align}
V_S &= \exp \left\{ - \frac{3\,\Gamma_0}{2\,\beta_0^2} \left[ \frac{1}{\TaW\big(\mu_S^0\big)} \left( \log r_S + \frac{1}{r_S} - 1 \right) + \left( \frac{\Gamma_1}{\Gamma_0} - \frac{\beta_1}{\beta_0} \right) \left( r_S - 1 - \log r_S \right) - \frac{\beta_1}{2\,\beta_0} \log^2 r_S \right] \right\}\,, \nn\\[5pt]
\omega_S &= \frac{3\,\Gamma_0}{2\,\beta_0} \left[ \log r_S + \TaW\big(\mu_S^0\big) \left( \frac{\Gamma_1}{\Gamma_0} - \frac{\beta_1}{\beta_0} \right) (r_S-1) \right] \Theta_S\,, \nn \\[5pt]
r_S &= \frac{\TaW(\mW)}{\TaW(\mu_S^0)} = \left[ 1 + 2\, \TaW\big(\mu_S^0\big)\, \beta_0 \,\log \left( \frac{\mW}{\mu_S^0} \right) \right]^{-1}\,, \nn \\[5pt]
\Theta_S &= \Theta \Big( \mu_S^0 - \mW \Big)\,.
\end{align}
Next we will state the functions that result from refactorizing the soft sector, solving the RGEs from $\mu_S^0$ to $\mu_S(s)$, and rediagonalizing after evolution.   Since the evolution is performed in Laplace space, the transformation back introduces a different set of functions that are relevant for the cumulative cross section than for the differential cross section -- therefore, we mark the functions with a ``$c\,$'' or a ``$d\,$'' subscript if they are relevant for the cumulative or the differential expressions respectively.\footnote{Note that we did not discuss this distinction in the main text because we did not explicitly evaluate the derivatives, which were left acting on different functions.}  We start with those that are relevant for the {\bf cumulative} cross section,
\begin{align}\label{eq:Lambdac}
\Lambda^a_c & = 1 + \TaW\big(\mu_J^0\big)\, \Gamma_0\, \Delta_{JSJ}^{c,(2)}\, \Theta_J + \TaW\big(\mu_J^0\big) \,\beta_0\, \Delta_{JS}^{c,(1)} \,\Theta_J - 12\, \TaW\,\big(\mu_S^0\big) \Big(\Delta_{JSS}^{c,(2)} - \Delta_{JS}^{c,(1)}\Big) \,\Theta_S\,, \nn \\
\Lambda^b_c & = 1 + \TaW\,\big(\mu_J^0\big)\, \Gamma_0\, \Delta_{JSJ}^{c,(2)}\, \Theta_J + \TaW\big(\mu_J^0\big)\, \beta_0\, \Delta_{JS}^{c,(1)}\, \Theta_J - 12\, \TaW\,\big(\mu_S^0\big) \,\Delta_{JSS}^{c,(2)}\, \Theta_S\,, \nn \\
\Lambda^c_c & = 1 + \TaW\,\big(\mu_J^0\big)\, \Gamma_0\, \Delta_J^{c,(2)}\, \Theta_J + \TaW\big(\mu_J^0\big)\, \beta_0\, \Delta_J^{c,(1)}\, \Theta_J + 24\, \TaW(\mu_S^0)\, \Delta_J^{c,(1)}\, \Theta_S\,, \nn \\
\Lambda^d_c & = 1 + \TaW\big(\mu_J^0\big)\, \Gamma_0\, \Delta_J^{c,(2)}\, \Theta_J + \TaW\big(\mu_J^0\big)\, \beta_0\, \Delta_J^{c,(1)}\, \Theta_J\,, 
\end{align}
see \Eq{eq:CapitalLambda}, and are built using 
\begin{align}
\Delta_J^{c,(1)} &= \gamma_E + \psi^{(0)}\big(1-\omega_J\big)\,, \nn \\
\Delta_J^{c,(2)} &= \left(\gamma_E + 2\, \log \left( \frac{\mu_J^0}{2\,M_{\chi}\,\sqrt{1-z}} \right) + \psi^{(0)}\big(1-\omega_J\big) \right)^2 - \psi^{(1)}\big(1-\omega_J\big)\,, \nn \\
\Delta_{JS}^{c,(1)} &= \gamma_E + \psi^{(0)}\big(1-\omega_J-2\,\omega_S\big)\,, \nn \\
\Delta_{JSJ}^{c,(2)} &= \left(\gamma_E + 2\, \log \left( \frac{\mu_J^0}{2\,M_{\chi}\,\sqrt{1-z}} \right) + \psi^{(0)}\big(1-\omega_J-2\,\omega_S\big) \right)^2 - \psi^{(1)}\big(1-\omega_J-2\,\omega_S\big)\,, \nn \\
\Delta_{JSS}^{c,(2)} &= \left(\gamma_E + \log \left( \frac{\mu_S^0}{2\,M_{\chi}\,(1-z)} \right) + \psi^{(0)}\big(1-\omega_J-2\,\omega_S\big) \right)^2 - \psi^{(1)}\big(1-\omega_J-2\,\omega_S\big)\,,
\end{align}
where $\psi^{(m)}$ is the polygamma function of order $m$.  For the {\bf differential} cross section, we have an associated $\Lambda^{a-d}_d$, which are identical to \Eq{eq:Lambdac}, but with the derivative substitutions replaced by
\begin{align}
\Delta_J^{d,(1)} &= \gamma_E + \psi^{(0)}\big(-\omega_J\big)\,, \nn \\
\Delta_J^{d,(2)} &= \left(\gamma_E + 2\, \log \left( \frac{\mu_J^0}{2\,M_{\chi}\,\sqrt{1-z}} \right) + \psi^{(0)}\big(-\omega_J\big) \right)^2 - \psi^{(1)}\big(-\omega_J\big)\,, \nn \\
\Delta_{JS}^{d,(1)} &= \gamma_E + \psi^{(0)}\big(-\omega_J-2\,\omega_S\big)\,, \nn \\
\Delta_{JSJ}^{d,(2)} &= \left(\gamma_E + 2\, \log \left( \frac{\mu_J^0}{2\,M_{\chi}\,\sqrt{1-z}} \right) + \psi^{(0)}\big(-\omega_J-2\,\omega_S\big) \right)^2 - \psi^{(1)}\big(-\omega_J-2\,\omega_S\big)\,, \nn \\
\Delta_{JSS}^{d,(2)} &= \left(\gamma_E + \log \left( \frac{\mu_S^0}{2\,M_{\chi}\,(1-z)} \right) + \psi^{(0)}\big(-\omega_J-2\,\omega_S\big) \right)^2 - \psi^{(1)}\big(-\omega_J-2\,\omega_S\big)\,.
\end{align}
Finally, the refactorization of the soft sector will introduce the ratio $r_H/r_S$, so we use the following notation to simplify expressions
\begin{align}
r_{HS} = \frac{r_H}{r_S}\,.
\end{align}
Now we have introduced all the ingredients relevant to give the cumulative cross section with scales not fixed at canonical values,
\begin{align}
\sigma^{\text{NLL}}(\zcut)\,&\!= \frac{\pi\, \aW^2\big(\mu_H^0\big) \sW^2\big(\mu_{\gamma}^0\big)}{9\, M_{\chi}^2\, v} C_{\gamma}\,C_H\,U_H\,\big((V_J-1)\, \Theta_J+1\big)\nn \\*
&\bigg \{\bigg(\hspace{5pt} \big|s_{00}\big|^2  \Big[ 
4\, \Lambda^d_c + 2\, r_{HS}^{12/\beta_0} \Lambda^c_c \Big] + \big|s_{0\pm}\big|^2 \bigg[ 
8\, \Lambda^d_c + r_{HS}^{12/\beta_0} \Lambda^c_c \bigg] \nn \\*
&\hspace{22pt}+ \sqrt{2} \operatorname{Re}\!\Big[s_{00} \,s_{0\pm}^*\Big]\bigg[ 
8\, \Lambda^d_c - 2\,r_{HS}^{12/\beta_0} \Lambda^c_c \bigg] \bigg) \frac{1}{\Gamma(1-\omega_J)} \left( \frac{\big(\mu_J^0\big)^2 e^{\gamma_E}}{4\, M_{\chi}^2\, (1-z)} \right)^{\omega_J} \nn \\
&\hspace{8pt}+ \big((V_S-1)\, \Theta_S+1\big) \, r_H^{6/\beta_0} \nn \\
&\hspace{22pt}\bigg(\hspace{5pt} \big|s_{00}\big|^2  \Big[2 \,r_{HS}^{6/\beta_0} \Lambda^a_c - 8\, c_H\, \Lambda^b_c \Big] 
+ \big|s_{0\pm}\big|^2 \bigg[ r_{HS}^{6/\beta_0}\, \Lambda^a_c+8\, c_H\, \Lambda^b_c \bigg]  \nn\\
&\hspace{32pt}+ \sqrt{2} \operatorname{Re}\!\Big[s_{00}\, s_{0\pm}^*\Big]\bigg[ - 2\,r_{HS}^{6/\beta_0}\, \Lambda^a_c -4\, c_H\, \Lambda^b_c \bigg] + \sqrt{2} \operatorname{Im}\!\Big[s_{00}\, s_{0\pm}^*\Big]\bigg[ -12\, s_H\, \Lambda^b_c \bigg] \bigg) \nn \\
&\hspace{32pt}\frac{1}{\Gamma(1-\omega_J-2\,\omega_S)} \left( \frac{\big(\mu_J^0\big)^2 e^{\gamma_E}}{4\, M_{\chi}^2\, (1-z)} \right)^{\omega_J}
\left( \frac{\mu_S^0\, e^{\gamma_E}}{2\, M_{\chi}\, (1-z)} \right)^{2\, \omega_S}
\bigg \}\,\Bigg|_{z\,\rightarrow\, \zcut}\,,
\end{align}
\noindent and the corresponding differential cross section
\begin{align}
\left(\frac{\text{d}\sigma}{\text{d}z}\right)^{\text{NLL}}& = \frac{\pi\, \aW^2\big(\mu_H^0\big)\, \sW^2\big(\mu_{\gamma}^0\big)}{9\, M_{\chi}^2\, v\,(1-z)}\, C_{\gamma}\,C_H\, U_H\big((V_J-1) \,\Theta_J+1\big)\nn \\*
&\Bigg \{\bigg(\hspace{5pt} \big|s_{00}\big|^2  \Big[ 
4\, \Lambda^d_d + 2\, r_{HS}^{12/\beta_0} \Lambda^c_d \Big] 
+ \big|s_{0\pm}\big|^2 \bigg[ 
8\, \Lambda^d_d + r_{HS}^{12/\beta_0} \Lambda^c_d \bigg] \nn \\*
&\hspace{22pt}+ \sqrt{2} \operatorname{Re}\!\Big[s_{00} s_{0\pm}^*\Big]\bigg[ 8\, \Lambda^d_d - 2\,r_{HS}^{12/\beta_0} \Lambda^c_d \bigg] \bigg) \frac{1}{\Gamma(-\omega_J)} 
\left( \frac{\big(\mu_J^0\big)^2 e^{\gamma_E}}{4\, M_{\chi}^2\, (1-z)} \right)^{\omega_J} \nn \\
&\hspace{8pt}+ \big((V_S-1) \,\Theta_S+1\big) \, r_H^{6/\beta_0} \nn \\
&\hspace{22pt}\bigg(\hspace{5pt} \big|s_{00}\big|^2  \Big[ 
2 \,r_{HS}^{6/\beta_0} \Lambda^a_d - 8\, c_H\, \Lambda^b_d \Big] 
+ \big|s_{0\pm}\big|^2 \bigg[ r_{HS}^{6/\beta_0}\, \Lambda^a_d
+8\, c_H\, \Lambda^b_d \bigg]  \nn\\
&\hspace{32pt}+ \sqrt{2} \operatorname{Re}\!\Big[s_{00} s_{0\pm}^*\Big]\bigg[ 
- 2\,r_{HS}^{6/\beta_0}\, \Lambda^a_d -4\, c_H\, \Lambda^b_d \bigg] \nn \\
&\hspace{32pt}+ \sqrt{2} \operatorname{Im}\!\Big[s_{00} s_{0\pm}^*\Big]\bigg[ -12\, s_H\, \Lambda^b_d \bigg] \bigg) \nn\\
&\hspace{32pt}\frac{1}{\Gamma(-\omega_J-2\,\omega_S)}
\left( \frac{\big(\mu_J^0\big)^2 e^{\gamma_E}}{4\, M_{\chi}^2\, (1-z)} \right)^{\omega_J}
\left( \frac{\mu_S^0\, e^{\gamma_E}}{2\, M_{\chi}\, (1-z)} \right)^{2\, \omega_S}
\Bigg \}\nn \\[5pt]
&+ \sigma_{\rm exc}^{\rm NLL} \delta(1-z)\,,
\end{align}
where
\begin{align}
\sigma_{\rm exc}^{\rm NLL} = &\,\frac{\pi\, \aW^2\big(2\,M_{\chi}\big) \sW^2\big(\mW\big)}{9\, M_{\chi}^2\, v}\, C_H\,U_H \nn \\[3pt]
&\times \bigg\{ \left[ 4 + 4\, r_H^{12/\beta_0} - 8\, r_H^{6/\beta_0}  c_H \right] \big|s_{00}\big|^2
+ \left[ 8 + 2\, r_H^{12/\beta_0} + 8\, r_H^{6/\beta_0}  c_H \right] \big|s_{0\pm}\big|^2 \nn \\
&+ \sqrt{2} \left[ 8 - 4\, r_H^{12/\beta_0} - 4\, r_H^{6/\beta_0}  c_H  \right] \operatorname{Re}\!\Big[s_{00}\, s_{0\pm}^*\Big] 
- 12 \sqrt{2}\, r_H^{6/\beta_0}  s_H \operatorname{Im}\!\Big[s_{00}\, s_{0\pm}^*\Big] \bigg\}\,.
\end{align}
The Sommerfeld factors are defined in \Eq{eq:wavefunction}, $\aW\big(\mu_H^0\big)$ is the weak fine structure constant evaluated at the hard scale, $\sW^2\big(\mu_\gamma^0\big)$ is sine squared of the weak mixing angle evaluated at the photon jet scale.  For concreteness, we use $1/\aW\big(m_{\scriptscriptstyle Z}\big) = 127.944$, and $\sW^2\big(m_{\scriptscriptstyle Z}\big) = 0.23126$~\cite{Patrignani:2016xqp}, and the running away from the canonical scale is computed using expressions from App.~D of~\cite{Ovanesyan:2016vkk}.

\section{Fixed Order Expansion}\label{app:intbrem}

In this Appendix, we show that expanding our resummed result in \Eq{eq:final_spectrum} to fixed order exactly reproduces the full NLO result of \cite{Bergstrom:2005ss} when expanded in the $z\to 1$ limit. This provides a highly non-trivial cross check of our formalism. To begin, we note that~\cite{Bergstrom:2005ss} did not include Sommerfeld enhancement -- this implies $s_{00}=1$ and $s_{0\pm}=0$ for the comparison. Next, we expand the endpoint contribution to \Eq{eq:final_spectrum} to leading order in $\aW$ and drop the $\Theta$-functions, which yields
\begin{align}
\frac{1}{\langle \sigma v \rangle_{\rm tree}}\left(\frac{\text{d} \langle \sigma v \rangle}{\text{d}z}\right)^{\text{NLL}} = \frac{\aW}{\pi} \left[ \frac{4}{1-z} \log \left( \frac{2\, M_{\chi}\,(1-z)}{\mW} \right) - \frac{2}{1-z} \right] + \mathcal{O}(\aW^2)\,,
\label{eq:NLLtoLO}
\end{align}
where the tree level annihilation cross section is
\be
\langle \sigma v \rangle_{\rm tree} = \frac{\pi\, \aW^2\big(2\, M_{\chi}\big) \sW^2\big(\mW\big)}{M_{\chi}^2}\,.
\ee

In order to compare this to the fixed order result, we expand Eq.~(3) of~\cite{Bergstrom:2005ss} to leading power in $1-z$,
\be\label{eq:FOtoLP}
\frac{1}{\langle \sigma v \rangle_{\rm tree}}\left(\frac{\text{d} \langle \sigma v \rangle}{\text{d}z}\right)^{\text{FO}} = \frac{\aW}{\pi} \left[ \frac{4}{1-z} \log \left( \frac{2\, M_{\chi}\,(1-z)}{\mW} \right) - \frac{2}{1-z} \right] + \mathcal{O}\Big((1-z)^0\Big)\,.
\ee
This result is clearly in exact agreement with \Eq{eq:NLLtoLO}, providing an explicit verification that our formalism captures all relevant leading power contributions in the endpoint. Note in~\cite{Baumgart:2017nsr} we performed a similar expansion of our LL result, which produced only the leading logarithmic term in \Eq{eq:NLLtoLO}.

A particular contribution to the dark matter photon spectrum that is often discussed in the literature is the so called ``internal bremsstrahlung'' process, where a photon is emitted off of an internal charged line, see \emph{e.g.}~\cite{Bergstrom:2005ss,Bringmann:2007nk,Cannoni:2010my,Ciafaloni:2010ti,Bringmann:2012vr,Garcia-Cely:2013zga,Garny:2013ama,Giacchino:2013bta,Ibarra:2014qma,Garcia-Cely:2015dda,Garcia-Cely:2015khw,Bambhaniya:2016cpr}. This contribution has, for example, motivated a search using the {\it Fermi} gamma-ray data~\cite{Bringmann:2012vr}. In particular, the fact that our result reproduces the full NLO result in the endpoint region implies that it correctly captures the internal bremsstrahlung diagrams. Since there is a clear interest in this effect, we feel it is worth elaborating briefly on how it is reproduced in our formalism. 

\begin{figure}[t!]
\begin{center}
\includegraphics[width=\columnwidth]{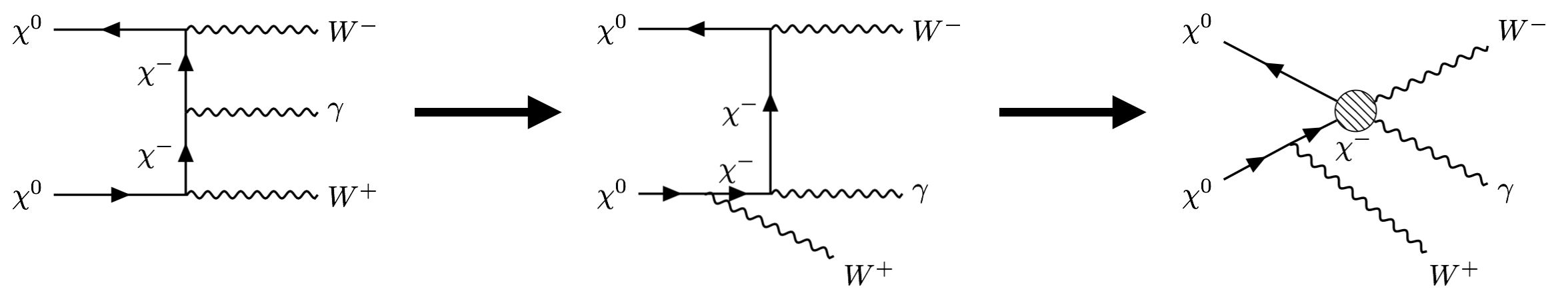}
\end{center}
\vspace{-15pt}
\caption{
A visualization of one example for how the internal bremsstrahlung diagram is captured within the EFT framework. Once the diagram on the left has been rearranged as it appears in the center, and taking the soft kinematics for the $W^+$, this process amounts to a $2 \to 2$ hard scattering accompanied by an additional $W$ emission as shown on the right.
}
\label{fig:IB}
\end{figure}

An illustrative internal bremsstrahlung diagram in the full theory is shown in the left of Fig.~\ref{fig:IB}. In general, to understand how this diagram is realized in the EFT, one must consider expanding it in all possible momentum regions. In the endpoint region, we have shown that all particles must be either collinear (with respect to the photon or the recoiling jet) or soft. For concreteness, we consider the soft limit. Since the diagram is symmetric in the $W^+$ and $W^-$, we take the $W^+$ to be soft. It is then useful to visualize the diagram as in the central panel of Fig.~\ref{fig:IB}, since the $W^+$, being soft, does not cause the wino DM to recoil. In our EFT setup, this particular diagram is then matched onto the local hard scattering operator shown in the right panel of Fig.~\ref{fig:IB}, and the soft $W^+$ emission is described by the Wilson line along the direction of the annihilating DM particle. The strength of the EFT is that this can be extended to an arbitrary number of emissions, giving rise to the schematic picture of \Fig{fig:csoft_intro_a}. This same logic can be applied to other full theory diagrams, and we leave it to the reader to make contact diagram-by-diagram with our formalism.

\addcontentsline{toc}{section}{\protect\numberline{}References}%
\bibliography{HDMA}
\bibliographystyle{JHEP}

\end{document}